\documentclass[12pt]{article}
\usepackage{latexsym}
\usepackage{amsmath}
\usepackage{epsfig,graphics}

\newcommand{\be}{\begin{equation}}
\newcommand{\ee}{\end{equation}}

\newlength{\figsize}
\begin{document}

\begin{titlepage}

\vspace*{0.7in}
 
\begin{center}
{\large\bf On the running of the bare coupling 
in SU(N) \\ lattice gauge theories \\}
\vspace*{1.0in}
{Chris Allton$^{a}$, Michael Teper$^{b}$ and  Aurora Trivini$^{a}$\\
\vspace*{.2in}
$^{a}$Department of Physics, University of Swansea\\
\vspace*{.1in}
$^{b}$ Rudolf Peierls Centre for Theoretical Physics, University of Oxford,\\
1 Keble Road, Oxford OX1 3NP, U.K.\\
}
\end{center}

\vspace*{0.55in}

\begin{center}
{\bf Abstract}
\end{center}

Interpreting the way that the SU(3) bare lattice coupling runs 
with the lattice spacing is complicated by the fact 
that there is a smooth  cross-over region in which the strong
coupling expansion transforms into a weak-coupling one.
For $N\geq 5$, however, there is a first order bulk transition
that cleanly separates the strong and weak coupling regimes.
We find that in this case the calculated string tension can be 
readily fitted throughout the weak coupling region by a standard 3-loop 
expression modified by lattice spacing corrections of the expected 
form. While our fits demand the presence of the latter, they do
not constrain the perturbative coupling scheme enough to enable
us to extract a usefully accurate value of $a(\beta)$ in units of
$\Lambda_{\overline{MS}}$. To resolve this ambiguity we turn to SU(3)
where we use the Schrodinger Functional coupling scheme to extract a value
of $r_0\Lambda_{SF}$ as a benchmark. We then find that the Parisi
mean-field improved coupling scheme closely reproduces this result. 
We also develop a comparison between different schemes that does not 
rely on the calculation of any physical quantity and which can
therefore be applied much further into weak coupling. Again
the Parisi scheme is favoured over the others that we compare.
Using the mean-field scheme we have fitted the values of the string tension 
$a^2\sigma$ that have been calculated for $2\leq N\leq 8$, to obtain 
$\Lambda_{\overline{MS}}/\surd\sigma = 0.503(2)(40) + 0.33(3)(3)/N^2$ 
for $N\geq 3$, where the first error is statistical and the 
second is our estimate of the systematic error from all sources.

\end{titlepage}

\setcounter{page}{1}
\newpage
\pagestyle{plain}

\section{Introduction}
\label{section_intro}

Consider SU($N$) gauge theories discretised 
onto a hypercubic lattice of spacing $a$, on a 4-torus, with 
SU($N$) matrices, $U_l$, assigned to the links $l$, and with 
the standard Wilson plaquette action. The partition function is
\be
Z = \int \prod_l dU_l \exp\left\{-
\beta \sum_{p}\left\{1-{1\over N}{\mathrm {ReTr}} U_p\right\}
\right\}
\label{eqn_Z}
\ee
where $U_p$ is  the ordered product of the SU($N$) matrices 
around the boundary of the plaquette $p$. (Although we shall 
be using the plaquette action in this paper, a parallel 
analysis could be carried out for any other lattice action.)
The parameter $\beta$ in the lattice partition function
is proportional to the inverse lattice bare coupling.
This defines a running coupling on the length scale $a$,
in what is often called the lattice scheme:
\be
\beta = \frac{2N}{g^2_L(a)}.
\label{eqn_betagl}
\ee

In practice one frequently wishes to compare physical quantities 
that have been calculated over a similar range of $\beta$,
but not at precisely the same values. For example the 
Sommer parameter $r_0$ 
\cite{sommer1} 
or the deconfining temperature $T_c$
\cite{Tc} 
in terms of the confining string tension $\sigma$. One 
then needs to interpolate such quantities to common values
of $\beta$. It is well-known that simple perturbative 
interpolations will not fit the calculated values. Thus one
is typically reduced to using, for example, interpolating 
polynomials 
\cite{sommer2} 
that bear no relationship to the weak-coupling expansion, or, 
if one tries to use a power series in $g^2_L \propto 1/\beta$,
one finds that the higher order fitted terms have large 
coefficients of oscillating signs, so that if one extrapolates 
further into weak coupling, the values diverge away from the 
expected 3 loop form until truly asymptotic values of the scale $1/a$ 
\cite{Tcfits}. 
This is, of course, a symptom of the well-known fact that
in the range where one currently performs calculations, no
plausibly simple perturbative expression for $g^2_L(a)$ even 
begins to be adequate. This is unfortunate because it
means that one cannot exploit dimensional transmutation
in a simple way, to transform the value of
the lattice bare coupling to a statement of
how large the lattice spacing $a$ is in units of the corresponding
$\Lambda$ parameter, which can then be expressed in terms of, say, 
$\Lambda_{\overline{MS}}$ in the theoretically and experimentally 
well-studied ${\overline{MS}}$ coupling scheme. (See
\cite{MS-theory} 
and 
\cite{MS-expt}
for recent reviews.)

This old problem has been approached both by stressing the need
for improved lattice coupling schemes
\cite{MF_Parisi,gimp_review} 
and the need for lattice spacing corrections
\cite{allton}. 
Determining what is important is, however, rendered ambiguous
by the fact that there is a smooth cross-over between strong
and weak-coupling in SU(3), so one does not really know
from which value of $\beta$ it is appropriate to attempt a
weak-coupling expansion in powers of $g^2$ and $a$. 

In this paper we use the fact that for SU($N\geq 5$) 
there is a first order `bulk' transition
\cite{blmt-bulk4}
separating the weak and strong coupling ranges, to remove
the ambiguity of where one might expect a weak coupling 
expansion to be applicable. This enables us to quantify
the importance of retaining $O(a^2)$ lattice corrections
in addition to the usual continuum perturbative variation.

While our SU(6) and SU(8) calculations prove accurate
enough to establish the need for $O(a^2)$ corrections
in the relationship between $a$ and $g^2(a)$, the range
in $a$ is not large enough to usefully constrain the coupling 
scheme and hence the value of $a\Lambda_{\overline{MS}}$.
In fact experimental studies are not 
accurate enough either. (See for example Fig.10 in
\cite{MS-expt2}.) 
Fortunately there exists in SU(3) an accurate calculation of the
running coupling in the `Schrodinger functional' (SF) scheme
that covers an energy range comparable to that of experiment, 
i.e. up to $\sim M_Z$, and with appreciably smaller errors
\cite{SF1,SF2}. 
We shall use this scheme to obtain from the values of $a/r_0$
calculated in 
\cite{sommer2,necco}
the continuum value of  $r_0\Lambda_{SF}$ and hence of
$r_0\Lambda_{\overline{MS}}$. We compare
this to what one obtains with some improved coupling 
extrapolations, and find that the
Parisi mean-field improved coupling scheme
\cite{MF_Parisi}
closely matches the SF result. We simultaneously perform a
comparison with the SF scheme that does not involve the
calculation of any physical quantity and therefore can be
carried out to much weaker coupling. This also points
to the `goodness' of the mean-field scheme. Motivated by this
we use this scheme for $N\not= 3$ to obtain continuum values for
$\Lambda_{\overline{MS}}/\sqrt\sigma$ for all $N$, and in particular
for $N\to\infty$.

In the next Section we discuss the weak coupling behaviour
of $a$ on $g^2(a)$ in more detail, and establish our notation.
We then move on to our calculations and then attempt to quantify
the systematic errors on our final results. We end with a
brief summary and our conclusions.

An abbreviated version of this work has been presented
at Lattice 2007
\cite{camtat_lat07}.
This study forms part of a more detailed and extensive 
analysis that will appear elsewhere
\cite{camtat_prep}.
\section{The lattice running coupling}
\label{section_coupling}

Ideally we would wish to be able to determine from the value of 
$g^2_L(a) \equiv 2N/\beta$ what is the lattice spacing $a$ 
as expressed in units of the corresponding
physical scale $\Lambda_L$ (which can then
be converted into a more familiar scale, such as 
$\Lambda_{\overline{MS}}$, using a one-loop calculation
\cite{Dashen-Gross}).
Although we know this to be possible in principle (dimensional 
transmutation) in practice doing so for the SU(3) gauge theory 
(and QCD) has proved notoriously ambiguous, despite the
fact that we know the $\beta$-function in the lattice scheme
to 3 loops
\cite{gL-3loop}.
There are some plausible reasons for this which will become
apparent in the following discussion.

To establish how $g^2_L(a)$
is related to $a$ we determine how $a$ varies with $g^2_L(a)$.
To do this we express $a$ in units of some physical mass $\mu$ 
that we calculate in lattice units, i.e. as $a\mu(a)$. Now we 
know that different choices for the quantity $\mu$ will have 
different lattice spacing corrections:
\be
\frac{a\mu^\prime(a)}{a\mu(a)}
=
\frac{\mu^\prime(0)}{\mu(0)}
\left( 1 + c_{\sigma}a^2\sigma(a) + O(a^4) \right)
\label{eqn_m1m2cor}
\ee
where we have chosen to use the string tension $\sigma$ to 
set the scale of these corrections. As an example, the ratio 
$m_{0^{++}}/\surd\sigma$, where $m_{0^{++}}$ is the lightest 
scalar glueball mass, has $c_{\sigma} \sim 2$ to 3 depending
on $N$, while $T_c/\surd\sigma$, where $T_c$ is the 
deconfining temperature, has $c_{\sigma} \sim 1/3$. 
Thus it is clear that the variation of $a\mu(a)$ with $g^2_L(a)$
{\it must} include a dependence on $a$ that has the functional
form shown in eqn(\ref{eqn_m1m2cor}), in addition to the
variation expected for a running coupling in the continuum theory. 
Generically we should expect $c_{\sigma} = O(1)$ as illustrated
by the examples just quoted. There is 
a similar dependence that follows from the fact that the
$\beta$-function for $g^2_L(a)$ will have lattice spacing 
corrections 
\be
\frac{\partial g^2_L}{\partial\log a^2}
=
\beta_0  g^4_L + \beta_1  g^6_L + \beta^L_2  g^8_L + ...
+ O(a^2).
\label{eqn_bfunction}
\ee
Here the $\beta_i$ are the coefficients of the continuum 
$\beta$-function; $\beta_0$ and $\beta_1$ are scheme-independent
while $\beta^L_2$ requires a 2-loop calculation relating  $g^2_L$
to a coupling for which  $\beta_2$ is already known 
\cite{gL-3loop}.
Thus, as has been emphasised in
\cite{allton},
the dependence of  $a$ on $g^2_L(a)$ will need to
incorporate such lattice spacing corrections in addition to
the perturbative expression that one obtains in the continuum. 
Using the string tension $\sigma$ for our scale this leads to
\begin{eqnarray}
a \sqrt\sigma(a)
& = &
\frac{\sqrt\sigma(0)}{\Lambda_s}
\left( 1 + c^s_{\sigma} a^2\sigma + O(a^4) \right) 
e^{-\frac{1}{2\beta_0 g_s^2}}
\left(\frac{\beta_1}{\beta_0^2}+\frac{1}{\beta_0 g_s^2}
\right)^\frac{\beta_1}{2\beta_0^2} \nonumber \\
&\times&
e^{-\frac{\beta^s_2}{2\beta_0^2}g_s^2
+\left(\frac{\beta_1\beta^s_2}{2\beta_0^3} - 
\frac{\beta^s_3}{4\beta_0^2}\right) g_s^4
+O(g_s^6)}
\label{eqn_ag}
\end{eqnarray}
where we make explicit the quantities that depend on the
coupling scheme $s$ being used (here $s=L$). 
We use the standard definition of the $\Lambda$ parameter,
where the constant term in the expansion of $1/g^2(a)$ 
is absorbed into the scale of the logarithm.
In eqn(\ref{eqn_ag}) the terms that involve only $\beta_0$ and 
$\beta_1$ constitute the exact 2-loop continuum result. (That is 
to say, it is the exact result when $\beta_{j\geq 2} = 0$.) For 
higher orders we are not aware of any such neat closed form in 
terms of elementary functions, so we present their contribution 
as a power series in $g^2$. Note also that although the
coefficient $c^s_{\sigma}$ is a power series in $g^2_s$, 
we shall, following usual practice, treat it as a constant in 
our fits, since $g^2_s$ does not vary very much in the region 
where the $O(a^2)$ correction is significant. Finally
we remark that we could introduce the lattice corrections
in different ways. A small change would be
to use, say, the mass gap $m_G$ rather than the string 
tension $\sigma$ to set the scale for the $O(a^2)$ corrections
in eqn(\ref{eqn_ag}). A more radical change would be to 
substitute for each occurence of $a^2\sigma$ on the RHS of 
eqn(\ref{eqn_ag}) the expression provided by  eqn(\ref{eqn_ag}) 
itself. After iteration this would
transform the lattice spacing corrections from a power series in 
$a^2\sigma$ into a power series in the perturbative factor in 
eqn(\ref{eqn_ag}). All such variations are $\grave{a}$ $priori$ equally 
valid.

In addition to these lattice spacing corrections, a further 
complication is that the $g_L(a)$ coupling scheme is expected 
to have large higher order perturbative corrections. This
follows from the large ratio between $\Lambda_L$ and
$\Lambda_{\overline{MS}}$
\cite{Dashen-Gross}:
\be
\frac{\Lambda_{\overline{MS}}}{\Lambda_L}
=
38.853 \exp\left\{-\frac{3\pi^2}{11N^2}\right\}
\label{eqn_lamLlamMS}
\ee
This implies that in the relationship between the two couplings, 
$g^2_{\overline{MS}} = g^2_L (1 + \gamma g^2_L + O(g^4_L))$,
the coefficient $\gamma$ must also be large, since one can
easily see that 
${\Lambda_{\overline{MS}}}/{\Lambda_L}=\exp(\gamma/2\beta_0)$.
So if $g^2_{\overline{MS}}$ is a `good' scheme with modest
higher order terms in the $\beta$-function, which is something
we shall assume from now on (see
\cite{MS-4loop}),
this will almost
certainly not be the case for $g^2_L$. This is confirmed
by explicit calculation of the three-loop coefficient 
\cite{gL-3loop}
where one finds
\be
\beta^L_2 \stackrel{N\to\infty}{\simeq} 6.9 \beta^{\overline{MS}}_2
\label{eqn_b2Lb2MS}
\ee
Since we only know the $\beta$-function to 3 loops, 
it would be wise to seek a lattice coupling scheme where
there is less reason to expect large higher order corrections.
This problem has of course been appreciated for a long time and 
there have been extensive efforts to improve the lattice coupling
(for a review see
\cite{gimp_review}.)
Perhaps the simplest and oldest suggestion is the `mean-field' 
improved coupling of Parisi
\cite{MF_Parisi}
\be
\frac{1}{g^2_I} = \frac{1}{g^2_L}
\langle \frac{1}{N}\mathrm{Tr}U_p \rangle
\equiv
\frac{u_p}{g^2_L}
\label{eqn_gI}
\ee
which has a nice physical motivation as the effective coupling
experienced by a background field (in a simple approximation)
\cite{MF_Parisi}.
Since the expansion of $\langle \mathrm{Tr}U_p \rangle$ 
in $g^2_L$ is known to 3-loops
\cite{plaq3loop}
we can substitute 
\be
\frac{1}{g^2_I} = \frac{1}{g^2_L}
\langle \frac{1}{N}\mathrm{Tr}U_p \rangle
= \frac{1}{g^2_L}
\left(1-\omega_1 g^2_L-\omega_2 g^4_L-\omega_3 g^6_L + O(g^8_L)\right)
\label{eqn_gIgL}
\ee
into the 3-loop $\beta$-function for $g^2_L$ to obtain the 
3-loop  $\beta$-function  for $g^2_I$, giving
\be
\beta^I_2
=
\beta^L_2 + \beta_0 \omega_2 - \beta_1 \omega_1
\stackrel{N\to\infty}{\simeq}
2.3 \beta^{\overline{MS}}_2
\label{eqn_b2Ib2MS}
\ee
which indeed promises smaller higher-order corrections than
eqn(\ref{eqn_b2Lb2MS}). Similarly one finds
\be
\frac{\Lambda_{\overline{MS}}}{\Lambda_I}
=
\frac{\Lambda_L}{\Lambda_I}
\frac{\Lambda_{\overline{MS}}}{\Lambda_L}
=
\exp\left\{ \frac{\omega_1}{2\beta_0} \right\}
\times
38.853 \exp\left\{-\frac{3\pi^2}{11N^2}\right\}
\simeq
2.633
\label{eqn_lamIlamMS}
\ee
using eqn(\ref{eqn_lamLlamMS}) and $\omega_1=(N^2-1)/8N$,
which again indicates that the mean-field improved coupling
scheme is a `good' one. Motivated by this we shall focus on this
coupling scheme, and some variations thereof, in our later calculations.

Although the need to improve the lattice coupling has long
been recognised, the suggestion 
\cite{allton}
that $O(a^2)$ lattice corrections are also needed has in
practice been more controversial. As demonstrated above,
there can be no question about the presence of such corrections.
But one might question their importance. How plausible is this?
As remarked above, we expect that generically $c_{\sigma}=O(1)$
in eqn(\ref{eqn_ag}), and since
$a\sqrt\sigma(a)$ and other masses are typically
calculated to a precision $\ll 1\%$, we would need
$a^2\sigma \ll 0.01$ for the $O(a^2)$ correction to be negligible.
In SU(3), for example, $a^2\sigma\simeq 0.01$ at $\beta=6.5$.
This coupling region is at the upper edge of the range in which 
useful calculations are usually performed. Thus generically we 
should expect $O(a^2)$ corrections to be important in current
calculations. Of course it might have been that 
the coefficient $c_{\sigma}$ just happened to be
unexpectedly  small, but the calculations in this paper 
will show that this is not the case.

\begin	{figure}[p]
\begin	{center}
\leavevmode
\begingroup%
  \makeatletter%
  \newcommand{\GNUPLOTspecial}{%
    \@sanitize\catcode`\%=14\relax\special}%
  \setlength{\unitlength}{0.1bp}%
{\GNUPLOTspecial{!
/gnudict 256 dict def
gnudict begin
/Color true def
/Solid false def
/gnulinewidth 5.000 def
/userlinewidth gnulinewidth def
/vshift -33 def
/dl {10 mul} def
/hpt_ 31.5 def
/vpt_ 31.5 def
/hpt hpt_ def
/vpt vpt_ def
/M {moveto} bind def
/L {lineto} bind def
/R {rmoveto} bind def
/V {rlineto} bind def
/vpt2 vpt 2 mul def
/hpt2 hpt 2 mul def
/Lshow { currentpoint stroke M
  0 vshift R show } def
/Rshow { currentpoint stroke M
  dup stringwidth pop neg vshift R show } def
/Cshow { currentpoint stroke M
  dup stringwidth pop -2 div vshift R show } def
/UP { dup vpt_ mul /vpt exch def hpt_ mul /hpt exch def
  /hpt2 hpt 2 mul def /vpt2 vpt 2 mul def } def
/DL { Color {setrgbcolor Solid {pop []} if 0 setdash }
 {pop pop pop Solid {pop []} if 0 setdash} ifelse } def
/BL { stroke userlinewidth 2 mul setlinewidth } def
/AL { stroke userlinewidth 2 div setlinewidth } def
/UL { dup gnulinewidth mul /userlinewidth exch def
      10 mul /udl exch def } def
/PL { stroke userlinewidth setlinewidth } def
/LTb { BL [] 0 0 0 DL } def
/LTa { AL [1 udl mul 2 udl mul] 0 setdash 0 0 0 setrgbcolor } def
/LT0 { PL [] 1 0 0 DL } def
/LT1 { PL [4 dl 2 dl] 0 1 0 DL } def
/LT2 { PL [2 dl 3 dl] 0 0 1 DL } def
/LT3 { PL [1 dl 1.5 dl] 1 0 1 DL } def
/LT4 { PL [5 dl 2 dl 1 dl 2 dl] 0 1 1 DL } def
/LT5 { PL [4 dl 3 dl 1 dl 3 dl] 1 1 0 DL } def
/LT6 { PL [2 dl 2 dl 2 dl 4 dl] 0 0 0 DL } def
/LT7 { PL [2 dl 2 dl 2 dl 2 dl 2 dl 4 dl] 1 0.3 0 DL } def
/LT8 { PL [2 dl 2 dl 2 dl 2 dl 2 dl 2 dl 2 dl 4 dl] 0.5 0.5 0.5 DL } def
/Pnt { stroke [] 0 setdash
   gsave 1 setlinecap M 0 0 V stroke grestore } def
/Dia { stroke [] 0 setdash 2 copy vpt add M
  hpt neg vpt neg V hpt vpt neg V
  hpt vpt V hpt neg vpt V closepath stroke
  Pnt } def
/Pls { stroke [] 0 setdash vpt sub M 0 vpt2 V
  currentpoint stroke M
  hpt neg vpt neg R hpt2 0 V stroke
  } def
/Box { stroke [] 0 setdash 2 copy exch hpt sub exch vpt add M
  0 vpt2 neg V hpt2 0 V 0 vpt2 V
  hpt2 neg 0 V closepath stroke
  Pnt } def
/Crs { stroke [] 0 setdash exch hpt sub exch vpt add M
  hpt2 vpt2 neg V currentpoint stroke M
  hpt2 neg 0 R hpt2 vpt2 V stroke } def
/TriU { stroke [] 0 setdash 2 copy vpt 1.12 mul add M
  hpt neg vpt -1.62 mul V
  hpt 2 mul 0 V
  hpt neg vpt 1.62 mul V closepath stroke
  Pnt  } def
/Star { 2 copy Pls Crs } def
/BoxF { stroke [] 0 setdash exch hpt sub exch vpt add M
  0 vpt2 neg V  hpt2 0 V  0 vpt2 V
  hpt2 neg 0 V  closepath fill } def
/TriUF { stroke [] 0 setdash vpt 1.12 mul add M
  hpt neg vpt -1.62 mul V
  hpt 2 mul 0 V
  hpt neg vpt 1.62 mul V closepath fill } def
/TriD { stroke [] 0 setdash 2 copy vpt 1.12 mul sub M
  hpt neg vpt 1.62 mul V
  hpt 2 mul 0 V
  hpt neg vpt -1.62 mul V closepath stroke
  Pnt  } def
/TriDF { stroke [] 0 setdash vpt 1.12 mul sub M
  hpt neg vpt 1.62 mul V
  hpt 2 mul 0 V
  hpt neg vpt -1.62 mul V closepath fill} def
/DiaF { stroke [] 0 setdash vpt add M
  hpt neg vpt neg V hpt vpt neg V
  hpt vpt V hpt neg vpt V closepath fill } def
/Pent { stroke [] 0 setdash 2 copy gsave
  translate 0 hpt M 4 {72 rotate 0 hpt L} repeat
  closepath stroke grestore Pnt } def
/PentF { stroke [] 0 setdash gsave
  translate 0 hpt M 4 {72 rotate 0 hpt L} repeat
  closepath fill grestore } def
/Circle { stroke [] 0 setdash 2 copy
  hpt 0 360 arc stroke Pnt } def
/CircleF { stroke [] 0 setdash hpt 0 360 arc fill } def
/C0 { BL [] 0 setdash 2 copy moveto vpt 90 450  arc } bind def
/C1 { BL [] 0 setdash 2 copy        moveto
       2 copy  vpt 0 90 arc closepath fill
               vpt 0 360 arc closepath } bind def
/C2 { BL [] 0 setdash 2 copy moveto
       2 copy  vpt 90 180 arc closepath fill
               vpt 0 360 arc closepath } bind def
/C3 { BL [] 0 setdash 2 copy moveto
       2 copy  vpt 0 180 arc closepath fill
               vpt 0 360 arc closepath } bind def
/C4 { BL [] 0 setdash 2 copy moveto
       2 copy  vpt 180 270 arc closepath fill
               vpt 0 360 arc closepath } bind def
/C5 { BL [] 0 setdash 2 copy moveto
       2 copy  vpt 0 90 arc
       2 copy moveto
       2 copy  vpt 180 270 arc closepath fill
               vpt 0 360 arc } bind def
/C6 { BL [] 0 setdash 2 copy moveto
      2 copy  vpt 90 270 arc closepath fill
              vpt 0 360 arc closepath } bind def
/C7 { BL [] 0 setdash 2 copy moveto
      2 copy  vpt 0 270 arc closepath fill
              vpt 0 360 arc closepath } bind def
/C8 { BL [] 0 setdash 2 copy moveto
      2 copy vpt 270 360 arc closepath fill
              vpt 0 360 arc closepath } bind def
/C9 { BL [] 0 setdash 2 copy moveto
      2 copy  vpt 270 450 arc closepath fill
              vpt 0 360 arc closepath } bind def
/C10 { BL [] 0 setdash 2 copy 2 copy moveto vpt 270 360 arc closepath fill
       2 copy moveto
       2 copy vpt 90 180 arc closepath fill
               vpt 0 360 arc closepath } bind def
/C11 { BL [] 0 setdash 2 copy moveto
       2 copy  vpt 0 180 arc closepath fill
       2 copy moveto
       2 copy  vpt 270 360 arc closepath fill
               vpt 0 360 arc closepath } bind def
/C12 { BL [] 0 setdash 2 copy moveto
       2 copy  vpt 180 360 arc closepath fill
               vpt 0 360 arc closepath } bind def
/C13 { BL [] 0 setdash  2 copy moveto
       2 copy  vpt 0 90 arc closepath fill
       2 copy moveto
       2 copy  vpt 180 360 arc closepath fill
               vpt 0 360 arc closepath } bind def
/C14 { BL [] 0 setdash 2 copy moveto
       2 copy  vpt 90 360 arc closepath fill
               vpt 0 360 arc } bind def
/C15 { BL [] 0 setdash 2 copy vpt 0 360 arc closepath fill
               vpt 0 360 arc closepath } bind def
/Rec   { newpath 4 2 roll moveto 1 index 0 rlineto 0 exch rlineto
       neg 0 rlineto closepath } bind def
/Square { dup Rec } bind def
/Bsquare { vpt sub exch vpt sub exch vpt2 Square } bind def
/S0 { BL [] 0 setdash 2 copy moveto 0 vpt rlineto BL Bsquare } bind def
/S1 { BL [] 0 setdash 2 copy vpt Square fill Bsquare } bind def
/S2 { BL [] 0 setdash 2 copy exch vpt sub exch vpt Square fill Bsquare } bind def
/S3 { BL [] 0 setdash 2 copy exch vpt sub exch vpt2 vpt Rec fill Bsquare } bind def
/S4 { BL [] 0 setdash 2 copy exch vpt sub exch vpt sub vpt Square fill Bsquare } bind def
/S5 { BL [] 0 setdash 2 copy 2 copy vpt Square fill
       exch vpt sub exch vpt sub vpt Square fill Bsquare } bind def
/S6 { BL [] 0 setdash 2 copy exch vpt sub exch vpt sub vpt vpt2 Rec fill Bsquare } bind def
/S7 { BL [] 0 setdash 2 copy exch vpt sub exch vpt sub vpt vpt2 Rec fill
       2 copy vpt Square fill
       Bsquare } bind def
/S8 { BL [] 0 setdash 2 copy vpt sub vpt Square fill Bsquare } bind def
/S9 { BL [] 0 setdash 2 copy vpt sub vpt vpt2 Rec fill Bsquare } bind def
/S10 { BL [] 0 setdash 2 copy vpt sub vpt Square fill 2 copy exch vpt sub exch vpt Square fill
       Bsquare } bind def
/S11 { BL [] 0 setdash 2 copy vpt sub vpt Square fill 2 copy exch vpt sub exch vpt2 vpt Rec fill
       Bsquare } bind def
/S12 { BL [] 0 setdash 2 copy exch vpt sub exch vpt sub vpt2 vpt Rec fill Bsquare } bind def
/S13 { BL [] 0 setdash 2 copy exch vpt sub exch vpt sub vpt2 vpt Rec fill
       2 copy vpt Square fill Bsquare } bind def
/S14 { BL [] 0 setdash 2 copy exch vpt sub exch vpt sub vpt2 vpt Rec fill
       2 copy exch vpt sub exch vpt Square fill Bsquare } bind def
/S15 { BL [] 0 setdash 2 copy Bsquare fill Bsquare } bind def
/D0 { gsave translate 45 rotate 0 0 S0 stroke grestore } bind def
/D1 { gsave translate 45 rotate 0 0 S1 stroke grestore } bind def
/D2 { gsave translate 45 rotate 0 0 S2 stroke grestore } bind def
/D3 { gsave translate 45 rotate 0 0 S3 stroke grestore } bind def
/D4 { gsave translate 45 rotate 0 0 S4 stroke grestore } bind def
/D5 { gsave translate 45 rotate 0 0 S5 stroke grestore } bind def
/D6 { gsave translate 45 rotate 0 0 S6 stroke grestore } bind def
/D7 { gsave translate 45 rotate 0 0 S7 stroke grestore } bind def
/D8 { gsave translate 45 rotate 0 0 S8 stroke grestore } bind def
/D9 { gsave translate 45 rotate 0 0 S9 stroke grestore } bind def
/D10 { gsave translate 45 rotate 0 0 S10 stroke grestore } bind def
/D11 { gsave translate 45 rotate 0 0 S11 stroke grestore } bind def
/D12 { gsave translate 45 rotate 0 0 S12 stroke grestore } bind def
/D13 { gsave translate 45 rotate 0 0 S13 stroke grestore } bind def
/D14 { gsave translate 45 rotate 0 0 S14 stroke grestore } bind def
/D15 { gsave translate 45 rotate 0 0 S15 stroke grestore } bind def
/DiaE { stroke [] 0 setdash vpt add M
  hpt neg vpt neg V hpt vpt neg V
  hpt vpt V hpt neg vpt V closepath stroke } def
/BoxE { stroke [] 0 setdash exch hpt sub exch vpt add M
  0 vpt2 neg V hpt2 0 V 0 vpt2 V
  hpt2 neg 0 V closepath stroke } def
/TriUE { stroke [] 0 setdash vpt 1.12 mul add M
  hpt neg vpt -1.62 mul V
  hpt 2 mul 0 V
  hpt neg vpt 1.62 mul V closepath stroke } def
/TriDE { stroke [] 0 setdash vpt 1.12 mul sub M
  hpt neg vpt 1.62 mul V
  hpt 2 mul 0 V
  hpt neg vpt -1.62 mul V closepath stroke } def
/PentE { stroke [] 0 setdash gsave
  translate 0 hpt M 4 {72 rotate 0 hpt L} repeat
  closepath stroke grestore } def
/CircE { stroke [] 0 setdash 
  hpt 0 360 arc stroke } def
/Opaque { gsave closepath 1 setgray fill grestore 0 setgray closepath } def
/DiaW { stroke [] 0 setdash vpt add M
  hpt neg vpt neg V hpt vpt neg V
  hpt vpt V hpt neg vpt V Opaque stroke } def
/BoxW { stroke [] 0 setdash exch hpt sub exch vpt add M
  0 vpt2 neg V hpt2 0 V 0 vpt2 V
  hpt2 neg 0 V Opaque stroke } def
/TriUW { stroke [] 0 setdash vpt 1.12 mul add M
  hpt neg vpt -1.62 mul V
  hpt 2 mul 0 V
  hpt neg vpt 1.62 mul V Opaque stroke } def
/TriDW { stroke [] 0 setdash vpt 1.12 mul sub M
  hpt neg vpt 1.62 mul V
  hpt 2 mul 0 V
  hpt neg vpt -1.62 mul V Opaque stroke } def
/PentW { stroke [] 0 setdash gsave
  translate 0 hpt M 4 {72 rotate 0 hpt L} repeat
  Opaque stroke grestore } def
/CircW { stroke [] 0 setdash 
  hpt 0 360 arc Opaque stroke } def
/BoxFill { gsave Rec 1 setgray fill grestore } def
end
}}%
\begin{picture}(4500,3240)(0,0)%
{\GNUPLOTspecial{"
gnudict begin
gsave
0 0 translate
0.100 0.100 scale
0 setgray
newpath
1.000 UL
LTb
800 400 M
63 0 V
3487 0 R
-63 0 V
800 1313 M
63 0 V
3487 0 R
-63 0 V
800 2227 M
63 0 V
3487 0 R
-63 0 V
800 3140 M
63 0 V
3487 0 R
-63 0 V
1786 400 M
0 63 V
0 2677 R
0 -63 V
2772 400 M
0 63 V
0 2677 R
0 -63 V
3758 400 M
0 63 V
0 2677 R
0 -63 V
1.000 UL
LTb
800 400 M
3550 0 V
0 2740 V
-3550 0 V
800 400 L
1.000 UP
1.000 UL
LT0
1194 2223 M
0 43 V
-31 -43 R
62 0 V
-62 43 R
62 0 V
167 -135 R
0 28 V
-31 -28 R
62 0 V
-62 28 R
62 0 V
166 -205 R
0 13 V
-31 -13 R
62 0 V
-62 13 R
62 0 V
68 -96 R
0 31 V
-31 -31 R
62 0 V
-62 31 R
62 0 V
67 -121 R
0 27 V
-31 -27 R
62 0 V
-62 27 R
62 0 V
68 -143 R
0 15 V
-31 -15 R
62 0 V
-62 15 R
62 0 V
67 -159 R
0 9 V
-31 -9 R
62 0 V
-62 9 R
62 0 V
68 -131 R
0 16 V
-31 -16 R
62 0 V
-62 16 R
62 0 V
18 -72 R
0 11 V
-31 -11 R
62 0 V
-62 11 R
62 0 V
4 -52 R
0 9 V
-31 -9 R
62 0 V
-62 9 R
62 0 V
-18 -17 R
0 8 V
-31 -8 R
62 0 V
-62 8 R
62 0 V
167 -186 R
0 6 V
-31 -6 R
62 0 V
-62 6 R
62 0 V
-30 -10 R
0 7 V
-31 -7 R
62 0 V
-62 7 R
62 0 V
2564 993 M
0 5 V
-31 -5 R
62 0 V
-62 5 R
62 0 V
2895 843 M
0 3 V
-31 -3 R
62 0 V
-62 3 R
62 0 V
241 -87 R
0 3 V
-31 -3 R
62 0 V
-62 3 R
62 0 V
241 -67 R
0 2 V
-31 -2 R
62 0 V
-62 2 R
62 0 V
318 -67 R
0 2 V
-31 -2 R
62 0 V
-62 2 R
62 0 V
1194 2245 CircleF
1392 2145 CircleF
1589 1961 CircleF
1688 1886 CircleF
1786 1794 CircleF
1885 1673 CircleF
1983 1525 CircleF
2082 1407 CircleF
2131 1348 CircleF
2166 1306 CircleF
2179 1298 CircleF
2377 1119 CircleF
2378 1115 CircleF
2564 995 CircleF
2895 844 CircleF
3167 761 CircleF
3439 696 CircleF
3788 631 CircleF
stroke
grestore
end
showpage
}}%
\put(2575,50){\makebox(0,0){\Large{$\beta=\frac{2N}{g^2}$}}}%
\put(100,2020){\makebox(0,0){\Large{$a\surd\sigma$}}}%
\put(3758,300){\makebox(0,0){\ {$6.5$}}}%
\put(2772,300){\makebox(0,0){\ {$6$}}}%
\put(1786,300){\makebox(0,0){\ {$5.5$}}}%
\put(750,3140){\makebox(0,0)[r]{\ \ {$1.2$}}}%
\put(750,2227){\makebox(0,0)[r]{\ \ {$0.8$}}}%
\put(750,1313){\makebox(0,0)[r]{\ \ {$0.4$}}}%
\put(750,400){\makebox(0,0)[r]{\ \ {$0$}}}%
\end{picture}%
\endgroup

\end	{center}
\vskip 0.15in
\caption{The SU(3) string tension versus the inverse lattice coupling,
including the region of the crossover between strong and weak coupling.}
\label{fig_bulkn3}
\end 	{figure}
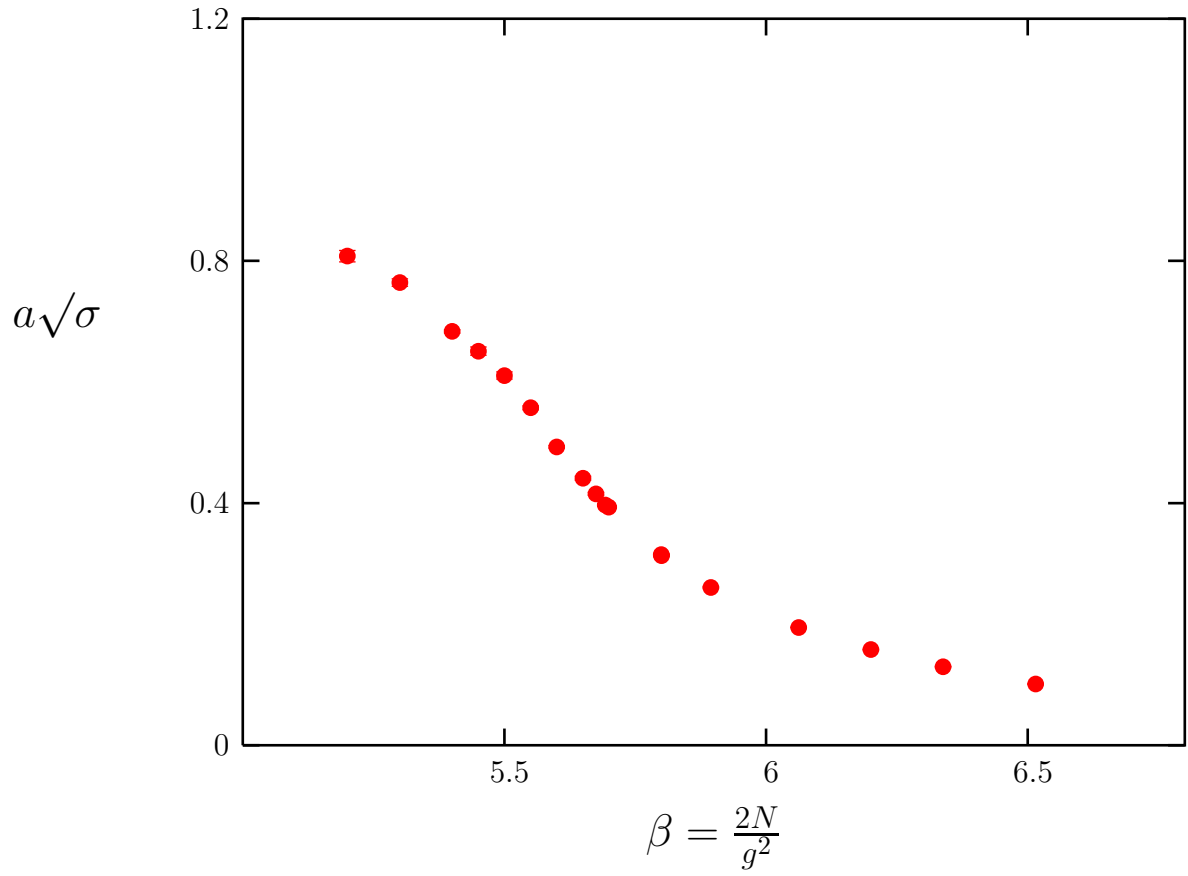

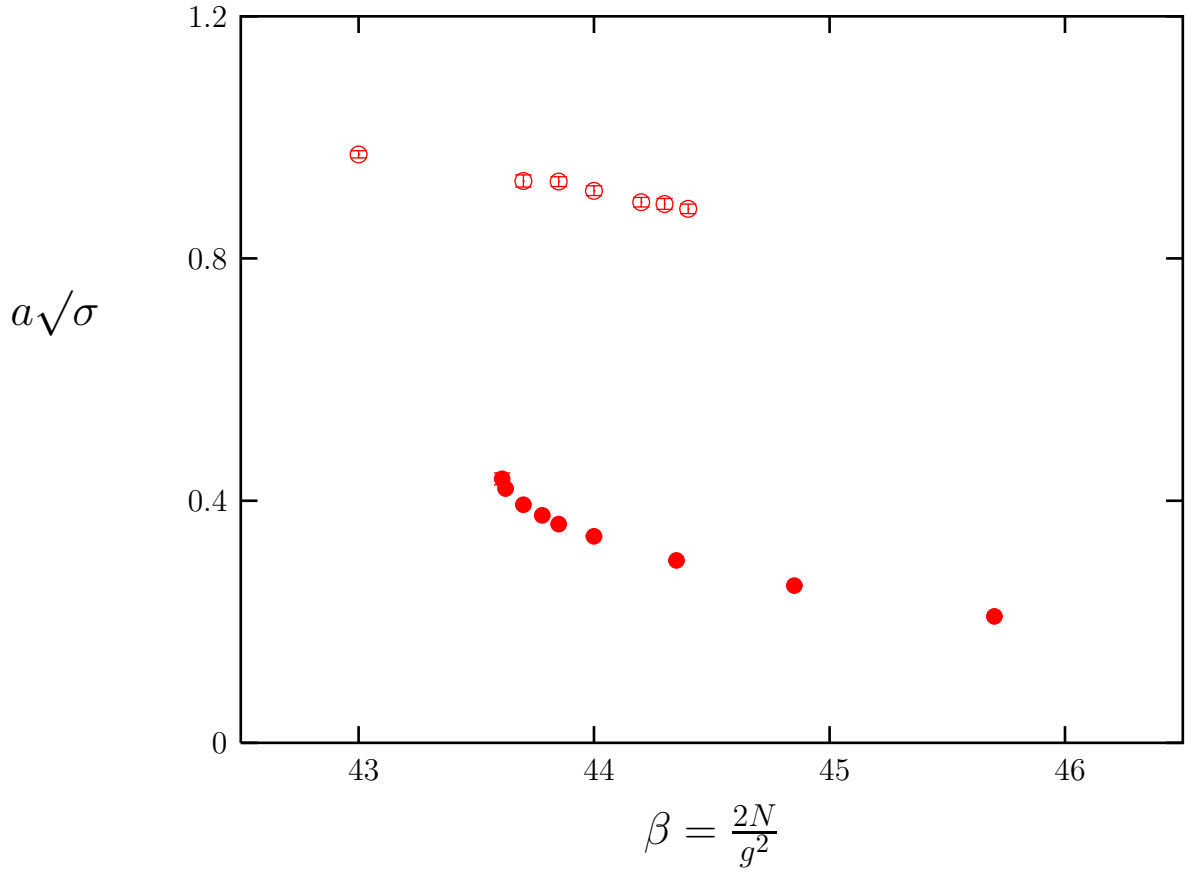
\begin	{figure}[p]
\begin	{center}
\leavevmode
\begingroup%
  \makeatletter%
  \newcommand{\GNUPLOTspecial}{%
    \@sanitize\catcode`\%=14\relax\special}%
  \setlength{\unitlength}{0.1bp}%
{\GNUPLOTspecial{!
/gnudict 256 dict def
gnudict begin
/Color true def
/Solid false def
/gnulinewidth 5.000 def
/userlinewidth gnulinewidth def
/vshift -33 def
/dl {10 mul} def
/hpt_ 31.5 def
/vpt_ 31.5 def
/hpt hpt_ def
/vpt vpt_ def
/M {moveto} bind def
/L {lineto} bind def
/R {rmoveto} bind def
/V {rlineto} bind def
/vpt2 vpt 2 mul def
/hpt2 hpt 2 mul def
/Lshow { currentpoint stroke M
  0 vshift R show } def
/Rshow { currentpoint stroke M
  dup stringwidth pop neg vshift R show } def
/Cshow { currentpoint stroke M
  dup stringwidth pop -2 div vshift R show } def
/UP { dup vpt_ mul /vpt exch def hpt_ mul /hpt exch def
  /hpt2 hpt 2 mul def /vpt2 vpt 2 mul def } def
/DL { Color {setrgbcolor Solid {pop []} if 0 setdash }
 {pop pop pop Solid {pop []} if 0 setdash} ifelse } def
/BL { stroke userlinewidth 2 mul setlinewidth } def
/AL { stroke userlinewidth 2 div setlinewidth } def
/UL { dup gnulinewidth mul /userlinewidth exch def
      10 mul /udl exch def } def
/PL { stroke userlinewidth setlinewidth } def
/LTb { BL [] 0 0 0 DL } def
/LTa { AL [1 udl mul 2 udl mul] 0 setdash 0 0 0 setrgbcolor } def
/LT0 { PL [] 1 0 0 DL } def
/LT1 { PL [4 dl 2 dl] 0 1 0 DL } def
/LT2 { PL [2 dl 3 dl] 0 0 1 DL } def
/LT3 { PL [1 dl 1.5 dl] 1 0 1 DL } def
/LT4 { PL [5 dl 2 dl 1 dl 2 dl] 0 1 1 DL } def
/LT5 { PL [4 dl 3 dl 1 dl 3 dl] 1 1 0 DL } def
/LT6 { PL [2 dl 2 dl 2 dl 4 dl] 0 0 0 DL } def
/LT7 { PL [2 dl 2 dl 2 dl 2 dl 2 dl 4 dl] 1 0.3 0 DL } def
/LT8 { PL [2 dl 2 dl 2 dl 2 dl 2 dl 2 dl 2 dl 4 dl] 0.5 0.5 0.5 DL } def
/Pnt { stroke [] 0 setdash
   gsave 1 setlinecap M 0 0 V stroke grestore } def
/Dia { stroke [] 0 setdash 2 copy vpt add M
  hpt neg vpt neg V hpt vpt neg V
  hpt vpt V hpt neg vpt V closepath stroke
  Pnt } def
/Pls { stroke [] 0 setdash vpt sub M 0 vpt2 V
  currentpoint stroke M
  hpt neg vpt neg R hpt2 0 V stroke
  } def
/Box { stroke [] 0 setdash 2 copy exch hpt sub exch vpt add M
  0 vpt2 neg V hpt2 0 V 0 vpt2 V
  hpt2 neg 0 V closepath stroke
  Pnt } def
/Crs { stroke [] 0 setdash exch hpt sub exch vpt add M
  hpt2 vpt2 neg V currentpoint stroke M
  hpt2 neg 0 R hpt2 vpt2 V stroke } def
/TriU { stroke [] 0 setdash 2 copy vpt 1.12 mul add M
  hpt neg vpt -1.62 mul V
  hpt 2 mul 0 V
  hpt neg vpt 1.62 mul V closepath stroke
  Pnt  } def
/Star { 2 copy Pls Crs } def
/BoxF { stroke [] 0 setdash exch hpt sub exch vpt add M
  0 vpt2 neg V  hpt2 0 V  0 vpt2 V
  hpt2 neg 0 V  closepath fill } def
/TriUF { stroke [] 0 setdash vpt 1.12 mul add M
  hpt neg vpt -1.62 mul V
  hpt 2 mul 0 V
  hpt neg vpt 1.62 mul V closepath fill } def
/TriD { stroke [] 0 setdash 2 copy vpt 1.12 mul sub M
  hpt neg vpt 1.62 mul V
  hpt 2 mul 0 V
  hpt neg vpt -1.62 mul V closepath stroke
  Pnt  } def
/TriDF { stroke [] 0 setdash vpt 1.12 mul sub M
  hpt neg vpt 1.62 mul V
  hpt 2 mul 0 V
  hpt neg vpt -1.62 mul V closepath fill} def
/DiaF { stroke [] 0 setdash vpt add M
  hpt neg vpt neg V hpt vpt neg V
  hpt vpt V hpt neg vpt V closepath fill } def
/Pent { stroke [] 0 setdash 2 copy gsave
  translate 0 hpt M 4 {72 rotate 0 hpt L} repeat
  closepath stroke grestore Pnt } def
/PentF { stroke [] 0 setdash gsave
  translate 0 hpt M 4 {72 rotate 0 hpt L} repeat
  closepath fill grestore } def
/Circle { stroke [] 0 setdash 2 copy
  hpt 0 360 arc stroke Pnt } def
/CircleF { stroke [] 0 setdash hpt 0 360 arc fill } def
/C0 { BL [] 0 setdash 2 copy moveto vpt 90 450  arc } bind def
/C1 { BL [] 0 setdash 2 copy        moveto
       2 copy  vpt 0 90 arc closepath fill
               vpt 0 360 arc closepath } bind def
/C2 { BL [] 0 setdash 2 copy moveto
       2 copy  vpt 90 180 arc closepath fill
               vpt 0 360 arc closepath } bind def
/C3 { BL [] 0 setdash 2 copy moveto
       2 copy  vpt 0 180 arc closepath fill
               vpt 0 360 arc closepath } bind def
/C4 { BL [] 0 setdash 2 copy moveto
       2 copy  vpt 180 270 arc closepath fill
               vpt 0 360 arc closepath } bind def
/C5 { BL [] 0 setdash 2 copy moveto
       2 copy  vpt 0 90 arc
       2 copy moveto
       2 copy  vpt 180 270 arc closepath fill
               vpt 0 360 arc } bind def
/C6 { BL [] 0 setdash 2 copy moveto
      2 copy  vpt 90 270 arc closepath fill
              vpt 0 360 arc closepath } bind def
/C7 { BL [] 0 setdash 2 copy moveto
      2 copy  vpt 0 270 arc closepath fill
              vpt 0 360 arc closepath } bind def
/C8 { BL [] 0 setdash 2 copy moveto
      2 copy vpt 270 360 arc closepath fill
              vpt 0 360 arc closepath } bind def
/C9 { BL [] 0 setdash 2 copy moveto
      2 copy  vpt 270 450 arc closepath fill
              vpt 0 360 arc closepath } bind def
/C10 { BL [] 0 setdash 2 copy 2 copy moveto vpt 270 360 arc closepath fill
       2 copy moveto
       2 copy vpt 90 180 arc closepath fill
               vpt 0 360 arc closepath } bind def
/C11 { BL [] 0 setdash 2 copy moveto
       2 copy  vpt 0 180 arc closepath fill
       2 copy moveto
       2 copy  vpt 270 360 arc closepath fill
               vpt 0 360 arc closepath } bind def
/C12 { BL [] 0 setdash 2 copy moveto
       2 copy  vpt 180 360 arc closepath fill
               vpt 0 360 arc closepath } bind def
/C13 { BL [] 0 setdash  2 copy moveto
       2 copy  vpt 0 90 arc closepath fill
       2 copy moveto
       2 copy  vpt 180 360 arc closepath fill
               vpt 0 360 arc closepath } bind def
/C14 { BL [] 0 setdash 2 copy moveto
       2 copy  vpt 90 360 arc closepath fill
               vpt 0 360 arc } bind def
/C15 { BL [] 0 setdash 2 copy vpt 0 360 arc closepath fill
               vpt 0 360 arc closepath } bind def
/Rec   { newpath 4 2 roll moveto 1 index 0 rlineto 0 exch rlineto
       neg 0 rlineto closepath } bind def
/Square { dup Rec } bind def
/Bsquare { vpt sub exch vpt sub exch vpt2 Square } bind def
/S0 { BL [] 0 setdash 2 copy moveto 0 vpt rlineto BL Bsquare } bind def
/S1 { BL [] 0 setdash 2 copy vpt Square fill Bsquare } bind def
/S2 { BL [] 0 setdash 2 copy exch vpt sub exch vpt Square fill Bsquare } bind def
/S3 { BL [] 0 setdash 2 copy exch vpt sub exch vpt2 vpt Rec fill Bsquare } bind def
/S4 { BL [] 0 setdash 2 copy exch vpt sub exch vpt sub vpt Square fill Bsquare } bind def
/S5 { BL [] 0 setdash 2 copy 2 copy vpt Square fill
       exch vpt sub exch vpt sub vpt Square fill Bsquare } bind def
/S6 { BL [] 0 setdash 2 copy exch vpt sub exch vpt sub vpt vpt2 Rec fill Bsquare } bind def
/S7 { BL [] 0 setdash 2 copy exch vpt sub exch vpt sub vpt vpt2 Rec fill
       2 copy vpt Square fill
       Bsquare } bind def
/S8 { BL [] 0 setdash 2 copy vpt sub vpt Square fill Bsquare } bind def
/S9 { BL [] 0 setdash 2 copy vpt sub vpt vpt2 Rec fill Bsquare } bind def
/S10 { BL [] 0 setdash 2 copy vpt sub vpt Square fill 2 copy exch vpt sub exch vpt Square fill
       Bsquare } bind def
/S11 { BL [] 0 setdash 2 copy vpt sub vpt Square fill 2 copy exch vpt sub exch vpt2 vpt Rec fill
       Bsquare } bind def
/S12 { BL [] 0 setdash 2 copy exch vpt sub exch vpt sub vpt2 vpt Rec fill Bsquare } bind def
/S13 { BL [] 0 setdash 2 copy exch vpt sub exch vpt sub vpt2 vpt Rec fill
       2 copy vpt Square fill Bsquare } bind def
/S14 { BL [] 0 setdash 2 copy exch vpt sub exch vpt sub vpt2 vpt Rec fill
       2 copy exch vpt sub exch vpt Square fill Bsquare } bind def
/S15 { BL [] 0 setdash 2 copy Bsquare fill Bsquare } bind def
/D0 { gsave translate 45 rotate 0 0 S0 stroke grestore } bind def
/D1 { gsave translate 45 rotate 0 0 S1 stroke grestore } bind def
/D2 { gsave translate 45 rotate 0 0 S2 stroke grestore } bind def
/D3 { gsave translate 45 rotate 0 0 S3 stroke grestore } bind def
/D4 { gsave translate 45 rotate 0 0 S4 stroke grestore } bind def
/D5 { gsave translate 45 rotate 0 0 S5 stroke grestore } bind def
/D6 { gsave translate 45 rotate 0 0 S6 stroke grestore } bind def
/D7 { gsave translate 45 rotate 0 0 S7 stroke grestore } bind def
/D8 { gsave translate 45 rotate 0 0 S8 stroke grestore } bind def
/D9 { gsave translate 45 rotate 0 0 S9 stroke grestore } bind def
/D10 { gsave translate 45 rotate 0 0 S10 stroke grestore } bind def
/D11 { gsave translate 45 rotate 0 0 S11 stroke grestore } bind def
/D12 { gsave translate 45 rotate 0 0 S12 stroke grestore } bind def
/D13 { gsave translate 45 rotate 0 0 S13 stroke grestore } bind def
/D14 { gsave translate 45 rotate 0 0 S14 stroke grestore } bind def
/D15 { gsave translate 45 rotate 0 0 S15 stroke grestore } bind def
/DiaE { stroke [] 0 setdash vpt add M
  hpt neg vpt neg V hpt vpt neg V
  hpt vpt V hpt neg vpt V closepath stroke } def
/BoxE { stroke [] 0 setdash exch hpt sub exch vpt add M
  0 vpt2 neg V hpt2 0 V 0 vpt2 V
  hpt2 neg 0 V closepath stroke } def
/TriUE { stroke [] 0 setdash vpt 1.12 mul add M
  hpt neg vpt -1.62 mul V
  hpt 2 mul 0 V
  hpt neg vpt 1.62 mul V closepath stroke } def
/TriDE { stroke [] 0 setdash vpt 1.12 mul sub M
  hpt neg vpt 1.62 mul V
  hpt 2 mul 0 V
  hpt neg vpt -1.62 mul V closepath stroke } def
/PentE { stroke [] 0 setdash gsave
  translate 0 hpt M 4 {72 rotate 0 hpt L} repeat
  closepath stroke grestore } def
/CircE { stroke [] 0 setdash 
  hpt 0 360 arc stroke } def
/Opaque { gsave closepath 1 setgray fill grestore 0 setgray closepath } def
/DiaW { stroke [] 0 setdash vpt add M
  hpt neg vpt neg V hpt vpt neg V
  hpt vpt V hpt neg vpt V Opaque stroke } def
/BoxW { stroke [] 0 setdash exch hpt sub exch vpt add M
  0 vpt2 neg V hpt2 0 V 0 vpt2 V
  hpt2 neg 0 V Opaque stroke } def
/TriUW { stroke [] 0 setdash vpt 1.12 mul add M
  hpt neg vpt -1.62 mul V
  hpt 2 mul 0 V
  hpt neg vpt 1.62 mul V Opaque stroke } def
/TriDW { stroke [] 0 setdash vpt 1.12 mul sub M
  hpt neg vpt 1.62 mul V
  hpt 2 mul 0 V
  hpt neg vpt -1.62 mul V Opaque stroke } def
/PentW { stroke [] 0 setdash gsave
  translate 0 hpt M 4 {72 rotate 0 hpt L} repeat
  Opaque stroke grestore } def
/CircW { stroke [] 0 setdash 
  hpt 0 360 arc Opaque stroke } def
/BoxFill { gsave Rec 1 setgray fill grestore } def
end
}}%
\begin{picture}(4500,3240)(0,0)%
{\GNUPLOTspecial{"
gnudict begin
gsave
0 0 translate
0.100 0.100 scale
0 setgray
newpath
1.000 UL
LTb
800 400 M
63 0 V
3487 0 R
-63 0 V
800 1313 M
63 0 V
3487 0 R
-63 0 V
800 2227 M
63 0 V
3487 0 R
-63 0 V
800 3140 M
63 0 V
3487 0 R
-63 0 V
1244 400 M
0 63 V
0 2677 R
0 -63 V
2131 400 M
0 63 V
0 2677 R
0 -63 V
3019 400 M
0 63 V
0 2677 R
0 -63 V
3906 400 M
0 63 V
0 2677 R
0 -63 V
1.000 UL
LTb
800 400 M
3550 0 V
0 2740 V
-3550 0 V
800 400 L
1.000 UP
1.000 UL
LT0
1785 1373 M
0 45 V
-31 -45 R
62 0 V
-62 45 R
62 0 V
-18 -64 R
0 10 V
-31 -10 R
62 0 V
-62 10 R
62 0 V
36 -71 R
0 9 V
-31 -9 R
62 0 V
-62 9 R
62 0 V
40 -48 R
0 8 V
-31 -8 R
62 0 V
-62 8 R
62 0 V
31 -40 R
0 6 V
-31 -6 R
62 0 V
-62 6 R
62 0 V
102 -52 R
0 6 V
-31 -6 R
62 0 V
-62 6 R
62 0 V
280 -97 R
0 6 V
-31 -6 R
62 0 V
-62 6 R
62 0 V
2886 991 M
0 4 V
-31 -4 R
62 0 V
-62 4 R
62 0 V
3640 875 M
0 4 V
-31 -4 R
62 0 V
-62 4 R
62 0 V
1785 1396 CircleF
1798 1359 CircleF
1865 1298 CircleF
1936 1258 CircleF
1998 1225 CircleF
2131 1179 CircleF
2442 1088 CircleF
2886 993 CircleF
3640 877 CircleF
1.000 UP
1.000 UL
LT0
1244 2606 M
0 27 V
-31 -27 R
62 0 V
-62 27 R
62 0 V
590 -137 R
0 46 V
-31 -46 R
62 0 V
-62 46 R
62 0 V
102 -44 R
0 37 V
-31 -37 R
62 0 V
-62 37 R
62 0 V
102 -71 R
0 37 V
-31 -37 R
62 0 V
-62 37 R
62 0 V
147 -80 R
0 36 V
-31 -36 R
62 0 V
-62 36 R
62 0 V
57 -45 R
0 41 V
-31 -41 R
62 0 V
-62 41 R
62 0 V
58 -57 R
0 36 V
-31 -36 R
62 0 V
-62 36 R
62 0 V
1244 2619 Circle
1865 2519 Circle
1998 2517 Circle
2131 2482 Circle
2309 2439 Circle
2397 2432 Circle
2486 2414 Circle
stroke
grestore
end
showpage
}}%
\put(2575,50){\makebox(0,0){\Large{$\beta=\frac{2N}{g^2}$}}}%
\put(100,2020){\makebox(0,0){\Large{$a\surd\sigma$}}}%
\put(3906,300){\makebox(0,0){\ {$46$}}}%
\put(3019,300){\makebox(0,0){\ {$45$}}}%
\put(2131,300){\makebox(0,0){\ {$44$}}}%
\put(1244,300){\makebox(0,0){\ {$43$}}}%
\put(750,3140){\makebox(0,0)[r]{\ \ {$1.2$}}}%
\put(750,2227){\makebox(0,0)[r]{\ \ {$0.8$}}}%
\put(750,1313){\makebox(0,0)[r]{\ \ {$0.4$}}}%
\put(750,400){\makebox(0,0)[r]{\ \ {$0$}}}%
\end{picture}%
\endgroup

\end	{center}
\vskip 0.15in
\caption{The SU(8) string tension versus the inverse lattice coupling,
including the region of the first order `bulk' transition between 
strong and weak coupling. Values $\circ$ are obtained coming
from strong coupling, while the values $\bullet$ are obtained coming
from weak coupling.}
\label{fig_bulkn8}
\end 	{figure}

Eqn(\ref{eqn_ag}) is only valid in weak coupling. This is
as true of the lattice corrections as of the perturbative
factor. In strong coupling the appropriate relationship
is a very different one,
\be
a^2\sigma
\stackrel{g^2_L\to\infty}{=}
-\log\frac{1}{g_L^2}
+\sum_{n=0} c_n \left(\frac{1}{g_L^2}\right)^n .
\label{eqn_sigSC}
\ee
(See 
\cite{MMbook}
for a detailed review.)  
A problem that now arises for small $N$, and in particular
for SU(3), is that there is
a smooth cross-over between the strong and weak coupling 
regimes, with a mid-point that occurs close to the value of 
$\beta$ where one typically starts to calculate interesting 
physical quantities. This is illustrated in Fig.~\ref{fig_bulkn3}
where we show how $a\surd\sigma$ varies with $\beta$ for SU(3). The
point of inflection, around $\beta \in [5.50,5.60]$, provides
an estimate of the mid-point of the cross-over. Clearly
there must be some range of $\beta$ beyond that where 
the functional form is still not pure weak coupling, 
but a priori we do not know how far that range extends.
This complication disappears for $N\geq 5$ where
the strong-weak coupling transition becomes a first order
phase transition. This is illustrated in Fig.~\ref{fig_bulkn8}
for SU(8). Here we see  a large hysteresis indicating a
strong first order transition. The simultaneous large
jump in the average plaquette, and hence action, shows
this to be a `bulk' transition where the physics changes on
all length scales. It is reasonable to expect 
that on the weak-coupling side of this bulk 
transition there are no strong-coupling artifacts. We shall
take advantage of this feature by first testing our 
weak-coupling fit in eqn(\ref{eqn_ag}) to the values
of $a\sqrt\sigma$ obtained on the whole  weak coupling branch of
the SU(8) theory, and then decreasing $N$ to $N=3$ (and $N=2$) 
having established in the much less ambiguous large-$N$ case
what kind of fit is needed.

Our assumption that there are no strong-coupling artifacts
on the weak-coupling side of the `bulk' transition is
not only plausible but is supported by analytic calculations
in a related but simpler context: SU($N$) lattice gauge theories 
in $1+1$ dimensions. Here (with the plaquette action) there is a 
strong-weak coupling cross-over at finite $N$ that becomes a 
third-order phase transition at $N=\infty$
\cite{gross-witten}.
In this limit one can calculate the string tension as a function
of $\beta$ analytically 
\cite{gross-witten}
and one finds a simple expansion in powers of $1/\beta$ with
no strong-coupling artifacts anywhere on the weak-coupling branch.  
While this does not prove that the same will be true in the case of
$D=3+1$, it provides an additional argument that such artifacts
will be either absent or highly suppressed.  

An interesting question concerns the functional form, for  $N\leq 4$,
in the extended cross-over region where the strong coupling expansion 
gradually transforms into a weak-coupling one. We can obtain
some intuition from the $D=1+1$ case where the whole problem
reduces to a single SU($N$) integral
\cite{gross-witten}.
In particular, for SU(2) the running of the coupling becomes quite 
elementary,
\be
u_p = \exp\{-a^2\sigma\} =
\frac{\int_{-1}^1 dz z (1-z^2)^{\frac{1}{2}}
\exp\left( \beta z \right) }
{\int_{-1}^1 dz (1-z^2)^{\frac{1}{2}}\exp\left(\beta z\right)}
\label{eqn_d2n2sigma}
\ee
and one can analyse the behaviour in the cross-over region.
We note that the terms in eqn(\ref{eqn_d2n2sigma}) are obtained by 
applying 
$\partial/\partial\beta (1-\partial^2/\partial\beta^2)^\frac{1}{2}$
or $(1-\partial^2/\partial\beta^2)^\frac{1}{2}$ to
\be
\int_{-1}^1 dz \exp\left( \beta z \right) 
=
\frac{1}{\beta}\left( e^{\beta} -  e^{-\beta}\right).
\label{eqn_int}
\ee
So on the weak coupling side we will have, in addition to
the expected weak coupling expansion in powers of $1/\beta$,
a correction term that is an expansion in  powers of $e^{-2\beta}$,
whose coefficients are themselves series in  powers of $1/\beta$.
It is these exponential terms that are our strong-coupling artifacts. 
They arise from the compact integration range of the plaquette
in eqn(\ref{eqn_d2n2sigma}), and this general origin suggests
that they will be present in some form in higher dimensions.
This is most convincing for the average plaquette, where such
$O(e^{-2\beta})$ corrections would create serious complications
for the standard methods of extracting the gluon condensate 
in $D=3+1$ since they (and their larger-$N$ homologues) would 
dominate over the $O(a^4)$ condensate contribution.
We note the close connection
of these corrections to `$Z_N$ vortex-instantons' in $D=1+1$.
In $D=2+1$ these will generalise to $Z_N$ `monopole-instantons' 
and in  $D=3+1$ to $Z_N$ `monopoles', as well as $Z_N$ vortex lines 
and sheets. Such ultraviolet fluctuations will disorder small
Wilson loops, and will affect the string tension in strong coupling.
Around the cross-over one might expect some complicated
contribution that is the ratio of powers and exponentials in $\beta$.
While it is not easy to be precise about this, it is clear that
the functional form suggested by our $D=1+1$ example is very
different to anything one might imagine on the basis of weak 
coupling arguments, and it will be no surprise if we find
that at small $N$ we have to go further into weak coupling to be 
able to apply standard weak coupling fits.

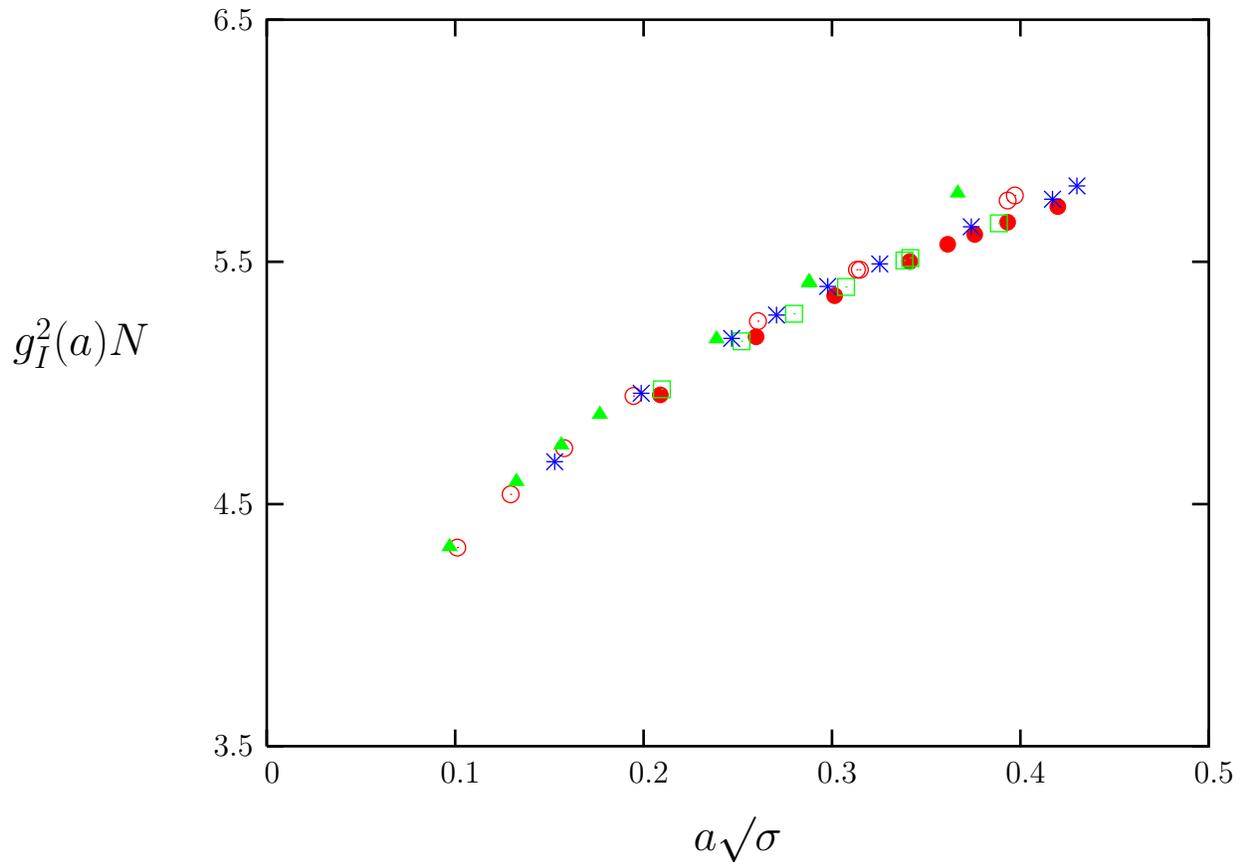
\begin	{figure}[p]
\begin	{center}
\leavevmode
\begingroup%
  \makeatletter%
  \newcommand{\GNUPLOTspecial}{%
    \@sanitize\catcode`\%=14\relax\special}%
  \setlength{\unitlength}{0.1bp}%
{\GNUPLOTspecial{!
/gnudict 256 dict def
gnudict begin
/Color true def
/Solid false def
/gnulinewidth 5.000 def
/userlinewidth gnulinewidth def
/vshift -33 def
/dl {10 mul} def
/hpt_ 31.5 def
/vpt_ 31.5 def
/hpt hpt_ def
/vpt vpt_ def
/M {moveto} bind def
/L {lineto} bind def
/R {rmoveto} bind def
/V {rlineto} bind def
/vpt2 vpt 2 mul def
/hpt2 hpt 2 mul def
/Lshow { currentpoint stroke M
  0 vshift R show } def
/Rshow { currentpoint stroke M
  dup stringwidth pop neg vshift R show } def
/Cshow { currentpoint stroke M
  dup stringwidth pop -2 div vshift R show } def
/UP { dup vpt_ mul /vpt exch def hpt_ mul /hpt exch def
  /hpt2 hpt 2 mul def /vpt2 vpt 2 mul def } def
/DL { Color {setrgbcolor Solid {pop []} if 0 setdash }
 {pop pop pop Solid {pop []} if 0 setdash} ifelse } def
/BL { stroke userlinewidth 2 mul setlinewidth } def
/AL { stroke userlinewidth 2 div setlinewidth } def
/UL { dup gnulinewidth mul /userlinewidth exch def
      10 mul /udl exch def } def
/PL { stroke userlinewidth setlinewidth } def
/LTb { BL [] 0 0 0 DL } def
/LTa { AL [1 udl mul 2 udl mul] 0 setdash 0 0 0 setrgbcolor } def
/LT0 { PL [] 1 0 0 DL } def
/LT1 { PL [4 dl 2 dl] 0 1 0 DL } def
/LT2 { PL [2 dl 3 dl] 0 0 1 DL } def
/LT3 { PL [1 dl 1.5 dl] 1 0 1 DL } def
/LT4 { PL [5 dl 2 dl 1 dl 2 dl] 0 1 1 DL } def
/LT5 { PL [4 dl 3 dl 1 dl 3 dl] 1 1 0 DL } def
/LT6 { PL [2 dl 2 dl 2 dl 4 dl] 0 0 0 DL } def
/LT7 { PL [2 dl 2 dl 2 dl 2 dl 2 dl 4 dl] 1 0.3 0 DL } def
/LT8 { PL [2 dl 2 dl 2 dl 2 dl 2 dl 2 dl 2 dl 4 dl] 0.5 0.5 0.5 DL } def
/Pnt { stroke [] 0 setdash
   gsave 1 setlinecap M 0 0 V stroke grestore } def
/Dia { stroke [] 0 setdash 2 copy vpt add M
  hpt neg vpt neg V hpt vpt neg V
  hpt vpt V hpt neg vpt V closepath stroke
  Pnt } def
/Pls { stroke [] 0 setdash vpt sub M 0 vpt2 V
  currentpoint stroke M
  hpt neg vpt neg R hpt2 0 V stroke
  } def
/Box { stroke [] 0 setdash 2 copy exch hpt sub exch vpt add M
  0 vpt2 neg V hpt2 0 V 0 vpt2 V
  hpt2 neg 0 V closepath stroke
  Pnt } def
/Crs { stroke [] 0 setdash exch hpt sub exch vpt add M
  hpt2 vpt2 neg V currentpoint stroke M
  hpt2 neg 0 R hpt2 vpt2 V stroke } def
/TriU { stroke [] 0 setdash 2 copy vpt 1.12 mul add M
  hpt neg vpt -1.62 mul V
  hpt 2 mul 0 V
  hpt neg vpt 1.62 mul V closepath stroke
  Pnt  } def
/Star { 2 copy Pls Crs } def
/BoxF { stroke [] 0 setdash exch hpt sub exch vpt add M
  0 vpt2 neg V  hpt2 0 V  0 vpt2 V
  hpt2 neg 0 V  closepath fill } def
/TriUF { stroke [] 0 setdash vpt 1.12 mul add M
  hpt neg vpt -1.62 mul V
  hpt 2 mul 0 V
  hpt neg vpt 1.62 mul V closepath fill } def
/TriD { stroke [] 0 setdash 2 copy vpt 1.12 mul sub M
  hpt neg vpt 1.62 mul V
  hpt 2 mul 0 V
  hpt neg vpt -1.62 mul V closepath stroke
  Pnt  } def
/TriDF { stroke [] 0 setdash vpt 1.12 mul sub M
  hpt neg vpt 1.62 mul V
  hpt 2 mul 0 V
  hpt neg vpt -1.62 mul V closepath fill} def
/DiaF { stroke [] 0 setdash vpt add M
  hpt neg vpt neg V hpt vpt neg V
  hpt vpt V hpt neg vpt V closepath fill } def
/Pent { stroke [] 0 setdash 2 copy gsave
  translate 0 hpt M 4 {72 rotate 0 hpt L} repeat
  closepath stroke grestore Pnt } def
/PentF { stroke [] 0 setdash gsave
  translate 0 hpt M 4 {72 rotate 0 hpt L} repeat
  closepath fill grestore } def
/Circle { stroke [] 0 setdash 2 copy
  hpt 0 360 arc stroke Pnt } def
/CircleF { stroke [] 0 setdash hpt 0 360 arc fill } def
/C0 { BL [] 0 setdash 2 copy moveto vpt 90 450  arc } bind def
/C1 { BL [] 0 setdash 2 copy        moveto
       2 copy  vpt 0 90 arc closepath fill
               vpt 0 360 arc closepath } bind def
/C2 { BL [] 0 setdash 2 copy moveto
       2 copy  vpt 90 180 arc closepath fill
               vpt 0 360 arc closepath } bind def
/C3 { BL [] 0 setdash 2 copy moveto
       2 copy  vpt 0 180 arc closepath fill
               vpt 0 360 arc closepath } bind def
/C4 { BL [] 0 setdash 2 copy moveto
       2 copy  vpt 180 270 arc closepath fill
               vpt 0 360 arc closepath } bind def
/C5 { BL [] 0 setdash 2 copy moveto
       2 copy  vpt 0 90 arc
       2 copy moveto
       2 copy  vpt 180 270 arc closepath fill
               vpt 0 360 arc } bind def
/C6 { BL [] 0 setdash 2 copy moveto
      2 copy  vpt 90 270 arc closepath fill
              vpt 0 360 arc closepath } bind def
/C7 { BL [] 0 setdash 2 copy moveto
      2 copy  vpt 0 270 arc closepath fill
              vpt 0 360 arc closepath } bind def
/C8 { BL [] 0 setdash 2 copy moveto
      2 copy vpt 270 360 arc closepath fill
              vpt 0 360 arc closepath } bind def
/C9 { BL [] 0 setdash 2 copy moveto
      2 copy  vpt 270 450 arc closepath fill
              vpt 0 360 arc closepath } bind def
/C10 { BL [] 0 setdash 2 copy 2 copy moveto vpt 270 360 arc closepath fill
       2 copy moveto
       2 copy vpt 90 180 arc closepath fill
               vpt 0 360 arc closepath } bind def
/C11 { BL [] 0 setdash 2 copy moveto
       2 copy  vpt 0 180 arc closepath fill
       2 copy moveto
       2 copy  vpt 270 360 arc closepath fill
               vpt 0 360 arc closepath } bind def
/C12 { BL [] 0 setdash 2 copy moveto
       2 copy  vpt 180 360 arc closepath fill
               vpt 0 360 arc closepath } bind def
/C13 { BL [] 0 setdash  2 copy moveto
       2 copy  vpt 0 90 arc closepath fill
       2 copy moveto
       2 copy  vpt 180 360 arc closepath fill
               vpt 0 360 arc closepath } bind def
/C14 { BL [] 0 setdash 2 copy moveto
       2 copy  vpt 90 360 arc closepath fill
               vpt 0 360 arc } bind def
/C15 { BL [] 0 setdash 2 copy vpt 0 360 arc closepath fill
               vpt 0 360 arc closepath } bind def
/Rec   { newpath 4 2 roll moveto 1 index 0 rlineto 0 exch rlineto
       neg 0 rlineto closepath } bind def
/Square { dup Rec } bind def
/Bsquare { vpt sub exch vpt sub exch vpt2 Square } bind def
/S0 { BL [] 0 setdash 2 copy moveto 0 vpt rlineto BL Bsquare } bind def
/S1 { BL [] 0 setdash 2 copy vpt Square fill Bsquare } bind def
/S2 { BL [] 0 setdash 2 copy exch vpt sub exch vpt Square fill Bsquare } bind def
/S3 { BL [] 0 setdash 2 copy exch vpt sub exch vpt2 vpt Rec fill Bsquare } bind def
/S4 { BL [] 0 setdash 2 copy exch vpt sub exch vpt sub vpt Square fill Bsquare } bind def
/S5 { BL [] 0 setdash 2 copy 2 copy vpt Square fill
       exch vpt sub exch vpt sub vpt Square fill Bsquare } bind def
/S6 { BL [] 0 setdash 2 copy exch vpt sub exch vpt sub vpt vpt2 Rec fill Bsquare } bind def
/S7 { BL [] 0 setdash 2 copy exch vpt sub exch vpt sub vpt vpt2 Rec fill
       2 copy vpt Square fill
       Bsquare } bind def
/S8 { BL [] 0 setdash 2 copy vpt sub vpt Square fill Bsquare } bind def
/S9 { BL [] 0 setdash 2 copy vpt sub vpt vpt2 Rec fill Bsquare } bind def
/S10 { BL [] 0 setdash 2 copy vpt sub vpt Square fill 2 copy exch vpt sub exch vpt Square fill
       Bsquare } bind def
/S11 { BL [] 0 setdash 2 copy vpt sub vpt Square fill 2 copy exch vpt sub exch vpt2 vpt Rec fill
       Bsquare } bind def
/S12 { BL [] 0 setdash 2 copy exch vpt sub exch vpt sub vpt2 vpt Rec fill Bsquare } bind def
/S13 { BL [] 0 setdash 2 copy exch vpt sub exch vpt sub vpt2 vpt Rec fill
       2 copy vpt Square fill Bsquare } bind def
/S14 { BL [] 0 setdash 2 copy exch vpt sub exch vpt sub vpt2 vpt Rec fill
       2 copy exch vpt sub exch vpt Square fill Bsquare } bind def
/S15 { BL [] 0 setdash 2 copy Bsquare fill Bsquare } bind def
/D0 { gsave translate 45 rotate 0 0 S0 stroke grestore } bind def
/D1 { gsave translate 45 rotate 0 0 S1 stroke grestore } bind def
/D2 { gsave translate 45 rotate 0 0 S2 stroke grestore } bind def
/D3 { gsave translate 45 rotate 0 0 S3 stroke grestore } bind def
/D4 { gsave translate 45 rotate 0 0 S4 stroke grestore } bind def
/D5 { gsave translate 45 rotate 0 0 S5 stroke grestore } bind def
/D6 { gsave translate 45 rotate 0 0 S6 stroke grestore } bind def
/D7 { gsave translate 45 rotate 0 0 S7 stroke grestore } bind def
/D8 { gsave translate 45 rotate 0 0 S8 stroke grestore } bind def
/D9 { gsave translate 45 rotate 0 0 S9 stroke grestore } bind def
/D10 { gsave translate 45 rotate 0 0 S10 stroke grestore } bind def
/D11 { gsave translate 45 rotate 0 0 S11 stroke grestore } bind def
/D12 { gsave translate 45 rotate 0 0 S12 stroke grestore } bind def
/D13 { gsave translate 45 rotate 0 0 S13 stroke grestore } bind def
/D14 { gsave translate 45 rotate 0 0 S14 stroke grestore } bind def
/D15 { gsave translate 45 rotate 0 0 S15 stroke grestore } bind def
/DiaE { stroke [] 0 setdash vpt add M
  hpt neg vpt neg V hpt vpt neg V
  hpt vpt V hpt neg vpt V closepath stroke } def
/BoxE { stroke [] 0 setdash exch hpt sub exch vpt add M
  0 vpt2 neg V hpt2 0 V 0 vpt2 V
  hpt2 neg 0 V closepath stroke } def
/TriUE { stroke [] 0 setdash vpt 1.12 mul add M
  hpt neg vpt -1.62 mul V
  hpt 2 mul 0 V
  hpt neg vpt 1.62 mul V closepath stroke } def
/TriDE { stroke [] 0 setdash vpt 1.12 mul sub M
  hpt neg vpt 1.62 mul V
  hpt 2 mul 0 V
  hpt neg vpt -1.62 mul V closepath stroke } def
/PentE { stroke [] 0 setdash gsave
  translate 0 hpt M 4 {72 rotate 0 hpt L} repeat
  closepath stroke grestore } def
/CircE { stroke [] 0 setdash 
  hpt 0 360 arc stroke } def
/Opaque { gsave closepath 1 setgray fill grestore 0 setgray closepath } def
/DiaW { stroke [] 0 setdash vpt add M
  hpt neg vpt neg V hpt vpt neg V
  hpt vpt V hpt neg vpt V Opaque stroke } def
/BoxW { stroke [] 0 setdash exch hpt sub exch vpt add M
  0 vpt2 neg V hpt2 0 V 0 vpt2 V
  hpt2 neg 0 V Opaque stroke } def
/TriUW { stroke [] 0 setdash vpt 1.12 mul add M
  hpt neg vpt -1.62 mul V
  hpt 2 mul 0 V
  hpt neg vpt 1.62 mul V Opaque stroke } def
/TriDW { stroke [] 0 setdash vpt 1.12 mul sub M
  hpt neg vpt 1.62 mul V
  hpt 2 mul 0 V
  hpt neg vpt -1.62 mul V Opaque stroke } def
/PentW { stroke [] 0 setdash gsave
  translate 0 hpt M 4 {72 rotate 0 hpt L} repeat
  Opaque stroke grestore } def
/CircW { stroke [] 0 setdash 
  hpt 0 360 arc Opaque stroke } def
/BoxFill { gsave Rec 1 setgray fill grestore } def
end
}}%
\begin{picture}(4500,3240)(0,0)%
{\GNUPLOTspecial{"
gnudict begin
gsave
0 0 translate
0.100 0.100 scale
0 setgray
newpath
1.000 UL
LTb
800 400 M
63 0 V
3487 0 R
-63 0 V
800 1313 M
63 0 V
3487 0 R
-63 0 V
800 2227 M
63 0 V
3487 0 R
-63 0 V
800 3140 M
63 0 V
3487 0 R
-63 0 V
800 400 M
0 63 V
0 2677 R
0 -63 V
1510 400 M
0 63 V
0 2677 R
0 -63 V
2220 400 M
0 63 V
0 2677 R
0 -63 V
2930 400 M
0 63 V
0 2677 R
0 -63 V
3640 400 M
0 63 V
0 2677 R
0 -63 V
4350 400 M
0 63 V
0 2677 R
0 -63 V
1.000 UL
LTb
800 400 M
3550 0 V
0 2740 V
-3550 0 V
800 400 L
1.000 UP
1.000 UL
LT0
3781 2435 CircleF
3592 2376 CircleF
3468 2330 CircleF
3366 2293 CircleF
3223 2228 CircleF
2940 2099 CircleF
2644 1944 CircleF
2283 1725 CircleF
1.000 UP
1.000 UL
LT0
3619 2477 Circle
3592 2458 Circle
3035 2197 Circle
3024 2197 Circle
2651 2003 Circle
2182 1721 Circle
1921 1524 Circle
1719 1350 Circle
1518 1149 Circle
1.000 UP
1.000 UL
LT1
3404 2485 TriUF
2844 2151 TriUF
2844 2146 TriUF
2495 1935 TriUF
2055 1650 TriUF
1910 1535 TriUF
1741 1397 TriUF
1489 1151 TriUF
1.000 UP
1.000 UL
LT1
3559 2372 Box
3225 2241 Box
3203 2232 Box
2983 2132 Box
2787 2031 Box
2589 1927 Box
2289 1746 Box
1.000 UP
1.000 UL
LT2
3853 2513 Star
3761 2463 Star
3455 2359 Star
3110 2219 Star
2914 2134 Star
2721 2026 Star
2552 1938 Star
2211 1731 Star
1885 1473 Star
stroke
grestore
end
showpage
}}%
\put(2575,50){\makebox(0,0){{\Large{$a\surd\sigma$}}}}%
\put(100,1920){\makebox(0,0){{\Large{$g^2_I(a)N$}}}}%
\put(4350,300){\makebox(0,0){\ {$0.5$}}}%
\put(3640,300){\makebox(0,0){\ {$0.4$}}}%
\put(2930,300){\makebox(0,0){\ {$0.3$}}}%
\put(2220,300){\makebox(0,0){\ {$0.2$}}}%
\put(1510,300){\makebox(0,0){\ {$0.1$}}}%
\put(800,300){\makebox(0,0){\ {$0$}}}%
\put(750,3140){\makebox(0,0)[r]{\ \ {$6.5$}}}%
\put(750,2227){\makebox(0,0)[r]{\ \ {$5.5$}}}%
\put(750,1313){\makebox(0,0)[r]{\ \ {$4.5$}}}%
\put(750,400){\makebox(0,0)[r]{\ \ {$3.5$}}}%
\end{picture}%
\endgroup

\end	{center}
\vskip 0.15in
\caption{The (mean-field improved) bare `t Hooft coupling
as a function ot the scale in units of the calculated string
tension, for $N=2$ ($\bigtriangleup$), $N=3$ ($\circ$),
$N=4$ ($\ast$), $N=6$ ($\Box$), and $N=8$ ($\bullet$).} 
\label{fig_gIN}
\end 	{figure}

Of course, using SU($N$) gauge theories at larger $N$ to teach
us something about SU(3) is only convincing if it is clear that
SU(3) is `close to' SU($\infty$). That this is so in the
context of the bare coupling was demonstrated in Fig.7 of
\cite{blmt-bulk4}
using calculations for $2\leq N \leq 5$. Here we repeat the exercise 
with calculations that not only go to larger $N$, but also 
to smaller values of $a$ at smaller $N$. 
So, for each value of $\beta$ at which we calculate
the string tension (see Section~\ref{subsection_results})
we extract the bare mean-field improved
`t Hooft coupling $g_I^2(a)N$ using eqn(\ref{eqn_gI}). We then
plot it against the length scale $a$ on which it is running,
with $a$ expressed in units of the string tension. Since the
physics has a smooth large-$N$ limit, this is a common unit 
up to corrections of $O(1/N^2)$ which we expect to be modest,
since this is what one finds for various mass ratios
\cite{blmt-bulk4}.
The resulting plot is shown in Fig.~\ref{fig_gIN}. This
provides convincing evidence that the weak-coupling
running of the bare coupling is in fact very similar for
all values of $N$. Thus it makes sense to extract lessons 
in this context from larger $N$ for all SU($N$) gauge theories.

\section{The calculation}
\label{section_calc}

\subsection{string tensions and ${\bf{\mathrm r_0}}$}
\label{subsection_results}

The string tensions used in this paper have been obtained
from calculations of the ground state energies of confining
flux loops wrapped around a spatial torus. We use the 
results of 
\cite{blmtuw}
supplemented in many cases by either higher statistics at
the same values of $\beta$ or new calculations at both 
higher and lower values of $\beta$. In Tables~\ref{table_su2K}-
\ref{table_su8K} we list these masses, $am_l$, together with 
the length, $L$, of the flux loop in each case, and the
average plaquette, $u_p$, which will be needed for 
transforming to the mean-field coupling scheme. 

To extract the string tension we use
\be
am_l(L) 
=
a^2\sigma L - \frac{\pi}{3L}.
\label{eqn_mtosig}
\ee
The linear term has been corrected by the universal
string correction
\cite{LSW}
with a coefficient that corresponds to a bosonic string 
theory (the universality class of the Nambu-Goto string).
That this is in fact the appropriate universality
class for the confining flux tube in D=3+1 non-Abelian
gauge theories, has received support from many lattice 
calculations in both SU(2) and SU(3) (see for example
\cite{csu2su3})
as well as at larger $N$
\cite{hmmt}.
The string lengths in Tables~\ref{table_su2K}-\ref{table_su8K}
have been chosen to be large enough, $aL\sqrt\sigma  \geq 3$,
that the string correction in eqn(\ref{eqn_mtosig})
provides only a $\sim 5\%$ shift in the calculated 
value of $a\sqrt\sigma$, so higher order corrections
in $1/a^2\sigma$ (the natural expansion parameter in
the relationship between $am_l(L)$ and $a^2\sigma$)
should be very small indeed. We shall estimate the
systematic error that this brings about below.

In the case of SU(3), the calculated values of the Sommer 
parameter, $r_0$, extend considerably further into weak coupling 
than the string tension. For this reason we will supplement
our fits of $a\surd\sigma$ with those of $a/r_0$, as listed in 
Table~\ref{table_su3r0}. These have been taken from Table 2.6 of
\cite{necco}
where, for the values at higher $\beta$, we have translated 
from $r_c$ to $r_0$ using eqn(2.73) therein. 
(We note from Fig.2.4 of
\cite{necco}
that the quoted error on the
conversion factor appears to encompass any possible dependence 
on $a$.) We remark that the lower $\beta$ values are from 
\cite{sommer2}
and the higher values from
\cite{necco2}.
We have calculated the values of the action listed in 
Table~\ref{table_su3r0}. These calculations have been
performed on $8^4$ lattices and will therefore differ slightly
from the large volume limit. However we have checked,
performing a calculation on $6^4$ lattices  for the values of
$\beta$ in Table~\ref{table_su3K}, that these minute finite 
volume corrections to the action, lead to errors that are
negligible compared to the relevant statistical errors.

\subsection{weak-coupling fits at all  N}
\label{subsection_fitsallN}

We begin by fitting the values of the SU(8) string tension that have
been calculated on the weak coupling branch in Fig.~\ref{fig_bulkn8}. 
We attempt to fit all the way to the point at which the system 
tunnels through to the strong coupling phase. Although at lower 
$\beta$ part of this branch is metastable, the 
barrier to the strong coupling phase is so large that its 
metastability is irrelevant. One can see this from the fact that
the distribution of values of the plaquette (averaged over
the whole volume for a given gauge field) shows no sign of
a tail developing towards the very different values that it
would take in the strong coupling phase.

Working in the  mean-field improved coupling scheme $s=I$,
we find that we can obtain an excellent fit to all the
values in Table~\ref{table_su8K}, using the $O(a^2)$, 3-loop 
truncation of eqn(\ref{eqn_ag})
\be
a \sqrt\sigma(a)
= 
\frac{\sqrt\sigma(0)}{\Lambda_I}
\left( 1 + c^I_\sigma a^2\sigma \right) 
e^{-\frac{1}{2\beta_0 g_I^2}}
\left(\frac{\beta_1}{\beta_0^2}+\frac{1}{\beta_0 g_I^2}
\right)^\frac{\beta_1}{2\beta_0^2}
e^{-\frac{\beta^I_2}{2\beta_0^2}g_I^2}.
\label{eqn_agfit}
\ee
The best fit is illustrated in Fig.~\ref{fig_gfitsu8} and the
fitted values of ${\sqrt\sigma}/{\Lambda_I}$ and the constant 
$c_I$ are given in Table~\ref{table_fitN}. We see that 
$c_I=1.18\pm 0.04$ is indeed $O(1)$ as naively expected. 
In fact any attempt to fit without
a lattice correction, i.e. with $c_I=0$, fails very badly,
even if we include in our fit only the values 
of $a\surd\sigma$ at the weakest couplings.
Thus for $N=8$ it is clear that the lattice spacing corrections
are important and once taken into account they allow a very good
fit with the usual 3-loop perturbative running coupling. This
turns out to be equally true for SU(6), and we list the fitted
parameters in Table~\ref{table_fitN}.

Having established at these larger $N$ that lattice spacing
corrections are indeed needed, we proceed to lower $N$
using the same functional form as in eqn(\ref{eqn_agfit}).
For SU(4), where the bulk transition has weakened to a 
relatively sharp cross-over
\cite{blmt-bulk4},
a good fit with eqn(\ref{eqn_agfit}) is still possible but
only if we restrict ourselves to somewhat weaker couplings.
This is displayed in Table~\ref{table_fitN} where we express 
the fitted range 
in terms of $a\surd\sigma$ so as to provide a common physical
measure of comparison for different $N$. In SU(3) the
bulk cross-over becomes smoother  and we have to move even
further into weak coupling to obtain an acceptable fit.
This is illustrated in Fig.~\ref{fig_gfitsu3}. Finally for
SU(2) the bulk crossover is smoother still and we find that
even if we include only the values at weakest coupling,
we obtain an uncomfortably large $\chi^2$ per degree of freedom,
as shown in  Table~\ref{table_fitN}. All this accords
with the naive expectation that the smoother the strong-to-weak
coupling crossover, the further into weak-coupling one
has to go before weak-coupling expansions become applicable. 

The computational cost of  SU($N$) lattice calculations
grows roughly $\propto N^3$, and so our SU(3) calculations 
extend to much smaller values of $a$ than
for SU(8). One might wonder if it is this greater range that
makes the fit to SU(3) so much more difficult than for SU(8).
In fact this is not the case. If we fit the SU(3) values
over a range of $a\surd\sigma$ values that is very similar
to that of the whole SU(8) range, we obtain a statistically
unacceptable $\chi^2/n_{df} \sim 8$.

We can now take each value of $\sqrt{\sigma}/\Lambda_I$
in Table~\ref{table_fitN} and transform the ratio to
$\sqrt{\sigma}/\Lambda_{\overline{MS}}$ using 
eqn(\ref{eqn_lamIlamMS}). If we then extrapolate to $N=\infty$ 
using the expected leading $O(1/N^2)$ correction, we obtain
a good fit for $N\geq 3$, giving
\be
\frac{\Lambda_{\overline{MS}}}{\sqrt\sigma}
= 
0.503(2)(?) + \frac{0.33(3)(?)}{N^2}  \qquad ; \qquad N\geq 3.
\label{eqn_siglamN}
\ee
The final result for $\Lambda_{\overline{MS}}$ is remarkably precise, 
but this appearance is quite misleading since the  quoted error is 
only statistical. The systematic errors, indicated by the `?' in 
eqn(\ref{eqn_siglamN}) are potentially very much larger.
The most significant uncertainty, the choice of coupling
scheme, is what we address first. There are some further smaller 
systematic errors (which are still much larger than the
statistical errors) which will be discussed later on in this Section.

One might have hoped hope that the accuracy of our calculation 
of $a\surd\sigma$ would prove sufficient to usefully constrain 
what are the best coupling schemes to use.
Unfortunately this turns out not to be the case. To show this
most graphically we consider three alternative schemes which
all have the same scale $\Lambda_s = \Lambda_I$. These are
obtained by replacing the value of the average plaquette
$u_p$ in eqn(\ref{eqn_gI}) by a truncation of its perturbative
expansion in eqn(\ref{eqn_gIgL}) to either $O(g^2)$, or $O(g^4)$, 
or $O(g^6)$. We label these schemes by $s=I_1,I_2,I_3$
respectively; i.e.
\be
\frac{1}{g^2_{I_j}} 
= \frac{1}{g^2_L}
\left(1-\omega_1 g^2_L- \cdots -\omega_j g^{2j}_L\right)
\label{eqn_gIjgL}
\ee
using the notation of  eqn(\ref{eqn_gIgL}).
For $j\geq 1$ these schemes are identical to the mean-field
scheme, $s=I$, to  $O(g^2)$, so they will have exactly the
same $\Lambda_s$ parameter. We now take eqn(\ref{eqn_agfit})
and fit our SU(8) string tensions, replacing the values of $g^2_I$ 
by the corrsponding values of $g^2_{I_i}$ for $i=1,2,3$.
We obtain perfectly good fits in all cases as shown in 
Table~\ref{table_fitN8scheme}. However, as we see there, the
fitted values of $\surd\sigma/\Lambda_I$ vary by almost a factor
of 2 between these four coupling schemes. (For completeness
we show the fit using the straightforward lattice
coupling scheme $s=L$, which, when translated to a value
of $\surd\sigma/\Lambda_{\overline{MS}}$, is similar to
the fit in the $I_2$ scheme.) We obtain very similar results for
$N=6$. Thus in the case of larger $N$, where the presence of 
a first order bulk transition defines for us where a weak 
coupling expansion should begin to be applicable, there 
is in practice  no useful constraint on the coupling scheme. 
At smaller $N$, where there are in fact marked differences in
where one can begin fitting with different schemes, 
the presence of a smooth strong-to-weak crossover 
means that it is not clear what significance one should
read into any particular comparison.

\subsection{SU(3) : choosing a scheme}
\label{subsection_fitsu3}

The simplest way to determine if any given bare coupling scheme 
is `good' is to compare it against a running coupling that 
has been calculated on the lattice with enough statistical and 
systematic precision to serve as a benchmark. A good example
is the Schrodinger Functional (SF) scheme of the 
Alpha Collaboration
\cite{SF0}.
In 
\cite{SF2}
the SU(3) calculation 
\cite{SF1}
has been extended so that it 
covers a range of energy scales comparable to that covered by
experimental measurements, but with much greater
precision. (Compare Fig.4 of
\cite{SF2}
with Fig.10 of
\cite{MS-expt2}.)
Its (continuum) running is well described by its
3-loop $\beta$-function
\cite{SF1,SF2},
while the size of the non-universal $\beta_2$ term is
known to be modest
\cite{SFbeta,SF2}.
All this encourages us to use it as our benchmark good
coupling scheme.

The $SF$ coupling is a coupling defined on a length scale
$la$ for a given $a$, where $l$ can take any integer value.
(Restrictions arise in practice.) The values of 
$\beta$ and $l$ at which calculations exist, with the 
corresponding values of $g^2_{SF}(al)$, are listed in 
Tables 3 and 6 of
\cite{SF2}.
Our strategy is to take some physical quantity  $\mu$ that
has been calculated over a large range of lattice couplings, 
interpolate from the values of $\beta$ at which it has been 
calculated to the values of $\beta$ at which
$g^2_{SF}(al)$ has been calculated for some $l$, and then
fit these values using a 3-loop perturbative expression,
modified by the expected lattice corrections, to obtain 
$\mu/\Lambda_{SF}$. (Note that in contrast to extrapolation, 
interpolation does not generate significant systematic errors.) 
We then fit these same values of $\mu$ using eqn(\ref{eqn_agfit})
with various improved lattice coupling schemes to obtain alternative
estimates of $\mu/\Lambda_{s}$ and hence of $\mu/\Lambda_{SF}$.
If the latter is consistent with the value obtained using
$g^2_{SF}(al)$, we have evidence that the corresponding 
`improvement' is indeed a `good' one.

Our string tension calculations in Table~\ref{table_su3K}
do not have enough overlap in $\beta$ with the $g^2_{SF}(la)$
calculations to be useful for this purpose. (Within
the relevant range $\beta \leq 6.515$, values of $g^2_{SF}$ 
have been calculated at only two values of $\beta$.)
Instead we turn to the $r_0$ calculations in Table~\ref{table_su3r0} 
which do have a useful overlap. We fit using 
\begin{eqnarray}
\frac{la}{r_0(a)}
& = & 
\frac{1}{r_0\Lambda_{SF}}
\left( 1 + c^{SF}_r \frac{a^2}{r^2_0}
+ d^{SF}_r \frac{1}{l^p}  \right)  \nonumber \\
&\times&
e^{-\frac{1}{2\beta_0 g_{SF}^2(la)}}
\left(\frac{\beta_1}{\beta_0^2}+\frac{1}{\beta_0 g_{SF}^2(la)}
\right)^\frac{\beta_1}{2\beta_0^2}
e^{-\frac{\beta^{SF}_2}{2\beta_0^2}g_{SF}^2(la)}.
\label{eqn_agfitSF}
\end{eqnarray}
Here there are two lattice spacing corrections. The usual $O(a^2)$
term arises from corrections to $r_0(a)$ etc. while the  
$O(1/l^p)$ term arises from lattice corrections to 
$g^2_{SF}(la)$ on the scale $l \times a$. As indicated in
\cite{SF1,SF2}
one expects the leading correction for the $SF$ scheme to have $p=1$. 
However the fact that the coupling is Symanzik-improved means that 
the dominant correction in our range of $a$ might in fact have $p=2$
\cite{SF1,SF2}.
We shall use fits with both values of $p$ taking the
difference as part of our estimate of the systematic error.
We expect (as usual) that $c_r \sim O(1)$ but we anticipate 
that $d_r$ will be small given that very small scaling violations
are seen in the step-scaling function in
\cite{SF1,SF2}.)

To obtain as strongly constrained a fit as possible, 
we want to maximise the number 
of values of the running coupling that we fit in the range 
$\beta\in [5.70,6.92]$ where we have calculations of $a/r_0$. 
We therefore use not only the values of $g^2_{SF}(la)$ 
and $g^2_{SF}(2la)$ in Table 3 of
\cite{SF2}
but also the values of $g^2_{SF}(la)$ listed in Table 6 therein.
(We exclude the last 4 rows of Table 6 since they use a different 
improvement.) We list in Table~\ref{table_fitr0N3SF} the results
of our various fits. We show separately the results of using
different interpolations in obtaining $a/r0$. As expected we
find that this makes a negligible difference. We show fits
for powers $p=1$ and $p=2$ and  see that this makes a small
$\sim 2\%$ difference. Finally we give an example of
a fit that uses values in the additional range
$\beta\in [6.92,7.26]$ where we obtain the values of $a/r_0$
by the less reliable process of extrapolation. Again there is
only a small change in the fit.

We note that the region of couplings $\beta \in [6.22,6.91]$ that
we actually use in our fit, corresponds to the range of scales 
\be
\frac{\mu}{\Lambda_{SF}} 
= 
\frac{1}{a \Lambda_{SF}}
= 
\frac{1}{r_0 \Lambda_{SF}} \times \frac{r_0}{a}
\in
[23.9,58.6].
\label{eqn_SFscales}
\ee
If we look at Fig.4 in
\cite{SF2}
we see that in this range, the 3-loop formula we use already 
appears to provide a very good approximation to the continuum
$\beta$-function. The lattice spacing corrections are less
well determined but by taking the range of results spanned by 
both the $p=1$ and the $p=2$ fits, our estimate of this systematic
error should be credible. Putting all this together we 
obtain 
\be
\frac{1}{r_0 \Lambda_{SF}} = 3.2(1)
\longrightarrow
r_0  \Lambda_{\overline{MS}} = 0.640(20)
\label{eqn_r0SF}
\ee
using $\Lambda_{SF} \simeq 0.48811 \Lambda_{\overline{MS}}$
\cite{SF1}.
If we now compare this with the values in
Table~\ref{table_fitr0N3scheme} that have been obtained by fitting 
$r_0/a$ using eqn(\ref{eqn_agfit}) with various bare coupling 
schemes, we see that the Mean-Field scheme produces values that
are consistent. It is therefore plausible to adopt 
the latter as our `good' scheme, while incorporating a systematic
error $\sim 6\%$ based on the difference between the value
in  Table~\ref{table_fitr0N3scheme} and the value $\sim 0.66$ 
that one gets at one standard deviation in eqn(\ref{eqn_r0SF}). 

As a by-product of these calculations we also obtain an updated
value for the relationship between the two standard scales
$r_0$ and $\surd\sigma$:
\be
r_0 \surd\sigma = 1.160(6)(6)   \quad : \quad  SU(3).
\label{eqn_r0sigma}
\ee
The first error is statistical and the second is systematic.
It includes a $\sim 0.5\%$ error on the value of $\sigma$
and a  $\sim 0.2\%$ error from $O(a^4)$ corrections. 
(See Section~\ref{subsection_syserr}.) We cannot
estimate, and therefore neglect, any systematic error on $r_0$.
Note that there are no perturbative uncertainties here, as
we see from the similarity of the various estimates of 
$r_0 \surd\sigma$ in Table~\ref{table_fitr0N3scheme}. What we are
doing is equivalent to interpolating $a/r_0$ and $a\surd\sigma$ 
to common values of $\beta$, taking the ratio, and performing
a continuum extrapolation with conventional lattice
corrections. This process is insensitive to the precise
interpolation as long as it is smooth.

\subsection{comparing schemes directly}
\label{subsection_direct}

Here we introduce a method for comparing different 
coupling schemes directly, without the use of any
physical quantity such as $r_0/a$ or $a\surd\sigma$.
This has the great advantage that it allows us to  perform 
comparisons much deeper into weak coupling.

For a scheme $s$ define the 3-loop perturbative factor
\be
F^s_3[g^2_s]
=
e^{-\frac{1}{2\beta_0 g_s^2}}
\left(\frac{\beta_1}{\beta_0^2}+\frac{1}{\beta_0 g_s^2}
\right)^\frac{\beta_1}{2\beta_0^2}
e^{-\frac{\beta^s_2}{2\beta_0^2}g_s^2}.
\label{eqn_F3}
\ee
Now we expect for the SF scheme
\be
la\Lambda_{SF}
\simeq
\left\{ 1+\frac{c_1}{l^p} \right\}
F^{SF}_3[g^2_{SF}(al)]
\label{eqn_aSF}
\ee
and for a `good' improved scheme $I^\prime$ based on the
bare lattice coupling
\be
a\Lambda_{I^\prime}
\simeq
\left\{ 1+ c^\prime a^2 \right\}
F^{I^\prime}_3[g^2_{I^\prime}(a)],
\label{eqn_aI}
\ee
up to the various higher order corrections. If we now replace the
$a^2$ on the RHS of eqn(\ref{eqn_aI}) by the expression for $a$ in  
eqn(\ref{eqn_aSF}), and if we then take the ratio of the two equations, 
we obtain
\be
\frac{\Lambda_{SF}}{\Lambda_{I^\prime}}
\simeq
c_0
=
\frac{1}{l}
\frac{F^{SF}_3[g^2_{SF}(al)]}{F^I_3[g^2_{I^\prime}(a)]}
\frac{\left\{ 1+\frac{c_1}{l^p} \right\}}
{ \left\{ 1+ c_2 \frac{1}{l^2} \left\{ 1+\frac{c_1}{l^p} \right\}^2 
\{F^{SF}_3[g^2_{SF}(al)]\}^2\right\} }
\label{eqn_SFI}
\ee
where $c_2 = c^\prime/\Lambda^2_{SF}$.
We can now perform a fit for the constants $c_0$, $c_1$ and $c_2$
over ranges of $\beta$ further and further into weak coupling,
and see how rapidly $c_0$ approaches the known value of
${\Lambda_{SF}}/{\Lambda_{I^\prime}}$. The more rapidly it
does so, the `better' we may judge the coupling scheme to be.

Before proceeding, some remarks. We do not expect
exact agreement between the fitted and known values of 
${\Lambda_{SF}}/{\Lambda_{I^\prime}}$ because we are missing
the 4-loop and higher contributions to the $\beta$-function.
Indeed it is precisely the discrepancy that will tell us how
`good' is our $I^\prime$ scheme.
In addition, over the wider range of $g^2(a)$ values that
becomes accessible with this method, we may well need
to worry about the fact that our supposed constants,
$c_1$ and $c_2$, are in fact power series in $g^2$. These,
and other issues that arise in such a calculation, we
shall ignore, with the caution that one should 
therefore regard our calculation as being less
than fully quantitative. On the other hand, we note that
the $O(a^2)$ corrections to eqns(\ref{eqn_aSF},\ref{eqn_aI})
are in fact negligible in such a comparison except at the
very smallest values of $\beta$, thus largely eliminating
one source of systematic error.

We have grouped the calculated values of $g^2_{SF}$
by the ranges of $\beta$ within which they have been 
calculated, and we have fitted each such group of
values separately. Each group has $\sim 10$ values
of $g^2_{SF}$ which is more than enough to constrain
the three parameters in the fit. We can thus see
how the fitted value of $c_0$ compares to the ratio
of the $\Lambda$ parameters as we go further into 
weak coupling. We do so separately for the mean-field 
coupling, $I^\prime=I$, for the variation $I^\prime=I_3$
where the plaquette is replaced by its 3-loop
approximation, and finally for the
lattice bare coupling, $I^\prime=L$. We choose to take a
power $p=2$ although in practice we get much the
same picture with $p=1$.

The results of our fits are shown in Fig.~\ref{fig_c0}.
We plot there the ratio of $c_0$ to its asymptotic value,
i.e. $\Lambda_{SF}/\Lambda_{I}$ for schemes
$I$ and $I_3$,   and  $\Lambda_{SF}/\Lambda_{L}$ for the
lattice scheme. We see quite clearly that 
Parisi's original mean-field scheme, using the
full plaquette in the improvement, is consistent
with converging remarkably quickly to the expected
value (with a small deviation that is consistent
with a slowly decreasing 4-loop term). 
If the $SF$ coupling is `good', 
then so, it would appear, is the mean-field scheme.
The other schemes considered fare much less
well by this criterion.

A final aside. We have emphasised that the $SF$ coupling possesses 
various desirable properties -- it has been calculated over a 
very large range of scales, it is very precise, it has
modest higher order perturbative corrections (as 
indicated by the results of
\cite{SF1,SF2}).
All this makes it an attractive benchmark coupling. However we 
also recall that $g^2_{SF}(l)$ is defined on a $l^4$ torus. Now, 
it is known that SU($N=\infty$) gauge theories suffer a sequence of 
phase transitions as $l\to 0$ (see
\cite{NN}
and references therein).
Each of these transitions involves the breaking of a $Z_N$ symmetry
in one of the space-time directions. The first transition is the usual 
deconfining transition and the remaining ones can be interpreted
as continuations of deconfining transitions on ever more
dimensionally reduced space-time volumes
\cite{FBMT}.
At finite $N$ these phase transitions will  become cross-overs.
Such cross-overs will, in general, contribute non-perturbatively
to the running of $g^2_{SF}(l)$. Although one might argue
that such contributions will be negligible at the level of
accuracy we are aiming for here, this needs to be checked
and, in any case, this raises interesting issues
(in its own right) which need to be addressed.

\subsection{other systematic errors}
\label{subsection_syserr}

There are a number of other, relatively straightforward, systematic 
errors that we now address. These include errors from neglecting 
higher order corrections in $1/l$ in  eqn(\ref{eqn_mtosig})
when extracting $a^2\sigma$ from the loop masses,
and also errors in the actual calculation of 
these loop masses. There is the error due to the neglect 
of  higher order corrections in $a^2$ in eqn(\ref{eqn_agfit}).
Then there is an error in using only a leading $O(1/N^2)$ 
correction in eqn(\ref{eqn_siglamN}). Finally there are
the errors arising from our ignorance of  
higher order perturbative corrections in eqn(\ref{eqn_agfit}),
within the mean-field improved coupling scheme. Assuming 
this to be a `good' scheme, as argued above, these last corrections 
can be plausibly bounded. 

We can obtain an estimate of the magnitude of the correction to
eqn(\ref{eqn_mtosig}) from the calculation in
\cite{hmmt}.
One finds that for SU(6), and SU(4), the extra correction can be 
fitted by a term that one can rewrite as $-1.23(21)/a^2\sigma L^3$. 
For our lattices $a^2\sigma L^2 \sim 10$, so this corresponds to a 
$\sim 10\%$ addition to the bosonic string correction, and 
hence a $\sim 0.5\%$ increase in the final estimate of
$a\surd\sigma$. This provides us with an estimate of our systematic 
error from this source.

We extract the loop masses by identifying effective mass plateaux
in appropriate correlators obtained from a variational calculation.
In practice the best variational ground state has an overlap of 
$\sim 99\%$ onto the true ground state so any error should be very 
small. In the technically very similar case of $D=2+1$ that was
analysed in
\cite{bbmtd3},
the shift in the extracted value of $a\surd\sigma$ induced by excited 
string states was shown to be less than  $0.5\%$. We can take this as 
a bound on the corresponding systematic error in the present
calculation. Note that this error will decrease the string tension
and will therefore partially cancel against the error discussed
in the previous paragraph.

The fitted values of $\surd\sigma/\Lambda_I$ turn out to be
very robust against the inclusion of additional $O(a^4)$ corrections
in eqn(\ref{eqn_agfit}), with shifts that are typically $\sim 0.25\%$.

The perturbative expression in  eqn(\ref{eqn_agfit}) is missing
an extra factor $\sim \left(1 + \sum d_n g^{2n}\right)$
that arises from the unknown higher order perturbative corrections. 
(Ignoring complications that arise from the - at best - asymptotic
nature of the expansion.) One might imagine that the simplest way to 
proceed is to try and fit the first couple of terms in this series.
This turns out not to work. The reason is that over our range of
scales, the coupling $g^2(a)$ varies very little; thus the 
correction term is to first approximation just a constant
$\left(1 + \sum d_n g^{2n}\right) 
\sim \left(1 + \sum d_n \overline{g^{2n}}\right) $ which
renormalises the fitted value of the overall constant coefficient
$\sqrt\sigma/\Lambda_I$. For example, if we remove from our fit
the 3-loop factor
$\exp\{-\beta^I_2 g_I^2/2\beta_0^2\}$ in eqn(\ref{eqn_agfit})
we find that we still get a perfectly good fit, but with a value 
for $\sqrt\sigma/\Lambda_I$ that is increased by a factor 
$\sim 1.14$. Now if we use, say, the value $g^2_I N \simeq 5.36$
that we get at $\beta=44.35$ in  Table~\ref{table_su8K},
we find that 
$\exp\{-\beta^s_2 g_I^2/2\beta_0^2\} \sim 1.17$ 
which is numerically  similar. What this means is that 
neglecting higher order perturbative corrections effectively shifts
$\sqrt\sigma/\Lambda_I$  away from its true value. In the
example above, neglecting the 3-loop contribution induces
a systematic error in the evaluation of $\sqrt\sigma/\Lambda_I$ 
that is roughly $15\%$. If we have a well-behaved coupling scheme,
one might hope that the error due to the neglect of 
4-loop and higher order corrections will be roughly the square of 
this, i.e. $\sim 2-3 \%$. In fact if we estimate, in the above way,
the shift induced
by the $O(g^4)$ term in eqn(\ref{eqn_ag}) in the $\overline{MS}$
scheme, where the necessary 4-loop calculation has been performed
\cite{MS-4loop},
we find a shift of this magnitude. Assuming that our mean-field
improved scheme is also a good one, it then seems safe to bound
the systematic error from this source by $\sim \pm 4\%$.

While our 2-loop expression in eqn(\ref{eqn_agfit}) is exact, 
the 3-loop expression is kept only to leading order in $g^2$.
This makes perfect sense since at higher orders in $g^2$
one obtains the unknown contributions from higher loops
discussed above. Nonetheless we have checked the difference it 
makes to our fits and extrapolations if we do employ an exact
3-loop expression (by using an expansion to much higher
order in $g^2$) and what we find is that the shift in calculated 
quantities is only at the $\sim 0.5\%$ level for both the
SF and mean-field coupling schemes.

Finally, we find that the inclusion of an additional $O(1/N^4)$ 
correction in the large-$N$ extrapolation in eqn(\ref{eqn_siglamN}) 
leads to a shift of no more than $\sim 0.5\%$. 

Note that all the systematic errors we discuss will to a first 
approximation be independent of $N$ and should therefore not affect 
the quality of the large-$N$ extrapolation, but will merely
add extra uncertainties to the fitted values.

We thus estimate a reasonable  bound on the systematic error 
from these sources to be $\pm 5\%$. Adding (in quadrature) the
$\sim 6\%$ systematic error coming from the choice of scheme
(as estimated in Section~\ref{subsection_fitsu3}) we shall
take   $\pm 8\%$ as our total systematic error.

\section{Conclusions}
\label{section_conc}

As a by-product of the calculations in
Section~\ref{subsection_fitsu3} we obtained an updated
value for the relationship between the two standard scales
$r_0$ and $\surd\sigma$ in the continuum limit : 
$r_0 \surd\sigma = 1.160(6)(6)$
where the first error is statistical and the second is systematic.

One of the main purposes of our calculation was to determine the
importance of lattice corrections in the relationship
between $g^2(a)$ and the lattice spacing $a$, where $g^2(a)$
is the bare lattice coupling, or some improvement thereof. 
The novelty of our approach is to do so for SU(6) and SU(8),
where the first-order strong-to-weak coupling transition
gives unambiguous guidance as to where a weak-coupling fit
should be applicable. By contrast, in SU(3) the presence
of a smooth strong-to-weak coupling cross-over makes
it difficult to evaluate the apparent success or failure
of different weak-coupling fits.

We found that, at larger $N$, the variation of $a\surd\sigma$ with  
$g^2(a)$ did indeed demand the 
presence of an $O(a^2)$ correction to the 3-loop perturbative
running, and that its coefficient is $O(1)\times \sigma$ as
one would naively expect. Indeed, such a fit accurately
describes the running of $g^2(a)$ along the whole weak-coupling
branch, including the metastable portion that extends 
well beyond the location of the bulk transition.

As we go to lower $N$, we find that we need to go deeper into 
weak-coupling before such weak coupling fits start working. 
Indeed for SU(2), 
it is not clear if there is a significant such region below
$\beta =2.70$ (the weakest coupling at which we have a
calculation of $a^2\sigma$). An interesting question concerns
the functional form in the extended cross-over region
where the strong coupling expansion gradually transforms into a 
weak-coupling one. As we discussed in Section~\ref{section_coupling}, 
a good place to begin such an analysis is in the $D=1+1$ case
where analytic arguments are possible
\cite{gross-witten}.

Having a well-motivated lattice correction
\cite{allton}
to the running coupling, invites us to address the more ambitious goal 
of calculating $\Lambda_{\overline{MS}}/\surd\sigma$ for various 
$N$, so as to interpolate and extrapolate to all $N$, including
$N=\infty$. However in the range of scales $a\surd\sigma$ where
calculations exist, different perturbative schemes give quite
different results. This well-known problem is usually
addressed by `improving' the lattice bare coupling.
However this turns not to be sufficient. Even if we use
variations on the improvement that leave the
$\Lambda$ parameter unchanged, one can easily have a factor of two
variation in the fitted $\Lambda_{\overline{MS}}/\surd\sigma$,
as we saw with the otherwise well-motivated mean-field improved 
coupling. Unfortunately our large-$N$ calculations turn out
not to be able to discriminate between these different coupling 
schemes -- typically they all work almost equally well as fits.
To finesse this impasse, we turned to SU(3) and compared a
number of variations on the mean-field scheme with the 
Schrodinger functional (SF) scheme 
\cite{SF2}
which has been calculated
over a range of energy scales comparable to that of
experimental determinations of $\alpha_s(Q^2)$ and with 
greater precision. Applied to calculated values of $r_0$ this 
comparison indicated that the original Parisi mean-field scheme
works well. Using this scheme for all our values of $N$,
we obtain the fit to all $N\geq 3$ shown in Fig.~\ref{fig_lamN}:
\be
\frac{\Lambda_{\overline{MS}}}{\sqrt\sigma}
= 
0.503(2)(40) + \frac{0.33(3)(3)}{N^2}  \qquad ; \qquad N\geq 3
\label{eqn_siglamNf}
\ee
where the first error is statistical and the second much
larger error is expected to provide a bound on the systematic
error from all sources. (A fit including the $N=2$ value is
not statistically excluded, but we prefer not to include
it given the marginal nature of the SU(2) calculation.) 
Note that the component systematic 
errors are independent of $N$ to a first approximation, and
we can therefore include that error as a common factor to
the best fit in eqn(\ref{eqn_siglamNf}).

We also presented a way of comparing different coupling schemes
that does not depend on the explicit calculation of any physical
quantities. This means that one can perform the comparison
much deeper into weak coupling. Here again we found 
evidence that the mean-field improved coupling
is a `good' one in the sense of possessing small higher order
corrections. 

Our calculations have, of course, have been limited to a small set
of improved lattice couplings. There will certainly be other `good' 
lattice couplings, and the methods described in this paper can help 
to identify them. We have also focussed on one particular lattice
action (albeit the one that has been most widely used); however
it is straightforward to construct a Mean-Field improved
coupling for other actions, in the spirit of
\cite{MF_Parisi}.
\section*{Acknowledgements}

One of us (MT) acknowledges useful discussions
with Barak Bringoltz.

\vfill\eject

\begin{table}
\begin{center}
\begin{tabular}{|c|c|c|c|}\hline
\multicolumn{4}{|c|}{SU(2)}  \\ \hline 
$\beta$ & $L$ & $am_l$ &     $u_p$  \\ \hline
2.1768  & 8  & 1.990(64)  &  0.56122 \\ 
2.2400  & 8  & 1.380(16)  &  0.58286 \\ 
2.2986  & 10 & 1.240(13)  &  0.60180 \\ 
2.3715  & 12 & 0.9071(94)  & 0.62272  \\ 
2.3726  & 12 & 0.9088(63)  & 0.62302  \\ 
2.4265  & 16 & 0.8470(68)  & 0.63632  \\ 
2.5115  & 20 & 0.5728(54)  & 0.65421  \\
2.5500  & 20 & 0.4363(40)  & 0.66137  \\
2.6000  & 24 & 0.3778(44)  & 0.67001  \\
2.7000  & 32 & 0.2688(30)  & 0.68557  \\ \hline
\end{tabular}
\caption{\label{table_su2K}
The mass, $am_l$, of a closed flux loop of length $L$, and the average 
plaquette, $u_p$, at the indicated values of $\beta$  in SU(2).}
\end{center}
\end{table}

\begin{table}
\begin{center}
\begin{tabular}{|c|c|c|c|}\hline
\multicolumn{4}{|c|}{SU(3)}  \\ \hline 
$\beta$ & $L$ & $am_l$ &     $u_p$  \\ \hline
5.6500 & 8  & 1.425(25)  & 0.53750  \\ 
5.6750 & 8  & 1.249(15)  & 0.54366  \\ 
5.6925 & 8  & 1.130(12)  & 0.54756  \\ 
5.6993 & 8  & 1.106(11)  & 0.54896  \\ 
5.7995 & 10 & 0.8860(79)  & 0.56755  \\ 
5.8000 & 10 & 0.8766(86)  & 0.56764  \\ 
5.8945 & 12 & 0.7283(67)  & 0.58111  \\
6.0625 & 16 & 0.5408(46)  & 0.60034  \\
6.2000 & 20 & 0.4465(32)  & 0.61362  \\
6.3380 & 24 & 0.3588(29)  & 0.62560  \\
6.5150 & 32 & 0.2943(35)  & 0.63948  \\ \hline
\end{tabular}
\caption{\label{table_su3K}
As in Table~\ref{table_su2K} but for SU(3).}
\end{center}
\end{table}

\begin{table}
\begin{center}
\begin{tabular}{|c|c|c|c|}\hline
\multicolumn{4}{|c|}{SU(4)}  \\ \hline 
$\beta$ & $L$ & $am_l$ &     $u_p$  \\ \hline
10.480 &  8 & 1.348(15)   & 0.5253   \\
10.500 &  8 & 1.234(13)   & 0.52920  \\
10.550 &  8 & 0.9878(87)  & 0.53732  \\
10.590 & 10 & 1.1296(94)  & 0.54259  \\
10.635 & 10 & 0.9587(105) & 0.54764  \\ 
10.637 & 10 & 0.9541(43)  & 0.54789  \\
10.700 & 10 & 0.7789(90)  & 0.55408  \\
10.789 & 12 & 0.7885(56)  & 0.56169  \\
10.870 & 12 & 0.6456(72)  & 0.56793  \\
11.085 & 16 & 0.5663(54)  & 0.58239  \\
11.400 & 20 & 0.0044(44)  & 0.60044  \\ \hline
\end{tabular}
\caption{\label{table_su4K}
As in Table~\ref{table_su2K} but for SU(4).}
\end{center}
\end{table}

\begin{table}
\begin{center}
\begin{tabular}{|c|c|c|c|}\hline
\multicolumn{4}{|c|}{SU(6)}  \\ \hline 
$\beta$ & $L$ & $am_l$ &     $u_p$  \\ \hline
24.300 & 8  & 1.246(14) & 0.51721 \\
24.350 & 8  & 1.077(11) & 0.52254 \\
24.425 & 10 & 1.210(12) & 0.52810 \\
24.500 & 10 & 1.063(12) & 0.53258 \\
24.515 & 10 & 1.041(11) &  0.53340 \\
24.670 & 10 & 0.8431(89) & 0.54089 \\
24.845 & 12 & 0.8540(81) & 0.54816 \\
25.050 & 12 & 0.6686(60) & 0.55570 \\
25.452 & 16 & 0.6396(55) & 0.56866 \\ \hline
\end{tabular}
\caption{\label{table_su6K}
As in Table~\ref{table_su2K} but for SU(6).}
\end{center}
\end{table}

\begin{table}
\begin{center}
\begin{tabular}{|c|c|c|c|}\hline
\multicolumn{4}{|c|}{SU(8)}  \\ \hline 
$\beta$ & $L$ & $am_l$ &     $u_p$  \\ \hline
43.625 & 8  & 1.279(15)   & 0.51220 \\
 43.70 & 8  & 1.106(12)   & 0.51713 \\
 43.78 & 8  & 0.9989(94)  & 0.52094 \\
 43.85 & 8  & 0.9139(78)  & 0.52374 \\
 44.00 & 10 & 1.0599(91)  & 0.52879 \\
 44.35 & 10 & 0.8036(77)  & 0.53848 \\
 44.85 & 12 & 0.7222(61)  & 0.54980 \\
 45.70 & 16 & 0.6326(45)  & 0.56571 \\ \hline
\end{tabular}
\caption{\label{table_su8K}
As in Table~\ref{table_su2K} but for SU(8).}
\end{center}
\end{table}

\begin{table}
\begin{center}
\begin{tabular}{|c|c|c|}\hline
\multicolumn{3}{|c|}{SU(3) ; $r_0$}  \\ \hline 
$\beta$ & $r_0/a$ & $u_p$  \\ \hline
5.70   & 2.922(9)  & 0.54939   \\
5.80   & 3.673(5)  & 0.56778   \\
5.95   & 4.898(12) & 0.58846   \\
6.07   & 6.033(17) & 0.60158   \\
6.20   & 7.380(26) & 0.61384   \\
6.40   & 9.740(50) & 0.63085   \\
6.57   & 12.18(10) & 0.64365   \\
6.69   & 14.20(12) & 0.65204   \\
6.81   & 16.54(13) & 0.65998   \\
6.92   & 19.13(15) & 0.66685   \\ \hline
\end{tabular}
\caption{\label{table_su3r0}
The value of $r_0$
\cite{necco},
and the average plaquette, $u_p$,
at the indicated values of $\beta$ in SU(3).}
\end{center}
\end{table}

\begin{table}
\begin{center}
\begin{tabular}{|c|c|c|c|c|}\hline
$N$ & $a\sqrt{\sigma}\in$ & $\sqrt{\sigma}/\Lambda_I$ & $c_\sigma$ & 
$\chi^2/n_{df}$ \\ \hline
2 & [0.177,0.097]  &  4.566(28) & 3.83(26) & 1.8 \\
3 & [0.261,0.101]  &  4.888(17) & 2.08(10) & 1.1 \\
4 & [0.374,0.153]  &  5.005(20) & 1.52(5)  & 1.0 \\
6 & [0.415,0.210]  &  5.131(25) & 1.30(5)  & 0.34 \\
8 & [0.420,0.209]  &  5.199(21) & 1.18(4)  & 0.27 \\ \hline
\end{tabular}
\caption{\label{table_fitN}
Results of weak-coupling fits using eqn(\ref{eqn_agfit}) over the
indicated ranges of $a\surd\sigma$ and for various $N$.}
\end{center}
\end{table}

\begin{table}
\begin{center}
\begin{tabular}{|c|c|c|c|c|c|}\hline
\multicolumn{6}{|c|}{SU(8)} \\ \hline
 $s$ & $a\sqrt{\sigma}\in$ & $\sqrt{\sigma}/\Lambda_s$ & $c_\sigma$ & 
$\chi^2/n_{df}$ & $\Lambda_{\overline{MS}}/\sqrt{\sigma}$ \\ \hline
latt         & [0.420,0.209]  &  96.80(47) & 3.71(5)  & 0.37  & 0.3848(19) \\ \hline
$u_p$ 1-loop & [0.420,0.209]  &  8.980(43) & 3.88(5)  & 0.54  & 0.2932(14) \\
$u_p$ 2-loop & [0.420,0.209]  &  6.832(43) & 3.67(5)  & 0.38  & 0.3854(25) \\
$u_p$ 3-loop & [0.420,0.209]  &  6.100(28) & 3.50(5)  & 0.32  & 0.4316(20) \\
$u_p$ meas   & [0.420,0.209]  &  5.199(21) & 1.18(4)  & 0.27  & 0.5064(21) \\ \hline
\end{tabular}
\caption{\label{table_fitN8scheme} 
Weak-coupling fit using eqn(\ref{eqn_agfit}) to the whole
weak-coupling branch in SU(8); for various coupling schemes $s$.}
\end{center}
\end{table}

\begin{table}
\begin{center}
\begin{tabular}{|c|c|c|c|c|c|c|}\hline
\multicolumn{7}{|c|}{SU(3) : SF scheme} \\ \hline
 interp. & p & $\beta \in$ & $1/r_0\Lambda_{SF}$ & $c_r$ &  $d_r$ & 
$\chi^2/n_{df}$ \\ \hline
latt         & 2 & [6.257,6.9079]  &  3.245(24) & 1.64(80)  & 1.05(26) & 0.52
 \\
$u_p$ 2-loop & 2 & [6.257,6.9079]  &  3.243(20) & 1.61(64)  & 1.10(28) & 0.54
 \\
$u_p$ 3-loop & 2 & [6.257,6.9079]  &  3.238(23) & 1.78(63)  & 1.10(28) & 0.38
 \\
$u_p$ meas   & 2 & [6.257,6.9079]  &  3.212(27) & 2.60(55)  & 1.13(28) & 0.25
 \\ \hline
$u_p$ 2-loop & 1 & [6.257,6.9079]  &  3.169(23) & 1.63(44)  & 0.34(8) & 0.45
 \\
$u_p$ meas   & 1 & [6.257,6.9079]  &  3.141(40) & 2.21(58)  & 0.36(09) & 0.18
 \\ \hline
$u_p$ meas   & 2 & [6.257,7.2611]  &  3.215(17) & 2.64(45)  & 1.09(23) & 0.28
 \\ \hline
\end{tabular}
\caption{\label{table_fitr0N3SF}
Weak coupling fits to $r_0$ in SU(3) using the $SF$ coupling scheme, as in
eqn(\ref{eqn_agfitSF}). }
\end{center}
\end{table}

\begin{table}
\begin{center}
\begin{tabular}{|c|c|c|c|c|c||c|}\hline
\multicolumn{7}{|c|}{SU(3) $\quad$ :$\quad$  } \\ \hline
 $s$ & $a/r_0 \in$ & $1/r_0\Lambda_s$ & $c_r$ & $\chi^2/n_{df}$
& $r_0\Lambda_{\overline{MS}}$ & $r_0\sqrt{\sigma}$ \\ \hline
latt         & [0.166,0.052]  &  53.26(21) & 7.03(20)  & 1.26  &
 0.5409(22)  &  1,143(11) \\ \hline
$u_p$ 1-loop & [0.103,0.052]  &  5.883(23) & 10.65(30) & 0.89  &
 0.4475(18)  &  1.166(11) \\
$u_p$ 2-loop & [0.166,0.052]  &  4.891(20) & 6.93(20)  & 1.28  &
 0.5383(22)  &  1.156(14) \\
$u_p$ 3-loop & [0.204,0.052]  &  4.567(16) & 5.87(13)  & 0.85  &
 0.5765(20)  &  1.159(7) \\
$u_p$ meas   & [0.204,0.052]  &  4.215(16) & 3.00(13)  & 0.48  &
 0.6246(24)  &  1.160(6) \\ \hline
\end{tabular}
\caption{\label{table_fitr0N3scheme}
Weak coupling fits to $r_0$ in SU(3) using the indicated coupling
schemes, $s$, in appropriate modifications of eqn(\ref{eqn_agfit}). }
\end{center}
\end{table}

\clearpage

\begin	{figure}[p]
\begin	{center}
\leavevmode
\begingroup%
  \makeatletter%
  \newcommand{\GNUPLOTspecial}{%
    \@sanitize\catcode`\%=14\relax\special}%
  \setlength{\unitlength}{0.1bp}%
{\GNUPLOTspecial{!
/gnudict 256 dict def
gnudict begin
/Color true def
/Solid false def
/gnulinewidth 5.000 def
/userlinewidth gnulinewidth def
/vshift -33 def
/dl {10 mul} def
/hpt_ 31.5 def
/vpt_ 31.5 def
/hpt hpt_ def
/vpt vpt_ def
/M {moveto} bind def
/L {lineto} bind def
/R {rmoveto} bind def
/V {rlineto} bind def
/vpt2 vpt 2 mul def
/hpt2 hpt 2 mul def
/Lshow { currentpoint stroke M
  0 vshift R show } def
/Rshow { currentpoint stroke M
  dup stringwidth pop neg vshift R show } def
/Cshow { currentpoint stroke M
  dup stringwidth pop -2 div vshift R show } def
/UP { dup vpt_ mul /vpt exch def hpt_ mul /hpt exch def
  /hpt2 hpt 2 mul def /vpt2 vpt 2 mul def } def
/DL { Color {setrgbcolor Solid {pop []} if 0 setdash }
 {pop pop pop Solid {pop []} if 0 setdash} ifelse } def
/BL { stroke userlinewidth 2 mul setlinewidth } def
/AL { stroke userlinewidth 2 div setlinewidth } def
/UL { dup gnulinewidth mul /userlinewidth exch def
      10 mul /udl exch def } def
/PL { stroke userlinewidth setlinewidth } def
/LTb { BL [] 0 0 0 DL } def
/LTa { AL [1 udl mul 2 udl mul] 0 setdash 0 0 0 setrgbcolor } def
/LT0 { PL [] 1 0 0 DL } def
/LT1 { PL [4 dl 2 dl] 0 1 0 DL } def
/LT2 { PL [2 dl 3 dl] 0 0 1 DL } def
/LT3 { PL [1 dl 1.5 dl] 1 0 1 DL } def
/LT4 { PL [5 dl 2 dl 1 dl 2 dl] 0 1 1 DL } def
/LT5 { PL [4 dl 3 dl 1 dl 3 dl] 1 1 0 DL } def
/LT6 { PL [2 dl 2 dl 2 dl 4 dl] 0 0 0 DL } def
/LT7 { PL [2 dl 2 dl 2 dl 2 dl 2 dl 4 dl] 1 0.3 0 DL } def
/LT8 { PL [2 dl 2 dl 2 dl 2 dl 2 dl 2 dl 2 dl 4 dl] 0.5 0.5 0.5 DL } def
/Pnt { stroke [] 0 setdash
   gsave 1 setlinecap M 0 0 V stroke grestore } def
/Dia { stroke [] 0 setdash 2 copy vpt add M
  hpt neg vpt neg V hpt vpt neg V
  hpt vpt V hpt neg vpt V closepath stroke
  Pnt } def
/Pls { stroke [] 0 setdash vpt sub M 0 vpt2 V
  currentpoint stroke M
  hpt neg vpt neg R hpt2 0 V stroke
  } def
/Box { stroke [] 0 setdash 2 copy exch hpt sub exch vpt add M
  0 vpt2 neg V hpt2 0 V 0 vpt2 V
  hpt2 neg 0 V closepath stroke
  Pnt } def
/Crs { stroke [] 0 setdash exch hpt sub exch vpt add M
  hpt2 vpt2 neg V currentpoint stroke M
  hpt2 neg 0 R hpt2 vpt2 V stroke } def
/TriU { stroke [] 0 setdash 2 copy vpt 1.12 mul add M
  hpt neg vpt -1.62 mul V
  hpt 2 mul 0 V
  hpt neg vpt 1.62 mul V closepath stroke
  Pnt  } def
/Star { 2 copy Pls Crs } def
/BoxF { stroke [] 0 setdash exch hpt sub exch vpt add M
  0 vpt2 neg V  hpt2 0 V  0 vpt2 V
  hpt2 neg 0 V  closepath fill } def
/TriUF { stroke [] 0 setdash vpt 1.12 mul add M
  hpt neg vpt -1.62 mul V
  hpt 2 mul 0 V
  hpt neg vpt 1.62 mul V closepath fill } def
/TriD { stroke [] 0 setdash 2 copy vpt 1.12 mul sub M
  hpt neg vpt 1.62 mul V
  hpt 2 mul 0 V
  hpt neg vpt -1.62 mul V closepath stroke
  Pnt  } def
/TriDF { stroke [] 0 setdash vpt 1.12 mul sub M
  hpt neg vpt 1.62 mul V
  hpt 2 mul 0 V
  hpt neg vpt -1.62 mul V closepath fill} def
/DiaF { stroke [] 0 setdash vpt add M
  hpt neg vpt neg V hpt vpt neg V
  hpt vpt V hpt neg vpt V closepath fill } def
/Pent { stroke [] 0 setdash 2 copy gsave
  translate 0 hpt M 4 {72 rotate 0 hpt L} repeat
  closepath stroke grestore Pnt } def
/PentF { stroke [] 0 setdash gsave
  translate 0 hpt M 4 {72 rotate 0 hpt L} repeat
  closepath fill grestore } def
/Circle { stroke [] 0 setdash 2 copy
  hpt 0 360 arc stroke Pnt } def
/CircleF { stroke [] 0 setdash hpt 0 360 arc fill } def
/C0 { BL [] 0 setdash 2 copy moveto vpt 90 450  arc } bind def
/C1 { BL [] 0 setdash 2 copy        moveto
       2 copy  vpt 0 90 arc closepath fill
               vpt 0 360 arc closepath } bind def
/C2 { BL [] 0 setdash 2 copy moveto
       2 copy  vpt 90 180 arc closepath fill
               vpt 0 360 arc closepath } bind def
/C3 { BL [] 0 setdash 2 copy moveto
       2 copy  vpt 0 180 arc closepath fill
               vpt 0 360 arc closepath } bind def
/C4 { BL [] 0 setdash 2 copy moveto
       2 copy  vpt 180 270 arc closepath fill
               vpt 0 360 arc closepath } bind def
/C5 { BL [] 0 setdash 2 copy moveto
       2 copy  vpt 0 90 arc
       2 copy moveto
       2 copy  vpt 180 270 arc closepath fill
               vpt 0 360 arc } bind def
/C6 { BL [] 0 setdash 2 copy moveto
      2 copy  vpt 90 270 arc closepath fill
              vpt 0 360 arc closepath } bind def
/C7 { BL [] 0 setdash 2 copy moveto
      2 copy  vpt 0 270 arc closepath fill
              vpt 0 360 arc closepath } bind def
/C8 { BL [] 0 setdash 2 copy moveto
      2 copy vpt 270 360 arc closepath fill
              vpt 0 360 arc closepath } bind def
/C9 { BL [] 0 setdash 2 copy moveto
      2 copy  vpt 270 450 arc closepath fill
              vpt 0 360 arc closepath } bind def
/C10 { BL [] 0 setdash 2 copy 2 copy moveto vpt 270 360 arc closepath fill
       2 copy moveto
       2 copy vpt 90 180 arc closepath fill
               vpt 0 360 arc closepath } bind def
/C11 { BL [] 0 setdash 2 copy moveto
       2 copy  vpt 0 180 arc closepath fill
       2 copy moveto
       2 copy  vpt 270 360 arc closepath fill
               vpt 0 360 arc closepath } bind def
/C12 { BL [] 0 setdash 2 copy moveto
       2 copy  vpt 180 360 arc closepath fill
               vpt 0 360 arc closepath } bind def
/C13 { BL [] 0 setdash  2 copy moveto
       2 copy  vpt 0 90 arc closepath fill
       2 copy moveto
       2 copy  vpt 180 360 arc closepath fill
               vpt 0 360 arc closepath } bind def
/C14 { BL [] 0 setdash 2 copy moveto
       2 copy  vpt 90 360 arc closepath fill
               vpt 0 360 arc } bind def
/C15 { BL [] 0 setdash 2 copy vpt 0 360 arc closepath fill
               vpt 0 360 arc closepath } bind def
/Rec   { newpath 4 2 roll moveto 1 index 0 rlineto 0 exch rlineto
       neg 0 rlineto closepath } bind def
/Square { dup Rec } bind def
/Bsquare { vpt sub exch vpt sub exch vpt2 Square } bind def
/S0 { BL [] 0 setdash 2 copy moveto 0 vpt rlineto BL Bsquare } bind def
/S1 { BL [] 0 setdash 2 copy vpt Square fill Bsquare } bind def
/S2 { BL [] 0 setdash 2 copy exch vpt sub exch vpt Square fill Bsquare } bind def
/S3 { BL [] 0 setdash 2 copy exch vpt sub exch vpt2 vpt Rec fill Bsquare } bind def
/S4 { BL [] 0 setdash 2 copy exch vpt sub exch vpt sub vpt Square fill Bsquare } bind def
/S5 { BL [] 0 setdash 2 copy 2 copy vpt Square fill
       exch vpt sub exch vpt sub vpt Square fill Bsquare } bind def
/S6 { BL [] 0 setdash 2 copy exch vpt sub exch vpt sub vpt vpt2 Rec fill Bsquare } bind def
/S7 { BL [] 0 setdash 2 copy exch vpt sub exch vpt sub vpt vpt2 Rec fill
       2 copy vpt Square fill
       Bsquare } bind def
/S8 { BL [] 0 setdash 2 copy vpt sub vpt Square fill Bsquare } bind def
/S9 { BL [] 0 setdash 2 copy vpt sub vpt vpt2 Rec fill Bsquare } bind def
/S10 { BL [] 0 setdash 2 copy vpt sub vpt Square fill 2 copy exch vpt sub exch vpt Square fill
       Bsquare } bind def
/S11 { BL [] 0 setdash 2 copy vpt sub vpt Square fill 2 copy exch vpt sub exch vpt2 vpt Rec fill
       Bsquare } bind def
/S12 { BL [] 0 setdash 2 copy exch vpt sub exch vpt sub vpt2 vpt Rec fill Bsquare } bind def
/S13 { BL [] 0 setdash 2 copy exch vpt sub exch vpt sub vpt2 vpt Rec fill
       2 copy vpt Square fill Bsquare } bind def
/S14 { BL [] 0 setdash 2 copy exch vpt sub exch vpt sub vpt2 vpt Rec fill
       2 copy exch vpt sub exch vpt Square fill Bsquare } bind def
/S15 { BL [] 0 setdash 2 copy Bsquare fill Bsquare } bind def
/D0 { gsave translate 45 rotate 0 0 S0 stroke grestore } bind def
/D1 { gsave translate 45 rotate 0 0 S1 stroke grestore } bind def
/D2 { gsave translate 45 rotate 0 0 S2 stroke grestore } bind def
/D3 { gsave translate 45 rotate 0 0 S3 stroke grestore } bind def
/D4 { gsave translate 45 rotate 0 0 S4 stroke grestore } bind def
/D5 { gsave translate 45 rotate 0 0 S5 stroke grestore } bind def
/D6 { gsave translate 45 rotate 0 0 S6 stroke grestore } bind def
/D7 { gsave translate 45 rotate 0 0 S7 stroke grestore } bind def
/D8 { gsave translate 45 rotate 0 0 S8 stroke grestore } bind def
/D9 { gsave translate 45 rotate 0 0 S9 stroke grestore } bind def
/D10 { gsave translate 45 rotate 0 0 S10 stroke grestore } bind def
/D11 { gsave translate 45 rotate 0 0 S11 stroke grestore } bind def
/D12 { gsave translate 45 rotate 0 0 S12 stroke grestore } bind def
/D13 { gsave translate 45 rotate 0 0 S13 stroke grestore } bind def
/D14 { gsave translate 45 rotate 0 0 S14 stroke grestore } bind def
/D15 { gsave translate 45 rotate 0 0 S15 stroke grestore } bind def
/DiaE { stroke [] 0 setdash vpt add M
  hpt neg vpt neg V hpt vpt neg V
  hpt vpt V hpt neg vpt V closepath stroke } def
/BoxE { stroke [] 0 setdash exch hpt sub exch vpt add M
  0 vpt2 neg V hpt2 0 V 0 vpt2 V
  hpt2 neg 0 V closepath stroke } def
/TriUE { stroke [] 0 setdash vpt 1.12 mul add M
  hpt neg vpt -1.62 mul V
  hpt 2 mul 0 V
  hpt neg vpt 1.62 mul V closepath stroke } def
/TriDE { stroke [] 0 setdash vpt 1.12 mul sub M
  hpt neg vpt 1.62 mul V
  hpt 2 mul 0 V
  hpt neg vpt -1.62 mul V closepath stroke } def
/PentE { stroke [] 0 setdash gsave
  translate 0 hpt M 4 {72 rotate 0 hpt L} repeat
  closepath stroke grestore } def
/CircE { stroke [] 0 setdash 
  hpt 0 360 arc stroke } def
/Opaque { gsave closepath 1 setgray fill grestore 0 setgray closepath } def
/DiaW { stroke [] 0 setdash vpt add M
  hpt neg vpt neg V hpt vpt neg V
  hpt vpt V hpt neg vpt V Opaque stroke } def
/BoxW { stroke [] 0 setdash exch hpt sub exch vpt add M
  0 vpt2 neg V hpt2 0 V 0 vpt2 V
  hpt2 neg 0 V Opaque stroke } def
/TriUW { stroke [] 0 setdash vpt 1.12 mul add M
  hpt neg vpt -1.62 mul V
  hpt 2 mul 0 V
  hpt neg vpt 1.62 mul V Opaque stroke } def
/TriDW { stroke [] 0 setdash vpt 1.12 mul sub M
  hpt neg vpt 1.62 mul V
  hpt 2 mul 0 V
  hpt neg vpt -1.62 mul V Opaque stroke } def
/PentW { stroke [] 0 setdash gsave
  translate 0 hpt M 4 {72 rotate 0 hpt L} repeat
  Opaque stroke grestore } def
/CircW { stroke [] 0 setdash 
  hpt 0 360 arc Opaque stroke } def
/BoxFill { gsave Rec 1 setgray fill grestore } def
end
}}%
\begin{picture}(4500,3240)(0,0)%
{\GNUPLOTspecial{"
gnudict begin
gsave
0 0 translate
0.100 0.100 scale
0 setgray
newpath
1.000 UL
LTb
700 857 M
63 0 V
3587 0 R
-63 0 V
700 1770 M
63 0 V
3587 0 R
-63 0 V
700 2683 M
63 0 V
3587 0 R
-63 0 V
700 400 M
0 63 V
0 2677 R
0 -63 V
1430 400 M
0 63 V
0 2677 R
0 -63 V
2160 400 M
0 63 V
0 2677 R
0 -63 V
2890 400 M
0 63 V
0 2677 R
0 -63 V
3620 400 M
0 63 V
0 2677 R
0 -63 V
4350 400 M
0 63 V
0 2677 R
0 -63 V
1.000 UL
LTb
700 400 M
3650 0 V
0 2740 V
-3650 0 V
700 400 L
1.000 UP
1.000 UL
LT0
3810 2453 M
146 0 V
-146 -31 R
0 62 V
146 -62 R
0 62 V
-208 -49 R
33 0 V
-33 -31 R
0 62 V
33 -62 R
0 62 V
-225 -90 R
28 0 V
-28 -31 R
0 62 V
28 -62 R
0 62 V
-152 -77 R
23 0 V
-23 -31 R
0 62 V
23 -62 R
0 62 V
-126 -68 R
19 0 V
-19 -31 R
0 62 V
19 -62 R
0 62 V
-166 -96 R
19 0 V
-19 -31 R
0 62 V
19 -62 R
0 62 V
2891 2099 M
19 0 V
-19 -31 R
0 62 V
19 -62 R
0 62 V
2589 1944 M
14 0 V
-14 -31 R
0 62 V
14 -62 R
0 62 V
2220 1725 M
10 0 V
-10 -31 R
0 62 V
10 -62 R
0 62 V
3883 2453 CircleF
3765 2435 CircleF
3570 2376 CircleF
3443 2330 CircleF
3338 2293 CircleF
3191 2228 CircleF
2900 2099 CircleF
2596 1944 CircleF
2225 1725 CircleF
3810 2453 M
146 0 V
-146 -31 R
0 62 V
146 -62 R
0 62 V
-208 -49 R
33 0 V
-33 -31 R
0 62 V
33 -62 R
0 62 V
-225 -90 R
28 0 V
-28 -31 R
0 62 V
28 -62 R
0 62 V
-152 -77 R
23 0 V
-23 -31 R
0 62 V
23 -62 R
0 62 V
-126 -68 R
19 0 V
-19 -31 R
0 62 V
19 -62 R
0 62 V
-166 -96 R
19 0 V
-19 -31 R
0 62 V
19 -62 R
0 62 V
2891 2099 M
19 0 V
-19 -31 R
0 62 V
19 -62 R
0 62 V
2589 1944 M
14 0 V
-14 -31 R
0 62 V
14 -62 R
0 62 V
2220 1725 M
10 0 V
-10 -31 R
0 62 V
10 -62 R
0 62 V
3883 2453 CircleF
3765 2435 CircleF
3570 2376 CircleF
3443 2330 CircleF
3338 2293 CircleF
3191 2228 CircleF
2900 2099 CircleF
2596 1944 CircleF
2225 1725 CircleF
0.500 UL
LTb
1324 948 M
80 91 V
86 92 V
93 91 V
99 91 V
107 92 V
114 91 V
122 91 V
131 92 V
140 91 V
150 91 V
162 92 V
176 91 V
190 91 V
209 92 V
231 91 V
260 91 V
296 92 V
350 91 V
30 6 V
stroke
grestore
end
showpage
}}%
\put(2525,50){\makebox(0,0){\Large{$a\surd\sigma$}}}%
\put(100,2020){\makebox(0,0){\Large{$g^2_I(a)N$}}}%
\put(4350,300){\makebox(0,0){\ {$0.5$}}}%
\put(3620,300){\makebox(0,0){\ {$0.4$}}}%
\put(2890,300){\makebox(0,0){\ {$0.3$}}}%
\put(2160,300){\makebox(0,0){\ {$0.2$}}}%
\put(1430,300){\makebox(0,0){\ {$0.1$}}}%
\put(700,300){\makebox(0,0){\ {$0$}}}%
\put(650,2683){\makebox(0,0)[r]{\ \ {$6$}}}%
\put(650,1770){\makebox(0,0)[r]{\ \ {$5$}}}%
\put(650,857){\makebox(0,0)[r]{\ \ {$4$}}}%
\end{picture}%
\endgroup

\end	{center}
\vskip 0.15in
\caption{The 't Hooft coupling, defined from the mean-field improved
lattice bare coupling, eqn(\ref{eqn_gI}), as a function of the scale $a$ 
in SU(8).
Shown is the 3-loop perturbative running modified by a $O(a^2)$
lattice correction.}
\label{fig_gfitsu8}
\end 	{figure}
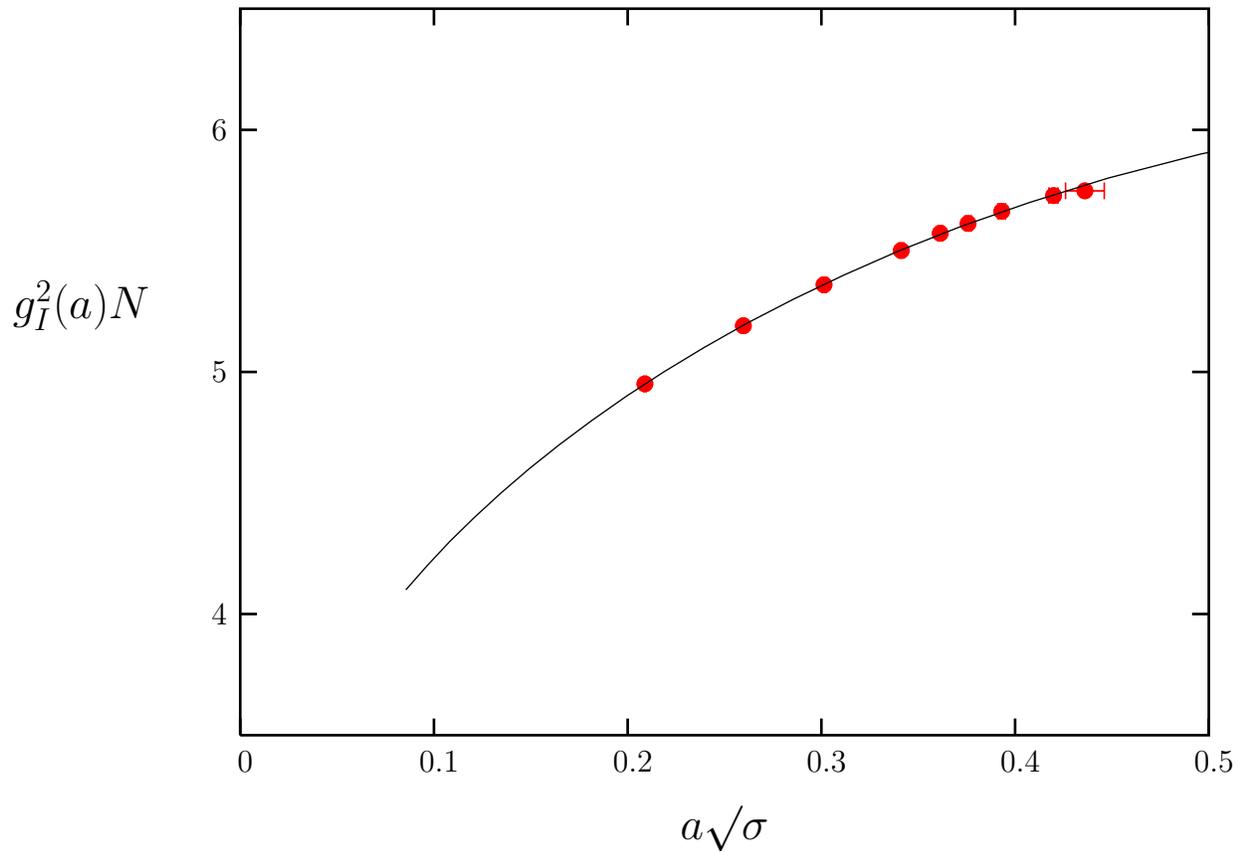

\begin	{figure}[p]
\begin	{center}
\leavevmode
\begingroup%
  \makeatletter%
  \newcommand{\GNUPLOTspecial}{%
    \@sanitize\catcode`\%=14\relax\special}%
  \setlength{\unitlength}{0.1bp}%
{\GNUPLOTspecial{!
/gnudict 256 dict def
gnudict begin
/Color true def
/Solid false def
/gnulinewidth 5.000 def
/userlinewidth gnulinewidth def
/vshift -33 def
/dl {10 mul} def
/hpt_ 31.5 def
/vpt_ 31.5 def
/hpt hpt_ def
/vpt vpt_ def
/M {moveto} bind def
/L {lineto} bind def
/R {rmoveto} bind def
/V {rlineto} bind def
/vpt2 vpt 2 mul def
/hpt2 hpt 2 mul def
/Lshow { currentpoint stroke M
  0 vshift R show } def
/Rshow { currentpoint stroke M
  dup stringwidth pop neg vshift R show } def
/Cshow { currentpoint stroke M
  dup stringwidth pop -2 div vshift R show } def
/UP { dup vpt_ mul /vpt exch def hpt_ mul /hpt exch def
  /hpt2 hpt 2 mul def /vpt2 vpt 2 mul def } def
/DL { Color {setrgbcolor Solid {pop []} if 0 setdash }
 {pop pop pop Solid {pop []} if 0 setdash} ifelse } def
/BL { stroke userlinewidth 2 mul setlinewidth } def
/AL { stroke userlinewidth 2 div setlinewidth } def
/UL { dup gnulinewidth mul /userlinewidth exch def
      10 mul /udl exch def } def
/PL { stroke userlinewidth setlinewidth } def
/LTb { BL [] 0 0 0 DL } def
/LTa { AL [1 udl mul 2 udl mul] 0 setdash 0 0 0 setrgbcolor } def
/LT0 { PL [] 1 0 0 DL } def
/LT1 { PL [4 dl 2 dl] 0 1 0 DL } def
/LT2 { PL [2 dl 3 dl] 0 0 1 DL } def
/LT3 { PL [1 dl 1.5 dl] 1 0 1 DL } def
/LT4 { PL [5 dl 2 dl 1 dl 2 dl] 0 1 1 DL } def
/LT5 { PL [4 dl 3 dl 1 dl 3 dl] 1 1 0 DL } def
/LT6 { PL [2 dl 2 dl 2 dl 4 dl] 0 0 0 DL } def
/LT7 { PL [2 dl 2 dl 2 dl 2 dl 2 dl 4 dl] 1 0.3 0 DL } def
/LT8 { PL [2 dl 2 dl 2 dl 2 dl 2 dl 2 dl 2 dl 4 dl] 0.5 0.5 0.5 DL } def
/Pnt { stroke [] 0 setdash
   gsave 1 setlinecap M 0 0 V stroke grestore } def
/Dia { stroke [] 0 setdash 2 copy vpt add M
  hpt neg vpt neg V hpt vpt neg V
  hpt vpt V hpt neg vpt V closepath stroke
  Pnt } def
/Pls { stroke [] 0 setdash vpt sub M 0 vpt2 V
  currentpoint stroke M
  hpt neg vpt neg R hpt2 0 V stroke
  } def
/Box { stroke [] 0 setdash 2 copy exch hpt sub exch vpt add M
  0 vpt2 neg V hpt2 0 V 0 vpt2 V
  hpt2 neg 0 V closepath stroke
  Pnt } def
/Crs { stroke [] 0 setdash exch hpt sub exch vpt add M
  hpt2 vpt2 neg V currentpoint stroke M
  hpt2 neg 0 R hpt2 vpt2 V stroke } def
/TriU { stroke [] 0 setdash 2 copy vpt 1.12 mul add M
  hpt neg vpt -1.62 mul V
  hpt 2 mul 0 V
  hpt neg vpt 1.62 mul V closepath stroke
  Pnt  } def
/Star { 2 copy Pls Crs } def
/BoxF { stroke [] 0 setdash exch hpt sub exch vpt add M
  0 vpt2 neg V  hpt2 0 V  0 vpt2 V
  hpt2 neg 0 V  closepath fill } def
/TriUF { stroke [] 0 setdash vpt 1.12 mul add M
  hpt neg vpt -1.62 mul V
  hpt 2 mul 0 V
  hpt neg vpt 1.62 mul V closepath fill } def
/TriD { stroke [] 0 setdash 2 copy vpt 1.12 mul sub M
  hpt neg vpt 1.62 mul V
  hpt 2 mul 0 V
  hpt neg vpt -1.62 mul V closepath stroke
  Pnt  } def
/TriDF { stroke [] 0 setdash vpt 1.12 mul sub M
  hpt neg vpt 1.62 mul V
  hpt 2 mul 0 V
  hpt neg vpt -1.62 mul V closepath fill} def
/DiaF { stroke [] 0 setdash vpt add M
  hpt neg vpt neg V hpt vpt neg V
  hpt vpt V hpt neg vpt V closepath fill } def
/Pent { stroke [] 0 setdash 2 copy gsave
  translate 0 hpt M 4 {72 rotate 0 hpt L} repeat
  closepath stroke grestore Pnt } def
/PentF { stroke [] 0 setdash gsave
  translate 0 hpt M 4 {72 rotate 0 hpt L} repeat
  closepath fill grestore } def
/Circle { stroke [] 0 setdash 2 copy
  hpt 0 360 arc stroke Pnt } def
/CircleF { stroke [] 0 setdash hpt 0 360 arc fill } def
/C0 { BL [] 0 setdash 2 copy moveto vpt 90 450  arc } bind def
/C1 { BL [] 0 setdash 2 copy        moveto
       2 copy  vpt 0 90 arc closepath fill
               vpt 0 360 arc closepath } bind def
/C2 { BL [] 0 setdash 2 copy moveto
       2 copy  vpt 90 180 arc closepath fill
               vpt 0 360 arc closepath } bind def
/C3 { BL [] 0 setdash 2 copy moveto
       2 copy  vpt 0 180 arc closepath fill
               vpt 0 360 arc closepath } bind def
/C4 { BL [] 0 setdash 2 copy moveto
       2 copy  vpt 180 270 arc closepath fill
               vpt 0 360 arc closepath } bind def
/C5 { BL [] 0 setdash 2 copy moveto
       2 copy  vpt 0 90 arc
       2 copy moveto
       2 copy  vpt 180 270 arc closepath fill
               vpt 0 360 arc } bind def
/C6 { BL [] 0 setdash 2 copy moveto
      2 copy  vpt 90 270 arc closepath fill
              vpt 0 360 arc closepath } bind def
/C7 { BL [] 0 setdash 2 copy moveto
      2 copy  vpt 0 270 arc closepath fill
              vpt 0 360 arc closepath } bind def
/C8 { BL [] 0 setdash 2 copy moveto
      2 copy vpt 270 360 arc closepath fill
              vpt 0 360 arc closepath } bind def
/C9 { BL [] 0 setdash 2 copy moveto
      2 copy  vpt 270 450 arc closepath fill
              vpt 0 360 arc closepath } bind def
/C10 { BL [] 0 setdash 2 copy 2 copy moveto vpt 270 360 arc closepath fill
       2 copy moveto
       2 copy vpt 90 180 arc closepath fill
               vpt 0 360 arc closepath } bind def
/C11 { BL [] 0 setdash 2 copy moveto
       2 copy  vpt 0 180 arc closepath fill
       2 copy moveto
       2 copy  vpt 270 360 arc closepath fill
               vpt 0 360 arc closepath } bind def
/C12 { BL [] 0 setdash 2 copy moveto
       2 copy  vpt 180 360 arc closepath fill
               vpt 0 360 arc closepath } bind def
/C13 { BL [] 0 setdash  2 copy moveto
       2 copy  vpt 0 90 arc closepath fill
       2 copy moveto
       2 copy  vpt 180 360 arc closepath fill
               vpt 0 360 arc closepath } bind def
/C14 { BL [] 0 setdash 2 copy moveto
       2 copy  vpt 90 360 arc closepath fill
               vpt 0 360 arc } bind def
/C15 { BL [] 0 setdash 2 copy vpt 0 360 arc closepath fill
               vpt 0 360 arc closepath } bind def
/Rec   { newpath 4 2 roll moveto 1 index 0 rlineto 0 exch rlineto
       neg 0 rlineto closepath } bind def
/Square { dup Rec } bind def
/Bsquare { vpt sub exch vpt sub exch vpt2 Square } bind def
/S0 { BL [] 0 setdash 2 copy moveto 0 vpt rlineto BL Bsquare } bind def
/S1 { BL [] 0 setdash 2 copy vpt Square fill Bsquare } bind def
/S2 { BL [] 0 setdash 2 copy exch vpt sub exch vpt Square fill Bsquare } bind def
/S3 { BL [] 0 setdash 2 copy exch vpt sub exch vpt2 vpt Rec fill Bsquare } bind def
/S4 { BL [] 0 setdash 2 copy exch vpt sub exch vpt sub vpt Square fill Bsquare } bind def
/S5 { BL [] 0 setdash 2 copy 2 copy vpt Square fill
       exch vpt sub exch vpt sub vpt Square fill Bsquare } bind def
/S6 { BL [] 0 setdash 2 copy exch vpt sub exch vpt sub vpt vpt2 Rec fill Bsquare } bind def
/S7 { BL [] 0 setdash 2 copy exch vpt sub exch vpt sub vpt vpt2 Rec fill
       2 copy vpt Square fill
       Bsquare } bind def
/S8 { BL [] 0 setdash 2 copy vpt sub vpt Square fill Bsquare } bind def
/S9 { BL [] 0 setdash 2 copy vpt sub vpt vpt2 Rec fill Bsquare } bind def
/S10 { BL [] 0 setdash 2 copy vpt sub vpt Square fill 2 copy exch vpt sub exch vpt Square fill
       Bsquare } bind def
/S11 { BL [] 0 setdash 2 copy vpt sub vpt Square fill 2 copy exch vpt sub exch vpt2 vpt Rec fill
       Bsquare } bind def
/S12 { BL [] 0 setdash 2 copy exch vpt sub exch vpt sub vpt2 vpt Rec fill Bsquare } bind def
/S13 { BL [] 0 setdash 2 copy exch vpt sub exch vpt sub vpt2 vpt Rec fill
       2 copy vpt Square fill Bsquare } bind def
/S14 { BL [] 0 setdash 2 copy exch vpt sub exch vpt sub vpt2 vpt Rec fill
       2 copy exch vpt sub exch vpt Square fill Bsquare } bind def
/S15 { BL [] 0 setdash 2 copy Bsquare fill Bsquare } bind def
/D0 { gsave translate 45 rotate 0 0 S0 stroke grestore } bind def
/D1 { gsave translate 45 rotate 0 0 S1 stroke grestore } bind def
/D2 { gsave translate 45 rotate 0 0 S2 stroke grestore } bind def
/D3 { gsave translate 45 rotate 0 0 S3 stroke grestore } bind def
/D4 { gsave translate 45 rotate 0 0 S4 stroke grestore } bind def
/D5 { gsave translate 45 rotate 0 0 S5 stroke grestore } bind def
/D6 { gsave translate 45 rotate 0 0 S6 stroke grestore } bind def
/D7 { gsave translate 45 rotate 0 0 S7 stroke grestore } bind def
/D8 { gsave translate 45 rotate 0 0 S8 stroke grestore } bind def
/D9 { gsave translate 45 rotate 0 0 S9 stroke grestore } bind def
/D10 { gsave translate 45 rotate 0 0 S10 stroke grestore } bind def
/D11 { gsave translate 45 rotate 0 0 S11 stroke grestore } bind def
/D12 { gsave translate 45 rotate 0 0 S12 stroke grestore } bind def
/D13 { gsave translate 45 rotate 0 0 S13 stroke grestore } bind def
/D14 { gsave translate 45 rotate 0 0 S14 stroke grestore } bind def
/D15 { gsave translate 45 rotate 0 0 S15 stroke grestore } bind def
/DiaE { stroke [] 0 setdash vpt add M
  hpt neg vpt neg V hpt vpt neg V
  hpt vpt V hpt neg vpt V closepath stroke } def
/BoxE { stroke [] 0 setdash exch hpt sub exch vpt add M
  0 vpt2 neg V hpt2 0 V 0 vpt2 V
  hpt2 neg 0 V closepath stroke } def
/TriUE { stroke [] 0 setdash vpt 1.12 mul add M
  hpt neg vpt -1.62 mul V
  hpt 2 mul 0 V
  hpt neg vpt 1.62 mul V closepath stroke } def
/TriDE { stroke [] 0 setdash vpt 1.12 mul sub M
  hpt neg vpt 1.62 mul V
  hpt 2 mul 0 V
  hpt neg vpt -1.62 mul V closepath stroke } def
/PentE { stroke [] 0 setdash gsave
  translate 0 hpt M 4 {72 rotate 0 hpt L} repeat
  closepath stroke grestore } def
/CircE { stroke [] 0 setdash 
  hpt 0 360 arc stroke } def
/Opaque { gsave closepath 1 setgray fill grestore 0 setgray closepath } def
/DiaW { stroke [] 0 setdash vpt add M
  hpt neg vpt neg V hpt vpt neg V
  hpt vpt V hpt neg vpt V Opaque stroke } def
/BoxW { stroke [] 0 setdash exch hpt sub exch vpt add M
  0 vpt2 neg V hpt2 0 V 0 vpt2 V
  hpt2 neg 0 V Opaque stroke } def
/TriUW { stroke [] 0 setdash vpt 1.12 mul add M
  hpt neg vpt -1.62 mul V
  hpt 2 mul 0 V
  hpt neg vpt 1.62 mul V Opaque stroke } def
/TriDW { stroke [] 0 setdash vpt 1.12 mul sub M
  hpt neg vpt 1.62 mul V
  hpt 2 mul 0 V
  hpt neg vpt -1.62 mul V Opaque stroke } def
/PentW { stroke [] 0 setdash gsave
  translate 0 hpt M 4 {72 rotate 0 hpt L} repeat
  Opaque stroke grestore } def
/CircW { stroke [] 0 setdash 
  hpt 0 360 arc Opaque stroke } def
/BoxFill { gsave Rec 1 setgray fill grestore } def
end
}}%
\begin{picture}(4500,3240)(0,0)%
{\GNUPLOTspecial{"
gnudict begin
gsave
0 0 translate
0.100 0.100 scale
0 setgray
newpath
1.000 UL
LTb
800 400 M
63 0 V
3487 0 R
-63 0 V
800 857 M
63 0 V
3487 0 R
-63 0 V
800 1313 M
63 0 V
3487 0 R
-63 0 V
800 1770 M
63 0 V
3487 0 R
-63 0 V
800 2227 M
63 0 V
3487 0 R
-63 0 V
800 2683 M
63 0 V
3487 0 R
-63 0 V
800 3140 M
63 0 V
3487 0 R
-63 0 V
800 400 M
0 63 V
0 2677 R
0 -63 V
1510 400 M
0 63 V
0 2677 R
0 -63 V
2220 400 M
0 63 V
0 2677 R
0 -63 V
2930 400 M
0 63 V
0 2677 R
0 -63 V
3640 400 M
0 63 V
0 2677 R
0 -63 V
4350 400 M
0 63 V
0 2677 R
0 -63 V
1.000 UL
LTb
800 400 M
3550 0 V
0 2740 V
-3550 0 V
800 400 L
1.000 UP
1.000 UL
LT0
3907 2617 M
48 0 V
-48 -31 R
0 62 V
48 -62 R
0 62 V
3725 2532 M
47 0 V
-47 -31 R
0 62 V
47 -62 R
0 62 V
-167 -86 R
27 0 V
-27 -31 R
0 62 V
27 -62 R
0 62 V
-51 -50 R
23 0 V
-23 -31 R
0 62 V
23 -62 R
0 62 V
3027 2197 M
17 0 V
-17 -31 R
0 62 V
17 -62 R
0 62 V
-29 -31 R
19 0 V
-19 -31 R
0 62 V
19 -62 R
0 62 V
2643 2003 M
16 0 V
-16 -31 R
0 62 V
16 -62 R
0 62 V
2177 1721 M
10 0 V
-10 -31 R
0 62 V
10 -62 R
0 62 V
1918 1524 M
7 0 V
-7 -31 R
0 62 V
7 -62 R
0 62 V
1716 1350 M
7 0 V
-7 -31 R
0 62 V
7 -62 R
0 62 V
1514 1149 M
7 0 V
-7 -31 R
0 62 V
7 -62 R
0 62 V
3931 2617 CircleF
3749 2532 CircleF
3619 2477 CircleF
3592 2458 CircleF
3035 2197 CircleF
3024 2197 CircleF
2651 2003 CircleF
2182 1721 CircleF
1921 1524 CircleF
1719 1350 CircleF
1518 1149 CircleF
3907 2617 M
48 0 V
-48 -31 R
0 62 V
48 -62 R
0 62 V
3725 2532 M
47 0 V
-47 -31 R
0 62 V
47 -62 R
0 62 V
-167 -86 R
27 0 V
-27 -31 R
0 62 V
27 -62 R
0 62 V
-51 -50 R
23 0 V
-23 -31 R
0 62 V
23 -62 R
0 62 V
3027 2197 M
17 0 V
-17 -31 R
0 62 V
17 -62 R
0 62 V
-29 -31 R
19 0 V
-19 -31 R
0 62 V
19 -62 R
0 62 V
2643 2003 M
16 0 V
-16 -31 R
0 62 V
16 -62 R
0 62 V
2177 1721 M
10 0 V
-10 -31 R
0 62 V
10 -62 R
0 62 V
1918 1524 M
7 0 V
-7 -31 R
0 62 V
7 -62 R
0 62 V
1716 1350 M
7 0 V
-7 -31 R
0 62 V
7 -62 R
0 62 V
1514 1149 M
7 0 V
-7 -31 R
0 62 V
7 -62 R
0 62 V
3931 2617 CircleF
3749 2532 CircleF
3619 2477 CircleF
3592 2458 CircleF
3035 2197 CircleF
3024 2197 CircleF
2651 2003 CircleF
2182 1721 CircleF
1921 1524 CircleF
1719 1350 CircleF
1518 1149 CircleF
0.500 UL
LTb
1358 948 M
71 91 V
78 92 V
84 91 V
90 91 V
98 92 V
105 91 V
113 91 V
123 92 V
133 91 V
146 91 V
161 92 V
179 91 V
202 91 V
233 92 V
279 91 V
352 91 V
499 92 V
stroke
grestore
end
showpage
}}%
\put(2575,50){\makebox(0,0){\Large{$a\surd\sigma$}}}%
\put(100,2020){\makebox(0,0){\Large{$g^2_I(a)N$}}}%
\put(4350,300){\makebox(0,0){\ {$0.5$}}}%
\put(3640,300){\makebox(0,0){\ {$0.4$}}}%
\put(2930,300){\makebox(0,0){\ {$0.3$}}}%
\put(2220,300){\makebox(0,0){\ {$0.2$}}}%
\put(1510,300){\makebox(0,0){\ {$0.1$}}}%
\put(800,300){\makebox(0,0){\ {$0$}}}%
\put(750,3140){\makebox(0,0)[r]{\ \ {$6.5$}}}%
\put(750,2683){\makebox(0,0)[r]{\ \ {$6$}}}%
\put(750,2227){\makebox(0,0)[r]{\ \ {$5.5$}}}%
\put(750,1770){\makebox(0,0)[r]{\ \ {$5$}}}%
\put(750,1313){\makebox(0,0)[r]{\ \ {$4.5$}}}%
\put(750,857){\makebox(0,0)[r]{\ \ {$4$}}}%
\put(750,400){\makebox(0,0)[r]{\ \ {$3.5$}}}%
\end{picture}%
\endgroup

\end	{center}
\vskip 0.15in
\caption{The 't Hooft coupling, defined from the mean-field improved
lattice bare coupling, eqn(\ref{eqn_gI}), as a function of the 
scale $a$ in SU(3).
Shown is the 3-loop perturbative running modified by a $O(a^2)$
lattice correction.}
\label{fig_gfitsu3}
\end 	{figure}

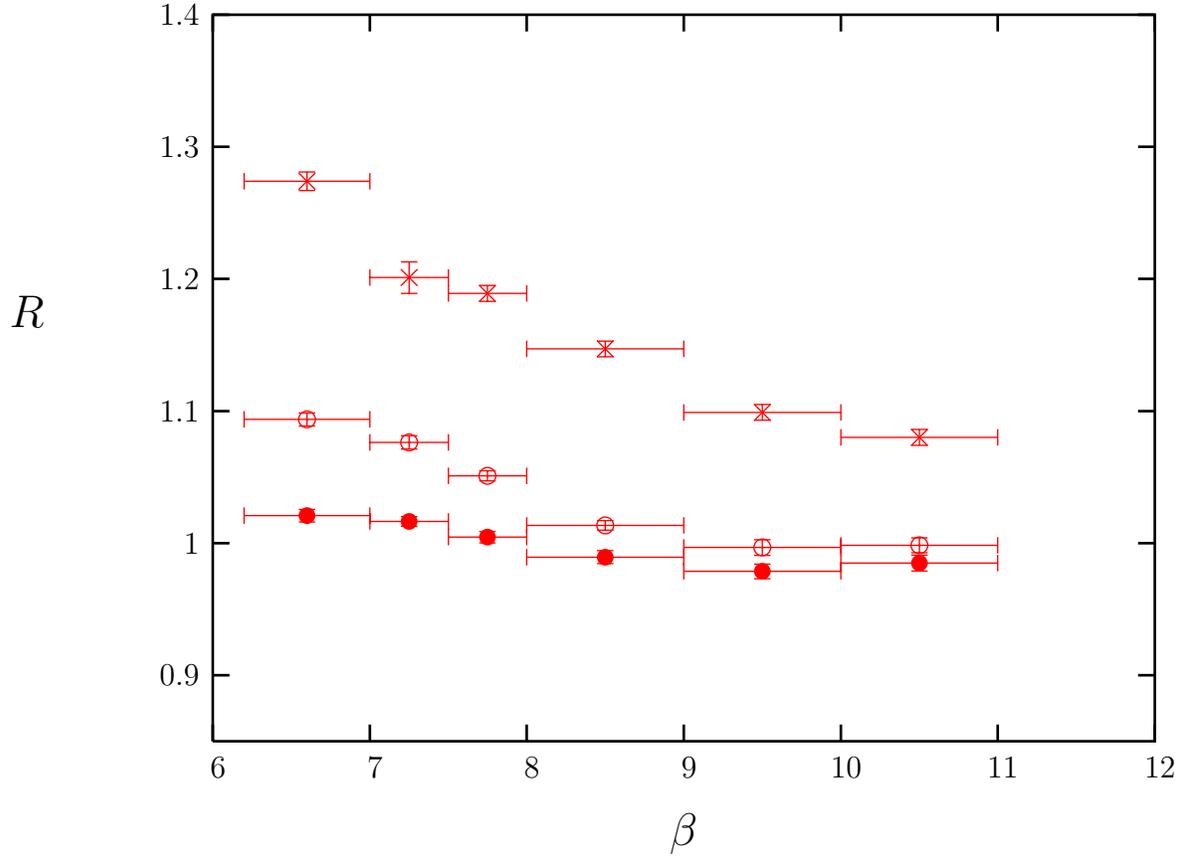
\begin	{figure}[p]
\begin	{center}
\leavevmode
\begingroup%
  \makeatletter%
  \newcommand{\GNUPLOTspecial}{%
    \@sanitize\catcode`\%=14\relax\special}%
  \setlength{\unitlength}{0.1bp}%
{\GNUPLOTspecial{!
/gnudict 256 dict def
gnudict begin
/Color true def
/Solid false def
/gnulinewidth 5.000 def
/userlinewidth gnulinewidth def
/vshift -33 def
/dl {10 mul} def
/hpt_ 31.5 def
/vpt_ 31.5 def
/hpt hpt_ def
/vpt vpt_ def
/M {moveto} bind def
/L {lineto} bind def
/R {rmoveto} bind def
/V {rlineto} bind def
/vpt2 vpt 2 mul def
/hpt2 hpt 2 mul def
/Lshow { currentpoint stroke M
  0 vshift R show } def
/Rshow { currentpoint stroke M
  dup stringwidth pop neg vshift R show } def
/Cshow { currentpoint stroke M
  dup stringwidth pop -2 div vshift R show } def
/UP { dup vpt_ mul /vpt exch def hpt_ mul /hpt exch def
  /hpt2 hpt 2 mul def /vpt2 vpt 2 mul def } def
/DL { Color {setrgbcolor Solid {pop []} if 0 setdash }
 {pop pop pop Solid {pop []} if 0 setdash} ifelse } def
/BL { stroke userlinewidth 2 mul setlinewidth } def
/AL { stroke userlinewidth 2 div setlinewidth } def
/UL { dup gnulinewidth mul /userlinewidth exch def
      10 mul /udl exch def } def
/PL { stroke userlinewidth setlinewidth } def
/LTb { BL [] 0 0 0 DL } def
/LTa { AL [1 udl mul 2 udl mul] 0 setdash 0 0 0 setrgbcolor } def
/LT0 { PL [] 1 0 0 DL } def
/LT1 { PL [4 dl 2 dl] 0 1 0 DL } def
/LT2 { PL [2 dl 3 dl] 0 0 1 DL } def
/LT3 { PL [1 dl 1.5 dl] 1 0 1 DL } def
/LT4 { PL [5 dl 2 dl 1 dl 2 dl] 0 1 1 DL } def
/LT5 { PL [4 dl 3 dl 1 dl 3 dl] 1 1 0 DL } def
/LT6 { PL [2 dl 2 dl 2 dl 4 dl] 0 0 0 DL } def
/LT7 { PL [2 dl 2 dl 2 dl 2 dl 2 dl 4 dl] 1 0.3 0 DL } def
/LT8 { PL [2 dl 2 dl 2 dl 2 dl 2 dl 2 dl 2 dl 4 dl] 0.5 0.5 0.5 DL } def
/Pnt { stroke [] 0 setdash
   gsave 1 setlinecap M 0 0 V stroke grestore } def
/Dia { stroke [] 0 setdash 2 copy vpt add M
  hpt neg vpt neg V hpt vpt neg V
  hpt vpt V hpt neg vpt V closepath stroke
  Pnt } def
/Pls { stroke [] 0 setdash vpt sub M 0 vpt2 V
  currentpoint stroke M
  hpt neg vpt neg R hpt2 0 V stroke
  } def
/Box { stroke [] 0 setdash 2 copy exch hpt sub exch vpt add M
  0 vpt2 neg V hpt2 0 V 0 vpt2 V
  hpt2 neg 0 V closepath stroke
  Pnt } def
/Crs { stroke [] 0 setdash exch hpt sub exch vpt add M
  hpt2 vpt2 neg V currentpoint stroke M
  hpt2 neg 0 R hpt2 vpt2 V stroke } def
/TriU { stroke [] 0 setdash 2 copy vpt 1.12 mul add M
  hpt neg vpt -1.62 mul V
  hpt 2 mul 0 V
  hpt neg vpt 1.62 mul V closepath stroke
  Pnt  } def
/Star { 2 copy Pls Crs } def
/BoxF { stroke [] 0 setdash exch hpt sub exch vpt add M
  0 vpt2 neg V  hpt2 0 V  0 vpt2 V
  hpt2 neg 0 V  closepath fill } def
/TriUF { stroke [] 0 setdash vpt 1.12 mul add M
  hpt neg vpt -1.62 mul V
  hpt 2 mul 0 V
  hpt neg vpt 1.62 mul V closepath fill } def
/TriD { stroke [] 0 setdash 2 copy vpt 1.12 mul sub M
  hpt neg vpt 1.62 mul V
  hpt 2 mul 0 V
  hpt neg vpt -1.62 mul V closepath stroke
  Pnt  } def
/TriDF { stroke [] 0 setdash vpt 1.12 mul sub M
  hpt neg vpt 1.62 mul V
  hpt 2 mul 0 V
  hpt neg vpt -1.62 mul V closepath fill} def
/DiaF { stroke [] 0 setdash vpt add M
  hpt neg vpt neg V hpt vpt neg V
  hpt vpt V hpt neg vpt V closepath fill } def
/Pent { stroke [] 0 setdash 2 copy gsave
  translate 0 hpt M 4 {72 rotate 0 hpt L} repeat
  closepath stroke grestore Pnt } def
/PentF { stroke [] 0 setdash gsave
  translate 0 hpt M 4 {72 rotate 0 hpt L} repeat
  closepath fill grestore } def
/Circle { stroke [] 0 setdash 2 copy
  hpt 0 360 arc stroke Pnt } def
/CircleF { stroke [] 0 setdash hpt 0 360 arc fill } def
/C0 { BL [] 0 setdash 2 copy moveto vpt 90 450  arc } bind def
/C1 { BL [] 0 setdash 2 copy        moveto
       2 copy  vpt 0 90 arc closepath fill
               vpt 0 360 arc closepath } bind def
/C2 { BL [] 0 setdash 2 copy moveto
       2 copy  vpt 90 180 arc closepath fill
               vpt 0 360 arc closepath } bind def
/C3 { BL [] 0 setdash 2 copy moveto
       2 copy  vpt 0 180 arc closepath fill
               vpt 0 360 arc closepath } bind def
/C4 { BL [] 0 setdash 2 copy moveto
       2 copy  vpt 180 270 arc closepath fill
               vpt 0 360 arc closepath } bind def
/C5 { BL [] 0 setdash 2 copy moveto
       2 copy  vpt 0 90 arc
       2 copy moveto
       2 copy  vpt 180 270 arc closepath fill
               vpt 0 360 arc } bind def
/C6 { BL [] 0 setdash 2 copy moveto
      2 copy  vpt 90 270 arc closepath fill
              vpt 0 360 arc closepath } bind def
/C7 { BL [] 0 setdash 2 copy moveto
      2 copy  vpt 0 270 arc closepath fill
              vpt 0 360 arc closepath } bind def
/C8 { BL [] 0 setdash 2 copy moveto
      2 copy vpt 270 360 arc closepath fill
              vpt 0 360 arc closepath } bind def
/C9 { BL [] 0 setdash 2 copy moveto
      2 copy  vpt 270 450 arc closepath fill
              vpt 0 360 arc closepath } bind def
/C10 { BL [] 0 setdash 2 copy 2 copy moveto vpt 270 360 arc closepath fill
       2 copy moveto
       2 copy vpt 90 180 arc closepath fill
               vpt 0 360 arc closepath } bind def
/C11 { BL [] 0 setdash 2 copy moveto
       2 copy  vpt 0 180 arc closepath fill
       2 copy moveto
       2 copy  vpt 270 360 arc closepath fill
               vpt 0 360 arc closepath } bind def
/C12 { BL [] 0 setdash 2 copy moveto
       2 copy  vpt 180 360 arc closepath fill
               vpt 0 360 arc closepath } bind def
/C13 { BL [] 0 setdash  2 copy moveto
       2 copy  vpt 0 90 arc closepath fill
       2 copy moveto
       2 copy  vpt 180 360 arc closepath fill
               vpt 0 360 arc closepath } bind def
/C14 { BL [] 0 setdash 2 copy moveto
       2 copy  vpt 90 360 arc closepath fill
               vpt 0 360 arc } bind def
/C15 { BL [] 0 setdash 2 copy vpt 0 360 arc closepath fill
               vpt 0 360 arc closepath } bind def
/Rec   { newpath 4 2 roll moveto 1 index 0 rlineto 0 exch rlineto
       neg 0 rlineto closepath } bind def
/Square { dup Rec } bind def
/Bsquare { vpt sub exch vpt sub exch vpt2 Square } bind def
/S0 { BL [] 0 setdash 2 copy moveto 0 vpt rlineto BL Bsquare } bind def
/S1 { BL [] 0 setdash 2 copy vpt Square fill Bsquare } bind def
/S2 { BL [] 0 setdash 2 copy exch vpt sub exch vpt Square fill Bsquare } bind def
/S3 { BL [] 0 setdash 2 copy exch vpt sub exch vpt2 vpt Rec fill Bsquare } bind def
/S4 { BL [] 0 setdash 2 copy exch vpt sub exch vpt sub vpt Square fill Bsquare } bind def
/S5 { BL [] 0 setdash 2 copy 2 copy vpt Square fill
       exch vpt sub exch vpt sub vpt Square fill Bsquare } bind def
/S6 { BL [] 0 setdash 2 copy exch vpt sub exch vpt sub vpt vpt2 Rec fill Bsquare } bind def
/S7 { BL [] 0 setdash 2 copy exch vpt sub exch vpt sub vpt vpt2 Rec fill
       2 copy vpt Square fill
       Bsquare } bind def
/S8 { BL [] 0 setdash 2 copy vpt sub vpt Square fill Bsquare } bind def
/S9 { BL [] 0 setdash 2 copy vpt sub vpt vpt2 Rec fill Bsquare } bind def
/S10 { BL [] 0 setdash 2 copy vpt sub vpt Square fill 2 copy exch vpt sub exch vpt Square fill
       Bsquare } bind def
/S11 { BL [] 0 setdash 2 copy vpt sub vpt Square fill 2 copy exch vpt sub exch vpt2 vpt Rec fill
       Bsquare } bind def
/S12 { BL [] 0 setdash 2 copy exch vpt sub exch vpt sub vpt2 vpt Rec fill Bsquare } bind def
/S13 { BL [] 0 setdash 2 copy exch vpt sub exch vpt sub vpt2 vpt Rec fill
       2 copy vpt Square fill Bsquare } bind def
/S14 { BL [] 0 setdash 2 copy exch vpt sub exch vpt sub vpt2 vpt Rec fill
       2 copy exch vpt sub exch vpt Square fill Bsquare } bind def
/S15 { BL [] 0 setdash 2 copy Bsquare fill Bsquare } bind def
/D0 { gsave translate 45 rotate 0 0 S0 stroke grestore } bind def
/D1 { gsave translate 45 rotate 0 0 S1 stroke grestore } bind def
/D2 { gsave translate 45 rotate 0 0 S2 stroke grestore } bind def
/D3 { gsave translate 45 rotate 0 0 S3 stroke grestore } bind def
/D4 { gsave translate 45 rotate 0 0 S4 stroke grestore } bind def
/D5 { gsave translate 45 rotate 0 0 S5 stroke grestore } bind def
/D6 { gsave translate 45 rotate 0 0 S6 stroke grestore } bind def
/D7 { gsave translate 45 rotate 0 0 S7 stroke grestore } bind def
/D8 { gsave translate 45 rotate 0 0 S8 stroke grestore } bind def
/D9 { gsave translate 45 rotate 0 0 S9 stroke grestore } bind def
/D10 { gsave translate 45 rotate 0 0 S10 stroke grestore } bind def
/D11 { gsave translate 45 rotate 0 0 S11 stroke grestore } bind def
/D12 { gsave translate 45 rotate 0 0 S12 stroke grestore } bind def
/D13 { gsave translate 45 rotate 0 0 S13 stroke grestore } bind def
/D14 { gsave translate 45 rotate 0 0 S14 stroke grestore } bind def
/D15 { gsave translate 45 rotate 0 0 S15 stroke grestore } bind def
/DiaE { stroke [] 0 setdash vpt add M
  hpt neg vpt neg V hpt vpt neg V
  hpt vpt V hpt neg vpt V closepath stroke } def
/BoxE { stroke [] 0 setdash exch hpt sub exch vpt add M
  0 vpt2 neg V hpt2 0 V 0 vpt2 V
  hpt2 neg 0 V closepath stroke } def
/TriUE { stroke [] 0 setdash vpt 1.12 mul add M
  hpt neg vpt -1.62 mul V
  hpt 2 mul 0 V
  hpt neg vpt 1.62 mul V closepath stroke } def
/TriDE { stroke [] 0 setdash vpt 1.12 mul sub M
  hpt neg vpt 1.62 mul V
  hpt 2 mul 0 V
  hpt neg vpt -1.62 mul V closepath stroke } def
/PentE { stroke [] 0 setdash gsave
  translate 0 hpt M 4 {72 rotate 0 hpt L} repeat
  closepath stroke grestore } def
/CircE { stroke [] 0 setdash 
  hpt 0 360 arc stroke } def
/Opaque { gsave closepath 1 setgray fill grestore 0 setgray closepath } def
/DiaW { stroke [] 0 setdash vpt add M
  hpt neg vpt neg V hpt vpt neg V
  hpt vpt V hpt neg vpt V Opaque stroke } def
/BoxW { stroke [] 0 setdash exch hpt sub exch vpt add M
  0 vpt2 neg V hpt2 0 V 0 vpt2 V
  hpt2 neg 0 V Opaque stroke } def
/TriUW { stroke [] 0 setdash vpt 1.12 mul add M
  hpt neg vpt -1.62 mul V
  hpt 2 mul 0 V
  hpt neg vpt 1.62 mul V Opaque stroke } def
/TriDW { stroke [] 0 setdash vpt 1.12 mul sub M
  hpt neg vpt 1.62 mul V
  hpt 2 mul 0 V
  hpt neg vpt -1.62 mul V Opaque stroke } def
/PentW { stroke [] 0 setdash gsave
  translate 0 hpt M 4 {72 rotate 0 hpt L} repeat
  Opaque stroke grestore } def
/CircW { stroke [] 0 setdash 
  hpt 0 360 arc Opaque stroke } def
/BoxFill { gsave Rec 1 setgray fill grestore } def
end
}}%
\begin{picture}(4500,3240)(0,0)%
{\GNUPLOTspecial{"
gnudict begin
gsave
0 0 translate
0.100 0.100 scale
0 setgray
newpath
1.000 UL
LTb
800 649 M
63 0 V
3487 0 R
-63 0 V
800 1147 M
63 0 V
3487 0 R
-63 0 V
800 1645 M
63 0 V
3487 0 R
-63 0 V
800 2144 M
63 0 V
3487 0 R
-63 0 V
800 2642 M
63 0 V
3487 0 R
-63 0 V
800 3140 M
63 0 V
3487 0 R
-63 0 V
800 400 M
0 63 V
0 2677 R
0 -63 V
1392 400 M
0 63 V
0 2677 R
0 -63 V
1983 400 M
0 63 V
0 2677 R
0 -63 V
2575 400 M
0 63 V
0 2677 R
0 -63 V
3167 400 M
0 63 V
0 2677 R
0 -63 V
3758 400 M
0 63 V
0 2677 R
0 -63 V
4350 400 M
0 63 V
0 2677 R
0 -63 V
1.000 UL
LTb
800 400 M
3550 0 V
0 2740 V
-3550 0 V
800 400 L
1.000 UP
1.000 UL
LT0
1155 1227 M
0 47 V
-31 -47 R
62 0 V
-62 47 R
62 0 V
354 -63 R
0 36 V
-31 -36 R
62 0 V
-62 36 R
62 0 V
264 -99 R
0 43 V
-31 -43 R
62 0 V
-62 43 R
62 0 V
413 -121 R
0 49 V
-31 -49 R
62 0 V
-62 49 R
62 0 V
561 -106 R
0 55 V
-31 -55 R
62 0 V
-62 55 R
62 0 V
561 -26 R
0 60 V
-31 -60 R
62 0 V
-62 60 R
62 0 V
918 1251 M
474 0 V
918 1220 M
0 62 V
474 -62 R
0 62 V
0 -53 R
296 0 V
-296 -31 R
0 62 V
296 -62 R
0 62 V
0 -90 R
295 0 V
-295 -31 R
0 62 V
295 -62 R
0 62 V
0 -107 R
592 0 V
-592 -31 R
0 62 V
592 -62 R
0 62 V
0 -84 R
592 0 V
-592 -31 R
0 62 V
592 -62 R
0 62 V
591 0 V
-591 -31 R
0 62 V
591 -62 R
0 62 V
1155 1251 CircleF
1540 1229 CircleF
1835 1170 CircleF
2279 1094 CircleF
2871 1041 CircleF
3463 1072 CircleF
1.000 UP
1.000 UL
LT0
1155 2477 M
0 70 V
-31 -70 R
62 0 V
-62 70 R
62 0 V
354 -458 R
0 119 V
-31 -119 R
62 0 V
-62 119 R
62 0 V
264 -149 R
0 60 V
-31 -60 R
62 0 V
-62 60 R
62 0 V
413 -269 R
0 59 V
-31 -59 R
62 0 V
-62 59 R
62 0 V
561 -298 R
0 59 V
-31 -59 R
62 0 V
-62 59 R
62 0 V
561 -154 R
0 60 V
-31 -60 R
62 0 V
-62 60 R
62 0 V
918 2512 M
474 0 V
918 2481 M
0 62 V
474 -62 R
0 62 V
0 -394 R
296 0 V
-296 -31 R
0 62 V
296 -62 R
0 62 V
0 -91 R
295 0 V
-295 -31 R
0 62 V
295 -62 R
0 62 V
0 -240 R
592 0 V
-592 -31 R
0 62 V
592 -62 R
0 62 V
0 -271 R
592 0 V
-592 -31 R
0 62 V
592 -62 R
0 62 V
0 -125 R
591 0 V
-591 -31 R
0 62 V
591 -62 R
0 62 V
1155 2512 Crs
1540 2149 Crs
1835 2089 Crs
2279 1880 Crs
2871 1640 Crs
3463 1546 Crs
1.000 UP
1.000 UL
LT0
1155 1589 M
0 49 V
-31 -49 R
62 0 V
-62 49 R
62 0 V
354 -136 R
0 50 V
-31 -50 R
62 0 V
-62 50 R
62 0 V
264 -170 R
0 38 V
-31 -38 R
62 0 V
-62 38 R
62 0 V
413 -224 R
0 36 V
-31 -36 R
62 0 V
-62 36 R
62 0 V
561 -131 R
0 59 V
-31 -59 R
62 0 V
-62 59 R
62 0 V
561 -49 R
0 56 V
-31 -56 R
62 0 V
-62 56 R
62 0 V
918 1614 M
474 0 V
918 1583 M
0 62 V
474 -62 R
0 62 V
0 -118 R
296 0 V
-296 -31 R
0 62 V
296 -62 R
0 62 V
0 -157 R
295 0 V
-295 -31 R
0 62 V
295 -62 R
0 62 V
0 -218 R
592 0 V
-592 -31 R
0 62 V
592 -62 R
0 62 V
0 -114 R
592 0 V
-592 -31 R
0 62 V
592 -62 R
0 62 V
0 -23 R
591 0 V
-591 -31 R
0 62 V
591 -62 R
0 62 V
1155 1614 Circle
1540 1527 Circle
1835 1401 Circle
2279 1214 Circle
2871 1131 Circle
3463 1139 Circle
stroke
grestore
end
showpage
}}%
\put(2575,50){\makebox(0,0){\Large{$\beta$}}}%
\put(100,2020){\makebox(0,0){\Large{$R$}}}%
\put(4350,300){\makebox(0,0){\ {$12$}}}%
\put(3758,300){\makebox(0,0){\ {$11$}}}%
\put(3167,300){\makebox(0,0){\ {$10$}}}%
\put(2575,300){\makebox(0,0){\ {$9$}}}%
\put(1983,300){\makebox(0,0){\ {$8$}}}%
\put(1392,300){\makebox(0,0){\ {$7$}}}%
\put(800,300){\makebox(0,0){\ {$6$}}}%
\put(750,3140){\makebox(0,0)[r]{\ \ {$1.4$}}}%
\put(750,2642){\makebox(0,0)[r]{\ \ {$1.3$}}}%
\put(750,2144){\makebox(0,0)[r]{\ \ {$1.2$}}}%
\put(750,1645){\makebox(0,0)[r]{\ \ {$1.1$}}}%
\put(750,1147){\makebox(0,0)[r]{\ \ {$1$}}}%
\put(750,649){\makebox(0,0)[r]{\ \ {$0.9$}}}%
\end{picture}%
\endgroup

\end	{center}
\vskip 0.15in
\caption{Calculated values of 
$R=c_0/(\Lambda_{SF}/\Lambda_{I^\prime})$, where $c_0$
comes from the fit in eqn(\ref{eqn_SFI}).
For the $I^\prime=I$, $\bullet$, $I^\prime=I_3$, $\circ$, and 
the $I^\prime=L$, $\times$, lattice coupling schemes.}
\label{fig_c0}
\end 	{figure}

\begin	{figure}[p]
\begin	{center}
\leavevmode
\begingroup%
  \makeatletter%
  \newcommand{\GNUPLOTspecial}{%
    \@sanitize\catcode`\%=14\relax\special}%
  \setlength{\unitlength}{0.1bp}%
{\GNUPLOTspecial{!
/gnudict 256 dict def
gnudict begin
/Color true def
/Solid false def
/gnulinewidth 5.000 def
/userlinewidth gnulinewidth def
/vshift -33 def
/dl {10 mul} def
/hpt_ 31.5 def
/vpt_ 31.5 def
/hpt hpt_ def
/vpt vpt_ def
/M {moveto} bind def
/L {lineto} bind def
/R {rmoveto} bind def
/V {rlineto} bind def
/vpt2 vpt 2 mul def
/hpt2 hpt 2 mul def
/Lshow { currentpoint stroke M
  0 vshift R show } def
/Rshow { currentpoint stroke M
  dup stringwidth pop neg vshift R show } def
/Cshow { currentpoint stroke M
  dup stringwidth pop -2 div vshift R show } def
/UP { dup vpt_ mul /vpt exch def hpt_ mul /hpt exch def
  /hpt2 hpt 2 mul def /vpt2 vpt 2 mul def } def
/DL { Color {setrgbcolor Solid {pop []} if 0 setdash }
 {pop pop pop Solid {pop []} if 0 setdash} ifelse } def
/BL { stroke userlinewidth 2 mul setlinewidth } def
/AL { stroke userlinewidth 2 div setlinewidth } def
/UL { dup gnulinewidth mul /userlinewidth exch def
      10 mul /udl exch def } def
/PL { stroke userlinewidth setlinewidth } def
/LTb { BL [] 0 0 0 DL } def
/LTa { AL [1 udl mul 2 udl mul] 0 setdash 0 0 0 setrgbcolor } def
/LT0 { PL [] 1 0 0 DL } def
/LT1 { PL [4 dl 2 dl] 0 1 0 DL } def
/LT2 { PL [2 dl 3 dl] 0 0 1 DL } def
/LT3 { PL [1 dl 1.5 dl] 1 0 1 DL } def
/LT4 { PL [5 dl 2 dl 1 dl 2 dl] 0 1 1 DL } def
/LT5 { PL [4 dl 3 dl 1 dl 3 dl] 1 1 0 DL } def
/LT6 { PL [2 dl 2 dl 2 dl 4 dl] 0 0 0 DL } def
/LT7 { PL [2 dl 2 dl 2 dl 2 dl 2 dl 4 dl] 1 0.3 0 DL } def
/LT8 { PL [2 dl 2 dl 2 dl 2 dl 2 dl 2 dl 2 dl 4 dl] 0.5 0.5 0.5 DL } def
/Pnt { stroke [] 0 setdash
   gsave 1 setlinecap M 0 0 V stroke grestore } def
/Dia { stroke [] 0 setdash 2 copy vpt add M
  hpt neg vpt neg V hpt vpt neg V
  hpt vpt V hpt neg vpt V closepath stroke
  Pnt } def
/Pls { stroke [] 0 setdash vpt sub M 0 vpt2 V
  currentpoint stroke M
  hpt neg vpt neg R hpt2 0 V stroke
  } def
/Box { stroke [] 0 setdash 2 copy exch hpt sub exch vpt add M
  0 vpt2 neg V hpt2 0 V 0 vpt2 V
  hpt2 neg 0 V closepath stroke
  Pnt } def
/Crs { stroke [] 0 setdash exch hpt sub exch vpt add M
  hpt2 vpt2 neg V currentpoint stroke M
  hpt2 neg 0 R hpt2 vpt2 V stroke } def
/TriU { stroke [] 0 setdash 2 copy vpt 1.12 mul add M
  hpt neg vpt -1.62 mul V
  hpt 2 mul 0 V
  hpt neg vpt 1.62 mul V closepath stroke
  Pnt  } def
/Star { 2 copy Pls Crs } def
/BoxF { stroke [] 0 setdash exch hpt sub exch vpt add M
  0 vpt2 neg V  hpt2 0 V  0 vpt2 V
  hpt2 neg 0 V  closepath fill } def
/TriUF { stroke [] 0 setdash vpt 1.12 mul add M
  hpt neg vpt -1.62 mul V
  hpt 2 mul 0 V
  hpt neg vpt 1.62 mul V closepath fill } def
/TriD { stroke [] 0 setdash 2 copy vpt 1.12 mul sub M
  hpt neg vpt 1.62 mul V
  hpt 2 mul 0 V
  hpt neg vpt -1.62 mul V closepath stroke
  Pnt  } def
/TriDF { stroke [] 0 setdash vpt 1.12 mul sub M
  hpt neg vpt 1.62 mul V
  hpt 2 mul 0 V
  hpt neg vpt -1.62 mul V closepath fill} def
/DiaF { stroke [] 0 setdash vpt add M
  hpt neg vpt neg V hpt vpt neg V
  hpt vpt V hpt neg vpt V closepath fill } def
/Pent { stroke [] 0 setdash 2 copy gsave
  translate 0 hpt M 4 {72 rotate 0 hpt L} repeat
  closepath stroke grestore Pnt } def
/PentF { stroke [] 0 setdash gsave
  translate 0 hpt M 4 {72 rotate 0 hpt L} repeat
  closepath fill grestore } def
/Circle { stroke [] 0 setdash 2 copy
  hpt 0 360 arc stroke Pnt } def
/CircleF { stroke [] 0 setdash hpt 0 360 arc fill } def
/C0 { BL [] 0 setdash 2 copy moveto vpt 90 450  arc } bind def
/C1 { BL [] 0 setdash 2 copy        moveto
       2 copy  vpt 0 90 arc closepath fill
               vpt 0 360 arc closepath } bind def
/C2 { BL [] 0 setdash 2 copy moveto
       2 copy  vpt 90 180 arc closepath fill
               vpt 0 360 arc closepath } bind def
/C3 { BL [] 0 setdash 2 copy moveto
       2 copy  vpt 0 180 arc closepath fill
               vpt 0 360 arc closepath } bind def
/C4 { BL [] 0 setdash 2 copy moveto
       2 copy  vpt 180 270 arc closepath fill
               vpt 0 360 arc closepath } bind def
/C5 { BL [] 0 setdash 2 copy moveto
       2 copy  vpt 0 90 arc
       2 copy moveto
       2 copy  vpt 180 270 arc closepath fill
               vpt 0 360 arc } bind def
/C6 { BL [] 0 setdash 2 copy moveto
      2 copy  vpt 90 270 arc closepath fill
              vpt 0 360 arc closepath } bind def
/C7 { BL [] 0 setdash 2 copy moveto
      2 copy  vpt 0 270 arc closepath fill
              vpt 0 360 arc closepath } bind def
/C8 { BL [] 0 setdash 2 copy moveto
      2 copy vpt 270 360 arc closepath fill
              vpt 0 360 arc closepath } bind def
/C9 { BL [] 0 setdash 2 copy moveto
      2 copy  vpt 270 450 arc closepath fill
              vpt 0 360 arc closepath } bind def
/C10 { BL [] 0 setdash 2 copy 2 copy moveto vpt 270 360 arc closepath fill
       2 copy moveto
       2 copy vpt 90 180 arc closepath fill
               vpt 0 360 arc closepath } bind def
/C11 { BL [] 0 setdash 2 copy moveto
       2 copy  vpt 0 180 arc closepath fill
       2 copy moveto
       2 copy  vpt 270 360 arc closepath fill
               vpt 0 360 arc closepath } bind def
/C12 { BL [] 0 setdash 2 copy moveto
       2 copy  vpt 180 360 arc closepath fill
               vpt 0 360 arc closepath } bind def
/C13 { BL [] 0 setdash  2 copy moveto
       2 copy  vpt 0 90 arc closepath fill
       2 copy moveto
       2 copy  vpt 180 360 arc closepath fill
               vpt 0 360 arc closepath } bind def
/C14 { BL [] 0 setdash 2 copy moveto
       2 copy  vpt 90 360 arc closepath fill
               vpt 0 360 arc } bind def
/C15 { BL [] 0 setdash 2 copy vpt 0 360 arc closepath fill
               vpt 0 360 arc closepath } bind def
/Rec   { newpath 4 2 roll moveto 1 index 0 rlineto 0 exch rlineto
       neg 0 rlineto closepath } bind def
/Square { dup Rec } bind def
/Bsquare { vpt sub exch vpt sub exch vpt2 Square } bind def
/S0 { BL [] 0 setdash 2 copy moveto 0 vpt rlineto BL Bsquare } bind def
/S1 { BL [] 0 setdash 2 copy vpt Square fill Bsquare } bind def
/S2 { BL [] 0 setdash 2 copy exch vpt sub exch vpt Square fill Bsquare } bind def
/S3 { BL [] 0 setdash 2 copy exch vpt sub exch vpt2 vpt Rec fill Bsquare } bind def
/S4 { BL [] 0 setdash 2 copy exch vpt sub exch vpt sub vpt Square fill Bsquare } bind def
/S5 { BL [] 0 setdash 2 copy 2 copy vpt Square fill
       exch vpt sub exch vpt sub vpt Square fill Bsquare } bind def
/S6 { BL [] 0 setdash 2 copy exch vpt sub exch vpt sub vpt vpt2 Rec fill Bsquare } bind def
/S7 { BL [] 0 setdash 2 copy exch vpt sub exch vpt sub vpt vpt2 Rec fill
       2 copy vpt Square fill
       Bsquare } bind def
/S8 { BL [] 0 setdash 2 copy vpt sub vpt Square fill Bsquare } bind def
/S9 { BL [] 0 setdash 2 copy vpt sub vpt vpt2 Rec fill Bsquare } bind def
/S10 { BL [] 0 setdash 2 copy vpt sub vpt Square fill 2 copy exch vpt sub exch vpt Square fill
       Bsquare } bind def
/S11 { BL [] 0 setdash 2 copy vpt sub vpt Square fill 2 copy exch vpt sub exch vpt2 vpt Rec fill
       Bsquare } bind def
/S12 { BL [] 0 setdash 2 copy exch vpt sub exch vpt sub vpt2 vpt Rec fill Bsquare } bind def
/S13 { BL [] 0 setdash 2 copy exch vpt sub exch vpt sub vpt2 vpt Rec fill
       2 copy vpt Square fill Bsquare } bind def
/S14 { BL [] 0 setdash 2 copy exch vpt sub exch vpt sub vpt2 vpt Rec fill
       2 copy exch vpt sub exch vpt Square fill Bsquare } bind def
/S15 { BL [] 0 setdash 2 copy Bsquare fill Bsquare } bind def
/D0 { gsave translate 45 rotate 0 0 S0 stroke grestore } bind def
/D1 { gsave translate 45 rotate 0 0 S1 stroke grestore } bind def
/D2 { gsave translate 45 rotate 0 0 S2 stroke grestore } bind def
/D3 { gsave translate 45 rotate 0 0 S3 stroke grestore } bind def
/D4 { gsave translate 45 rotate 0 0 S4 stroke grestore } bind def
/D5 { gsave translate 45 rotate 0 0 S5 stroke grestore } bind def
/D6 { gsave translate 45 rotate 0 0 S6 stroke grestore } bind def
/D7 { gsave translate 45 rotate 0 0 S7 stroke grestore } bind def
/D8 { gsave translate 45 rotate 0 0 S8 stroke grestore } bind def
/D9 { gsave translate 45 rotate 0 0 S9 stroke grestore } bind def
/D10 { gsave translate 45 rotate 0 0 S10 stroke grestore } bind def
/D11 { gsave translate 45 rotate 0 0 S11 stroke grestore } bind def
/D12 { gsave translate 45 rotate 0 0 S12 stroke grestore } bind def
/D13 { gsave translate 45 rotate 0 0 S13 stroke grestore } bind def
/D14 { gsave translate 45 rotate 0 0 S14 stroke grestore } bind def
/D15 { gsave translate 45 rotate 0 0 S15 stroke grestore } bind def
/DiaE { stroke [] 0 setdash vpt add M
  hpt neg vpt neg V hpt vpt neg V
  hpt vpt V hpt neg vpt V closepath stroke } def
/BoxE { stroke [] 0 setdash exch hpt sub exch vpt add M
  0 vpt2 neg V hpt2 0 V 0 vpt2 V
  hpt2 neg 0 V closepath stroke } def
/TriUE { stroke [] 0 setdash vpt 1.12 mul add M
  hpt neg vpt -1.62 mul V
  hpt 2 mul 0 V
  hpt neg vpt 1.62 mul V closepath stroke } def
/TriDE { stroke [] 0 setdash vpt 1.12 mul sub M
  hpt neg vpt 1.62 mul V
  hpt 2 mul 0 V
  hpt neg vpt -1.62 mul V closepath stroke } def
/PentE { stroke [] 0 setdash gsave
  translate 0 hpt M 4 {72 rotate 0 hpt L} repeat
  closepath stroke grestore } def
/CircE { stroke [] 0 setdash 
  hpt 0 360 arc stroke } def
/Opaque { gsave closepath 1 setgray fill grestore 0 setgray closepath } def
/DiaW { stroke [] 0 setdash vpt add M
  hpt neg vpt neg V hpt vpt neg V
  hpt vpt V hpt neg vpt V Opaque stroke } def
/BoxW { stroke [] 0 setdash exch hpt sub exch vpt add M
  0 vpt2 neg V hpt2 0 V 0 vpt2 V
  hpt2 neg 0 V Opaque stroke } def
/TriUW { stroke [] 0 setdash vpt 1.12 mul add M
  hpt neg vpt -1.62 mul V
  hpt 2 mul 0 V
  hpt neg vpt 1.62 mul V Opaque stroke } def
/TriDW { stroke [] 0 setdash vpt 1.12 mul sub M
  hpt neg vpt 1.62 mul V
  hpt 2 mul 0 V
  hpt neg vpt -1.62 mul V Opaque stroke } def
/PentW { stroke [] 0 setdash gsave
  translate 0 hpt M 4 {72 rotate 0 hpt L} repeat
  Opaque stroke grestore } def
/CircW { stroke [] 0 setdash 
  hpt 0 360 arc Opaque stroke } def
/BoxFill { gsave Rec 1 setgray fill grestore } def
end
}}%
\begin{picture}(4500,3240)(0,0)%
{\GNUPLOTspecial{"
gnudict begin
gsave
0 0 translate
0.100 0.100 scale
0 setgray
newpath
1.000 UL
LTb
800 400 M
63 0 V
3487 0 R
-63 0 V
800 1085 M
63 0 V
3487 0 R
-63 0 V
800 1770 M
63 0 V
3487 0 R
-63 0 V
800 2455 M
63 0 V
3487 0 R
-63 0 V
800 3140 M
63 0 V
3487 0 R
-63 0 V
800 400 M
0 63 V
0 2677 R
0 -63 V
1983 400 M
0 63 V
0 2677 R
0 -63 V
3167 400 M
0 63 V
0 2677 R
0 -63 V
4350 400 M
0 63 V
0 2677 R
0 -63 V
1.000 UL
LTb
800 400 M
3550 0 V
0 2740 V
-3550 0 V
800 400 L
1.000 UP
1.000 UL
LT0
3758 2363 M
0 24 V
-31 -24 R
62 0 V
-62 24 R
62 0 V
2115 2239 M
0 13 V
-31 -13 R
62 0 V
-62 13 R
62 0 V
-606 -57 R
0 14 V
-31 -14 R
62 0 V
-62 14 R
62 0 V
-442 -60 R
0 17 V
-31 -17 R
62 0 V
-62 17 R
62 0 V
985 2128 M
0 14 V
-31 -14 R
62 0 V
-62 14 R
62 0 V
3758 2375 CircleF
2115 2245 CircleF
1540 2202 CircleF
1129 2158 CircleF
985 2135 CircleF
0.500 UL
LTb
800 2123 M
36 3 V
36 4 V
36 3 V
35 3 V
36 4 V
36 3 V
36 4 V
36 3 V
36 4 V
36 3 V
35 3 V
36 4 V
36 3 V
36 4 V
36 3 V
36 4 V
36 3 V
35 3 V
36 4 V
36 3 V
36 4 V
36 3 V
36 4 V
36 3 V
35 3 V
36 4 V
36 3 V
36 4 V
36 3 V
36 4 V
36 3 V
35 3 V
36 4 V
36 3 V
36 4 V
36 3 V
36 4 V
36 3 V
35 3 V
36 4 V
36 3 V
36 4 V
36 3 V
36 3 V
36 4 V
35 3 V
36 4 V
36 3 V
36 4 V
36 3 V
36 3 V
36 4 V
36 3 V
35 4 V
36 3 V
36 4 V
36 3 V
36 3 V
36 4 V
36 3 V
35 4 V
36 3 V
36 4 V
36 3 V
36 3 V
36 4 V
36 3 V
35 4 V
36 3 V
36 4 V
36 3 V
36 3 V
36 4 V
36 3 V
35 4 V
36 3 V
36 4 V
36 3 V
36 3 V
36 4 V
36 3 V
35 4 V
36 3 V
36 3 V
36 4 V
36 3 V
36 4 V
36 3 V
35 4 V
36 3 V
36 3 V
36 4 V
36 3 V
36 4 V
36 3 V
35 4 V
36 3 V
36 3 V
36 4 V
stroke
grestore
end
showpage
}}%
\put(2575,50){\makebox(0,0){\Large{$1/N^2$}}}%
\put(100,2520){\makebox(0,0){\Large{${\Lambda_{\overline{MS}}}\over{\surd\sigma}$}}}%
\put(4350,300){\makebox(0,0){\ {$0.3$}}}%
\put(3167,300){\makebox(0,0){\ {$0.2$}}}%
\put(1983,300){\makebox(0,0){\ {$0.1$}}}%
\put(800,300){\makebox(0,0){\ {$0$}}}%
\put(750,3140){\makebox(0,0)[r]{\ \ {$0.8$}}}%
\put(750,2455){\makebox(0,0)[r]{\ \ {$0.6$}}}%
\put(750,1770){\makebox(0,0)[r]{\ \ {$0.4$}}}%
\put(750,1085){\makebox(0,0)[r]{\ \ {$0.2$}}}%
\put(750,400){\makebox(0,0)[r]{\ \ {$0$}}}%
\end{picture}%
\endgroup

\end	{center}
\vskip 0.15in
\caption{Calculated values of $\Lambda_{\overline{MS}}/\surd\sigma$
versus $1/N^2$ with a linear extrapolation to $N=\infty$ shown.}
\label{fig_lamN}
\end 	{figure}
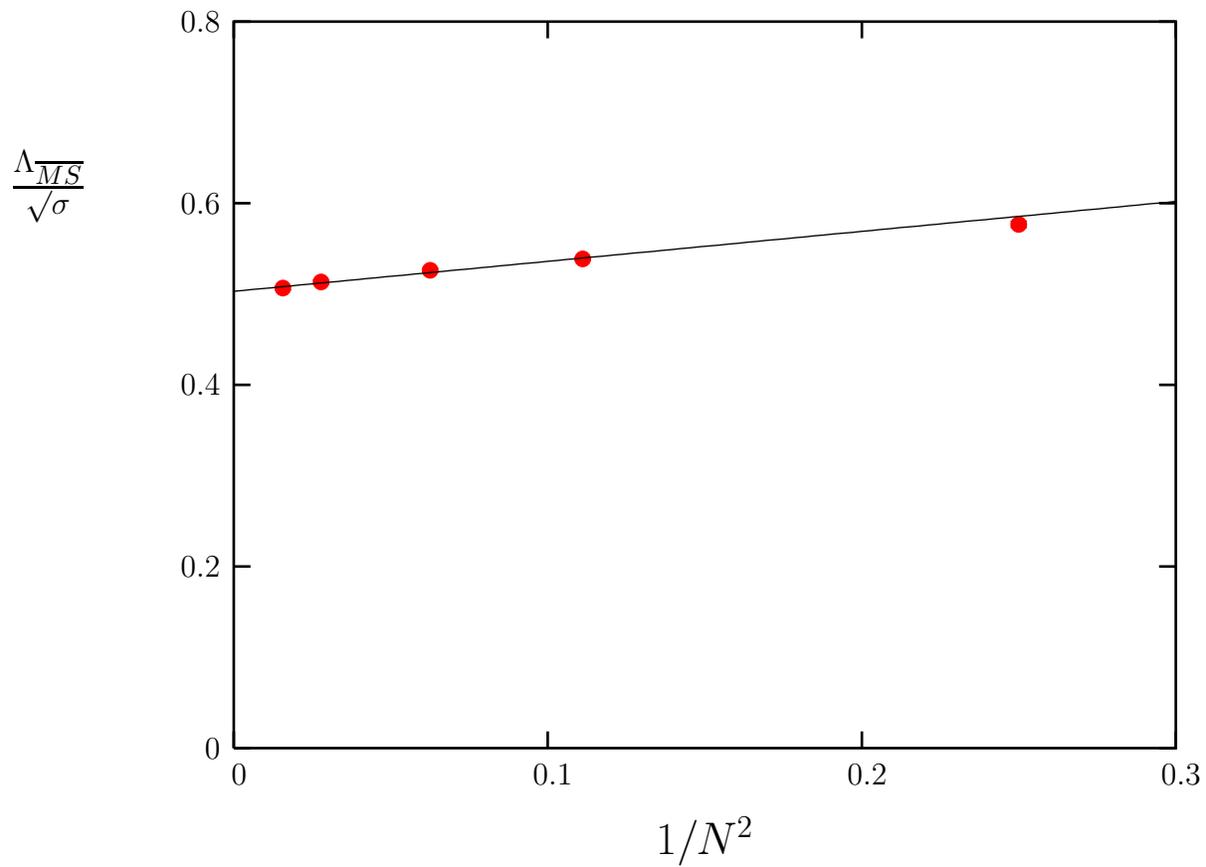

\end{document}